\documentclass[12pt]{article}
\pdfoutput=1 
\usepackage[T1]{fontenc}
\usepackage[utf8]{inputenc}
\usepackage[english]{babel}
\usepackage{hyperref}
\usepackage{siunitx}
\usepackage{multicol}
\usepackage{graphicx}
\usepackage{amsmath}
\usepackage{amssymb}
\usepackage{multirow}
\usepackage{bbm}
\usepackage{bm}
\usepackage{authblk}
\usepackage{color}
\usepackage[round]{natbib}
\usepackage{fancyhdr} 
\usepackage{subfig}
\usepackage{float}
\usepackage{textgreek}
\usepackage{mathtools}
\usepackage{paralist}
\usepackage{adjustbox}
\usepackage{bbm}
\usepackage{booktabs,ctable,threeparttable}
\usepackage{enumitem}
\usepackage{amsthm}

\newcommand{\blind}{0}
\newcommand{\diagdots}[3][-25]{%
	\rotatebox{#1}{\makebox[0pt]{\makebox[#2]{\xleaders\hbox{$\cdot$\hskip#3}\hfill\kern0pt}}}}

\DeclareMathOperator*{\argmaxA}{arg\,max} 

\newtheoremstyle{plains}
{0 pt}   
{0 pt}   
{\itshape}  
{0pt}       
{\bfseries} 
{.}         
{5pt plus 1pt minus 1pt} 
{}          

\theoremstyle{plains} 
\theoremstyle{plains} \newtheorem{proposition}{Proposition}
\theoremstyle{plains} 
\theoremstyle{plains} 
\theoremstyle{plains} \newtheorem{corollary}{Corollary}
\theoremstyle{remark}\newtheorem{remark}{Remark}

\usepackage{paralist} 
\DeclareMathAlphabet{\mathpzc}{OT1}{pzc}{m}{it}

\allowdisplaybreaks

\date{\vspace{-5ex}}
\begin{document}
	\def\spacingset#1{\renewcommand{\baselinestretch}%
		{#1}\small\normalsize} \spacingset{1}
	\if0\blind
	{
		\title{\bf Simultaneous Inference for Empirical Best Predictors with a Poverty Study in Small Areas}
		
		\author{Katarzyna Reluga\thanks{Katarzyna Reluga is a Research Associate at the University of Cambridge, DPMMS, United Kingdom. E-mail: katarzyna.reluga@dpmms.cam.ac.uk.},\; Mar\'{i}a-Jos\'{e} Lombard\'{i}a\thanks{Mar\'{i}a-Jos\'{e} Lombard\'{i}a is a Professor at the University of A Coru\~{n}a, CITIC, Spain. E-mail: maria.jose.lombardia@udc.es.}\; and\;Stefan Sperlich\thanks{Stefan Sperlich is a Professor at the University of Geneva, GSEM,  Switzerland. E-mail: stefan.sperlich@unige.ch.\\ {{The authors gratefully acknowledge support from} the MINECO grants MTM2017-82724-R and MTM2014-52876-R, the Xunta de Galicia (Grupos de Referencia Competitiva ED431C-2016-015 and Centro Singular de Investigaci\'on de Galicia ED431G/01), all of them through the ERDF.\\
		The computations were performed at University of Geneva on the Baobab cluster.
			}}
		}
		\maketitle
	} \fi	
	
	\if1\blind
	{
		\bigskip
		\bigskip
		\bigskip
		\begin{center}
{\LARGE\bf Simultaneous Inference for Empirical Best Predictors with a Poverty Study in Small Areas\par}
		\end{center}
	} \fi
	\begin{abstract}
		\noindent
Today, generalized linear mixed models are broadly used in many fields. However, the development of tools for performing simultaneous inference has been largely neglected in this domain. A framework for joint inference is indispensable to carry out statistically valid multiple comparisons of parameters of interest between all or several clusters. We therefore develop simultaneous confidence intervals and multiple testing procedures for empirical best predictors under generalized linear mixed models. In addition, we implement our methodology to study widely employed examples of mixed models, that is, the unit-level binomial, the area-level Poisson-gamma and the area-level Poisson-lognormal mixed models. The asymptotic results are accompanied by extensive simulations. A case study on predicting poverty rates illustrates applicability and advantages of our simultaneous inference tools.
	\end{abstract}
	
	\noindent%
	
	\vfill
	
	\newpage
	\spacingset{1} 
	
	\section{Introduction}\label{sec:intro}
	
Generalized linear mixed models (GLMM) are suitable for modeling clustered and correlated data with categorical or count outcomes. They are ubiquitous in applied statistics, 
e.g., in biometrics or small area estimation (SAE). In the latter, they serve to analyze surveys on a disaggregated level. 
Despite an increasing interest, e.g., to guide resource allocation, the development of methods for simultaneous inference for predictors is missing. It is surprising as only those would make joint considerations of clusters  valid. Available $(1-\alpha)$-confidence intervals (CI) for mixed parameters \citep[except the credibility intervals of][]{ganesh2009simultaneous} are constructed such that for each study at least $\alpha 100\%$ of them do not contain the true value. Undoubtedly, practitioners do compare, but so far without valid statistical tools. We aim to close this distressing gap, not to improve any existing method. 
	
Specifically, we introduce simultaneous confidence interval (SCI) and multiple test procedure (MTP) for the empirical best predictor (EBP) of \cite{jiang2003empirical}. They are based on max-type statistics combined with extreme value theory. We prove asymptotic convergence of SCI and MTP for nested (or hierarchical) GLMM within the exponential family. We study the numerical performance of our SCI and MTP for two area-level and one unit-level mixed models that are widely used, e.g., for studying local poverty rates \citep{pratesi2016analysis}. All introduced methods show a satisfactory performance within considered modeling frameworks.
Even though our estimates under the area-level models appear to be less volatile, one can argue that EBPs are not directly comparable because different methods and model classes are used. Finally, under area-level Poisson-gamma model, we derive a new mean squared error (MSE) estimator  which is of crucial interest in SAE. 
The amount of literature on estimation and testing under GLMM is considerable, see, i.a., the review of \cite{tuerlinckx2006statistical}, the monograph of \cite{jiang2007linear}, and the article of \cite{ghosh1998} which is particularly interesting within the context of SAE. 
Furthermore, researchers put forward several methodologies broadly used in the analysis of count data. 
\cite{molina2007small} and \cite{scealy2010small} study the estimation of labor force status using multinomial logistic models, whereas \cite{saei2012labour} focus on the same target parameter, and examine the performance of a bivariate random components model. \cite{chandra2012} and \cite{franco2015borrowing} provide extensions for modeling proportions using logistic unit- and area-level models. \cite{hobza2016empirical} implement the EBP for unit-level, and \cite{boubeta2016empirical} for area-level GLMM to study poverty in small areas. \cite{chambers2012m} and \cite{tzavidis2015robust} extend the M-quantile inference for robust estimation and prediction of count data. Yet, to the best of our knowledge, no one addresses the issue of simultaneous inference for clusters-level parameters when applying GLMM. Likewise, little research has been carried out on simultaneous inference for cluster level parameters in linear mixed effects models (LMM). \cite{ganesh2009simultaneous} develops credibility intervals for a mixed parameter in a particular area-level model. \cite{reluga2019} propose bootstrap SCI and MTP for mixed parameter under LMM, whereas \cite{kramlinger2018marginal} develop a framework for marginal and conditional inference with quadratic forms.  
	
After an introduction of model and estimators in Section \ref{sec:model}, we propose the construction of SCI and MTP for a general EBP, followed by the theoretical justifications in Section \ref{sec:SCI_EBP}. Sections \ref{sec:simul} and \ref{sec:data_example} present simulations and a case study. Conclusions are drawn in Section \ref{sec:conc}. 
More details are deferred to Appendix 
and our supplementary material (SM). 

\section{Best Prediction For Generalized Linear Mixed Models}  \label{sec:model}

Let $D$ be the number of clusters or areas with $d\in [D]$, $n_d$ the number of sampled units in each area $j\in [n_d]$ with $n=\sum_{d=1}^{D}n_d$, $N_d$ the known population sizes with $N=\sum_{d=1}^{D}N_d$, $[A]=\{1,\dots,A\}$. Since in our context the notion of cluster and area can be used synonymously, we proceed with the latter. Suppose that $\{v_d:d=1,\dots,D\}$ is a set of independent and identically distributed (i.i.d.) random effects with unknown variance $\delta^2$, $\delta>0$, which 
is often parametrized as $v_d=\delta u_d$ with $u_d\sim N(0,1)$. The target variable $y_{dj}$ represents the $j^{th}$ sample observation from the $d^{th}$ area. Furthermore, we consider nested data structures such that $y_{dj}\neq y_{d'j}$ for $d\neq d'$. In full generalization, we assume that random variables $Y_{dj}$, conditionally on a random effect $u_d$, are independent with a probability density function (p.d.f.) from the exponential family $y_{dj}|u_d\sim Exp. Family (\bm{\theta})$ 
\begin{equation*}\label{eq:GLMM}
y_{dj}|u_d\sim \text{indep. }g_{Y_{dj}|u_d}(y_{dj}|u_d), \quad
g_{dj}(y_{dj}|u_d,\bm{\theta})=\exp\{\varphi^{-1}[y_{dj}\gamma_{dj}-b(\gamma_{dj})] +c(y_{dj},\varphi)\},
\end{equation*}
where $\bm{\theta}=(\bm{\beta}^t,\delta,\varphi)^t$ with $\delta$ the variability parameter, $\bm{\beta}=(\beta_1, \dots, \beta_p)^t$ regression parameters of auxiliary variables $\bm{x}_{dj}=(x_{dj1}, \dots, x_{djp})^t$ for which typically $x_{dj1}=1$, $\forall j \in [n_d]$ $\forall d \in [D]$, and $\gamma_{dj}$, $\varphi$ are canonical and scale parameters respectively. Link function $M$ relates $\mathbb{E}(Y_{dj}|u_d)$ to a linear mixed model such that $\gamma_{dj}=M(\mathbb{E}[Y_{dj}|u_d])=\bm{x}_{dj}^t\bm{\beta}+\delta u_d$. 

\subsection{Estimation and computation}\label{sec:estimation_comp}
Let $\bm{y}_d=(y_{d1}, \dots, y_{dn_d})^t$ for all $d\in [D]$ be the vector of outcomes, and $\bm{y}=(\bm{y}^t_{1}, \dots, \bm{y}^t_{D})^t$. 
A conditional p.d.f.\ of $\bm{y}$ and the likelihood contribution from each area $d$ are given by 
\begin{equation}\label{eq:cont_int}
\mathcal{L}_d(\bm{\theta})\coloneqq f_d(\bm{y}_d|\bm{\theta})=\int g_d(\bm{y}_d|u_d,\bm{\theta})h(u_d)\mathrm{d}u_d=
\int \prod_{j=1}^{n_d}g_{dj}(y_{dj}|u_d,\bm{\theta})h(u_d)\mathrm{d}u_d, 
\end{equation}
where $\bm{\theta}$ can be derived from 
$
\mathcal{L}(\bm{\theta}) \coloneqq  \prod_{d=1}^{D} \mathcal{L}_d(\bm{\theta})=\prod_{d=1}^{D} \int \prod_{j=1}^{n_d}g_{dj}(y_{dj}|u_d,\bm{\theta})h(u_d)\mathrm{d}u_d 
$  .
In case of area-level models, $n_d=1$, and \eqref{eq:cont_int} simplifies accordingly. For a concise presentation, we assume that there is a single random effect for each area such that the integral in \eqref{eq:cont_int} is one-dimensional. Extensions to multidimensional random effects follow immediately with some changes of notations and more complicated computation. Finding an analytical solution to \eqref{eq:cont_int} is difficult unless the integral can be  simplified. Often one evaluates the integral numerically by Laplace approximation (LA) \citep{de1981asymptotic}, Gaussian quadrature (GQ) \citep{naylor1982applications} or adaptive GQ (AGQ) \citep{pinheiro1995approximations}. In what follows, we proceed with AGQ as it is a higher order version of LA, i.e., it gives smaller approximation errors \citep{bianconcini2014asymptotic}. An alternative is the quasi-likelihood \citep{breslow1993approximate} which suffers from a non-decreasing bias  \citep{tuerlinckx2006statistical}, and the method of moments \citep{jiang1998consistent}. In addition, researchers considerably advanced in developing methods to compute maximum likelihood (ML) estimators under GLMM. \cite{jiang2007linear}, Section 4.1, proposes an expectation-maximization algorithm, whereas \cite{lele2010} 
develop a so-called data cloning subsequently implemented by \cite{torabi2012likelihood}. 

Since we consider a prediction problem of possibly non-linear mixed effects $\zeta_d=\zeta_d(\bm{\beta},u_d)$, we use the best predictor (BP) $\tilde{\zeta_d}$ 
in the sense of  minimizing the area-specific MSE in \eqref{eq:MSE_zeta}
which is actually the area-specific mean squared prediction error: 
\begin{equation}\label{eq:EP_general}
\tilde{\zeta}_d= \tilde{\zeta_d}(\bm{\theta})\coloneqq\mathbb{E}\{ \zeta_d(\bm{\beta},u_d)|\bm{y} \} =\mathbb{E}\{ \zeta_d(\bm{\beta},u_d)|\bm{y}_d \}
=\frac{\int \zeta_d(\bm{\beta},u_d)  g_d(\bm{y}_d|u_d,\bm{\theta})h(u_d)\mathrm{d} u_d}{\int g_d(\bm{y}_d|u_d,\bm{\theta})h(u_d)\mathrm{d} u_d}   \ .
\end{equation}
Simplification of \eqref{eq:EP_general} is possible by choosing the p.d.f.\ of $u_d$ accordingly. If we replace $\bm{\theta}$ by a consistent estimator, we obtain the empirical best predictor (EBP) $\hat\zeta_d\coloneqq\tilde{\zeta}_d(\hat{\bm{\theta}})$.  Note that in order to obtain the consistency for random effects, one needs to assume that $n_d\rightarrow \infty $ for each $\hat{\zeta}_d$, $d=1,\dots,D$ \citep{jiang2001empirical}.
	 	
Regarding the estimation of the variability of the EBP $\hat{\zeta}_d$, MSE is by far the most popular measure. Well known techniques are the analytical approximation based on a Taylor expansion \citep{jiang2003empirical}, and parametric bootstrap approaches \citep{boubeta2016empirical,hobza2016empirical}. Consider the following MSE decomposition
\begin{equation}\label{eq:MSE_zeta}
\mathrm{MSE}(\hat{\zeta}_d)=\mathbb{E}[\{\tilde{\zeta}_d(\hat{\bm{\theta}}) -\zeta_d\}^2]=
\mathbb{E}[\{\tilde{\zeta}_d(\hat{\bm{\theta}})-\tilde{\zeta}_d(\bm{\theta})\}^2]
+\mathbb{E}[\{\tilde{\zeta}_d(\bm{\theta})-\zeta_d\}^2]\eqqcolon g_{2d}+g_{1d}  \ ,
\end{equation}
which can be derived applying the law of iterated expectations \citep[for details, see][and our SM]{jiang2003empirical}. The analytical formulas of MSE estimators are model dependent. Bootstrapping permits to obtain estimators that do not vary with the model assumed. In what follows, we denote with $\mathbb{E}^*$, $\mathbb{V}ar^*$ and $\mathrm{MSE}^*$, the corresponding bootstrap operators for expected value, variance and mean squared error and define 
\begin{equation}\label{eq:MSE_zeta_boot}
\mathrm{MSE}^*_{B}(\hat{\zeta}_d)=
\mathbb{E}^*\{(\hat{\zeta}^*_d -\zeta^*_d)^2\} \approx B_1^{-1}\sum_{b_1=1}^{B_1} \left(\hat{\zeta}^{*(b_1)}_d-\zeta^{*(b_1)}_d\right)^2\eqqcolon \mathrm{mse}_{B}(\hat{\zeta}_d),
\end{equation}	
which is a bootstrap equivalent of \eqref{eq:MSE_zeta}. In their paper, \cite{hall2006parametric} point out that \eqref{eq:MSE_zeta_boot} tends to underestimate the MSE, and propose a double-bootstrap bias-correction 
\begin{equation}\label{eq:MSE_bc1}
\mathrm{MSE}^*_{BC}(\hat{\zeta}_d) = 2 \mathrm{MSE}^*_{B}(\hat{\zeta}_d)- \mathrm{MSE}^{*}_{B2}(\hat{\zeta}_d)
\approx 2mse_{B}(\hat{\zeta}_d)-mse_{B2}(\hat{\zeta}_d),
\end{equation}
where $\mathrm{MSE}^{*}_{B2}(\hat{\zeta}_d)$ is a second-stage bootstrap MSE estimator, that is
\begin{equation*}
\mathrm{MSE}^{*}_{B2}(\hat{\zeta}_d)=\mathbb{E}^{**}\{(\hat{\zeta}^{**}_d-\zeta^{**}_d)^2\}\approx B_1^{-1}B_2^{-1}\sum_{b_1=1}^{B_1} \sum_{b_2=1}^{B_2}\left(\hat{\zeta}^{**(b_1b_2)}_d-\zeta^{**(b_1b_2)}_d\right)^2\eqqcolon \mathrm{mse}_{B2}(\hat{\zeta}_d)  \ .
\end{equation*}
The computation of $\mathrm{MSE}^{*}_{B2}(\hat{\zeta}_d)$ involves selecting $B_2$ bootstrap replicates from each first-stage bootstrap. In this article we do not aim for a precise estimation of the variability of EBP, but the construction of narrow SCIs and reliable MTPs. It turns out that for doing this, the use of an estimate of $g_{1d}$ as in \eqref{eq:MSE_zeta} yields better results than using an estimate of the entire MSE (see Section \ref{sec:simul}), similarly as in \cite{chatterjee2008parametric} under LMMs.

\subsection{Popular Examples of GLMM and their properties} \label{sec:Models}

\subsubsection{Poisson-gamma area-level model} \label{sec:Poisson_area}
%
The Poisson-gamma model is widely applied for modeling counts in the presence of overdispersion \citep[see][Section 4.2.2]{cameron2013regression}. Within the SAE context, \cite{chen2015} investigate  observed best prediction and bootstrap MSE estimation for small area mean counts. Among others, they also consider a Poisson-gamma specification. We propose a different model formulation, focus on the EBP of $\zeta_d \coloneqq \mu^{PG}_d$ and develop a plug-in MSE estimator. Let    
$y_d|u_d\sim Poiss(\mu^{PG}_d)$, $d=1,\dots, D$, where $\mu^{PG}_d>0$, $n_d=1$ $\forall d\in [D]$, with canonical parameter $\log \mu^{PG}_d=\bm{x}_d^t\bm{\beta}+u_d$, and $w_d\coloneqq\exp(u_d)\sim Gamma(\delta, \delta)$ such that $\mathbb{E}(y_d|u_d )=\mu_d^{PG}=\lambda_dw_d=\exp(\bm{x}^t_d\bm{\beta})w_d=\exp(\bm{x}^t_d\bm{\beta}+u_d)$. 
Since Gamma p.d.f. is conjugate to the Poisson, their mixture yields a negative binomial $y_d\sim NB(\lambda_d,\delta^{-1})$ with likelihood  
\begin{equation}\label{eq:likelihood_Poisson}
\mathcal{L}^{PG}(\bm{\theta}) \coloneqq 	f^{PG}(\bm{y}|\bm{\theta})= \prod_{d=1}^{D} \frac{\Gamma(y_d+\delta)}{\Gamma(y_d+1) \Gamma(\delta)}\left(\frac{\delta}{\delta+\lambda_d}\right)^{\delta} \left( \frac{\lambda_d}{\delta+\lambda_d} \right)^{y_d},
\end{equation}
where $\mathbb{E}(y_d)=\lambda_d$ and $\mathbb{V}ar(y_d)=\lambda_d+\delta^{-1}\lambda_d^2$. The marginal mean of $y_d$ is the same as in the Poisson case, but the random effect increases the variance. Suppose that this model holds for all areas of population $\mathcal{P}$ of size $N$ partitioned into subpopulations $\mathcal{P}_1, \mathcal{P}_2, \dots, \mathcal{P}_D$ of sizes $N_1, N_2, \dots, N_D$. We can show that the BP for counts $\tilde{\mu}^{PG}_d(\bm{\theta}):=\mathbb{E}(\mu^{PG}_d|y_d)$ is  
\begin{equation}\label{eq:BP_Poisson}
\mathbb{E}(\mu^{PG}_d|y_d)= \frac{\int_{0}^{\infty} \lambda_d w_d g(y_d|w_d)h(w_d) dw_d}
{\int_{0}^{\infty}g(y_d|w_d)h(w_d) dw_d} =\frac{A_d^{PG}(y_d, \bm{\theta})}{C_d^{PG}(y_d,\bm{\theta})}=\frac{\lambda_d(y_d+\delta)}{(\lambda_d+\delta)}\eqqcolon\psi_d^{PG}(y_d,\bm{\theta})  \ . 
\end{equation}
Equation \eqref{eq:BP_Poisson} follows from the conjugation of the Gamma p.d.f. to Poisson p.d.f., while 
\begin{equation*}\label{eq:A_Poisson}
A_d^{PG}=\int_{0}^{\infty}\lambda_d w_d
\frac{\exp(-\lambda_d w_d) \lambda_d^{y_d} w_d^{y_d} \delta^{\delta} w_d^{\delta-1} \exp( -w_d\delta)}{ y_d!\Gamma(\delta)}
dw_d
=\frac{\lambda_d^{y_d+1} \delta^{\delta} \Gamma(y_d+1+\delta)}{\Gamma(\delta)y_d! (\lambda_d +\delta)^{y_d+1+\delta}} \  .
\end{equation*}
The EBP $\hat{\mu}^{PG}_d$ is obtained by replacing the vector of unknown parameters $\bm{\theta}$ in \eqref{eq:BP_Poisson} with a consistent estimator $\hat{\bm{\theta}}$. Under the Poisson-gamma model $\varphi=1$ and $\bm{\theta}=(\bm{\beta}, \delta)$. We derive an analytical plug-in MSE estimator to measure the variability of our EBP. 
\begin{proposition}
Let $\mathbb{V}ar_{d} (\bm{\theta}) =D\ \mathbb{E}\{ ( \hat{\bm{\theta}} -\bm{\theta}) ( \hat{\bm{\theta}} -\bm{\theta})^t \} $. An analytical MSE decomposition with its corresponding practical plug-in estimator are given by
\begin{eqnarray}
\mathrm{MSE}_{PG}(\tilde{\mu}^{PG}_d) &=& g_{PG1d}+\frac{1}{D}c_d(\bm{\theta})+o(1/D)	\quad \text{and} \quad \mathrm{mse}_{PG}(\hat{\mu}^{PG}_d)=\hat{g}_{PG1d}+\frac{1}{D}\hat{c}_d(\hat{\bm{\theta}}), \label{eq:MSE_Poisson}
\\   g_{PG1d}&\coloneqq& \kappa_{1d}(\bm{\theta})-\kappa_{2d}(\bm{\theta}) 
\quad  \hat{g}_{PG1d}=\kappa_{1d}(\hat{\bm{\theta}})- \hat{\kappa}_{2d}(\hat{\bm{\theta}}),\quad d\in [D], \label{eq:g1_Poisson}
\\ \nonumber \mbox{	with } && \kappa_{1d}(\bm{\theta})=\frac{\lambda_d^2 (\delta+1) }{ \delta} \mbox{  and  }\ 
\kappa_{2d}(\bm{\theta})=\sum_{j=0}^{\infty}\frac{\lambda^2_d(j+\delta)^2}{(\lambda_d+\delta)^2}P(y_d=j),
\\ \nonumber \mbox{	as well as  } &&
c_d(\bm{\theta}) =\sum_{j=1}^{\infty} \left\{\frac{\partial}{\partial \bm{\theta}} \psi^{PG}_d(y_d,\bm{\theta})
\right\}^t\mathbb{V}ar_{d}(\bm{\theta})\left\{\frac{\partial}{\partial \bm{\theta}} 
\psi^{PG}_d(y_d,\bm{\theta}) \right\}P(y_d=j). 
\end{eqnarray}
$\hat{c}_d(\bm{\theta})$ is a Monte Carlo approximation of $c_d(\bm{\theta})$, $\hat{\kappa}_{2d}$ refers to $\kappa_{2d}$ with an infinite series truncated at a large term and $\bm{\theta}$ replaced by $\hat{\bm{\theta}}$. To estimate $\kappa_{1d}$ we need only the latter. 
\end{proposition}

One can estimate $\mathbb{V}ar_{d}(\bm{\theta})$ using any reasonable method. In Section \ref{sec:simul} we use bootstrap estimators defined in \eqref{eq:var-boot-est}. Details on the derivation of \eqref{eq:likelihood_Poisson} and \eqref{eq:MSE_Poisson} are deferred to the SM.  

\subsubsection{Poisson-lognormal area-level Model} \label{sec:Poisson_normal}
%
The Poisson-lognormal model has been thoroughly examined by, among others, \cite{cameron2013regression}, Section 4.2.4, \cite{franco2015borrowing} and \cite{boubeta2016empirical}. For $u_d\sim N(0,1)$, let
$y_d|u_d\sim Poiss(\mu^{PL}_d)$, $d=1,\dots, D$, where $\mu^{PL}_d>0$, $n_d=1$ $\forall d\in [D]$, with $\mu^{PL}_d=\nu_d \rho_d$, where $\nu_d$ is a known size variable, and $\rho_d$ a binomial probability. The canonical parameter is $\log \mu^{PL}_d=\log \nu_d + \bm{x}_d\bm{\beta}+\delta u_d$, for all $d=1,\dots, D$. Typically $\zeta_d\coloneqq\rho_d$ for which we have $\rho_d=\exp(\bm{x}_d\beta+\delta u_d)$ with $\bm{\theta}=(\bm{\beta}^t,\delta)$.  The likelihood is 
\begin{equation*}
\mathcal{L}^{PL}(\bm{\theta}) \coloneqq 	f^{PL}(\bm{y}|\bm{\theta})=(2\pi)^{-D/2}\prod_{d=1}^{D}\int_{\mathbb{R}}^{} 
\frac{\exp(-\nu_d\rho_d)\nu_d^{y_d}\exp\{ y_d (\bm{x}_d\bm{\beta} +\delta u_d)\}}{y_d!}
\exp\left(\frac{-u^2_d}{2}\right) du        .
\end{equation*}
Once $\bm{\theta}$ is estimated, we obtain for BPs $\tilde{\mu}^{PL}_d$ and $\tilde{\rho}_d$ the EBPs $\hat{\mu}^{PL}_d$ and $\hat{\rho}_d$, using the formulas from \cite{boubeta2016empirical}. In Section \ref{sec:pois_sim} we estimate its MSE by bootstrap.

\subsubsection{Logit unit-level  model} \label{sec:binomial_unit}
%
%
A unit-level logit model is a popular choice for binary responses, comprehensively discussed   by \cite{hobza2016empirical}. 
Under this setting, $
\  y_{dj}|u_{d}\sim Bin(m_{dj}, p_{dj}), \  u_d\sim N(0,1)
$
with $m_{dj}$ a known size parameter for a logistic regression. The natural parameter is ${p_{dj}} / ( {1-p_{dj}} ) =\bm{x}^t_{dj}\bm{\beta}+\delta u_d$, $d=1, \dots, D$, $j=1,\dots, n_d$, where $p_{dj}=\{\exp(\bm{x}^t_{dj}\bm{\beta}+\delta u_{d})\}/\{1+\exp(\bm{x}^t_{dj}\bm{\beta}+\delta u_{d})\}$. We assume that logit unit-level model holds for all units of population $\mathcal{P}$ of size $N$, partitioned into $D$ subpopulations $\mathcal{P}_d$ of sizes $N_d$, $d=1,\dots,D$. Let $\zeta_d \coloneqq \mu^U_d=\sum_{j=1}^{N_d} p_{dj}$. As for the Poisson models, we have $\varphi=1$, and therefore $\bm{\theta}=(\bm{\beta}^t,\delta)$. The likelihood is given by
\begin{eqnarray} \nonumber 
\mathcal{L}^U(\bm{\theta})&\coloneqq& f^U(\bm{y}|\bm{\beta}, \delta)
=(2\pi)^{-D/2}\prod_{d=1}^{D}\int_{\mathbb{R}}^{} 
\exp\left[ \sum_{j=1}^{n_d}\log{m_{dj}\choose y_{dj}} +\sum_{j=1}^{n_d} y_{dj}(  \bm{x}_{dj}^t\bm{\beta}+\delta u_{d}   ) \right.   \\  \label{eq:log_bin2}
& &\left.- \frac{u_{d}^2}{2}- \sum_{j=1}^{n_d}m_{dj}\log\left\{
1+\exp\left( \bm{x}_{dj}^t\bm{\beta}+\delta u_d   \right)   \right\} \right]\mathrm{d}u_d  . 
\end{eqnarray} 

We can proceed with the estimation of the BP $\tilde{p}_{dj}(\bm{\theta})$ and  $\tilde{\mu}^U_d=\sum_{j=1}^{N_d}\tilde{p}_{dj}$ only if we have access to the information on each population unit. In practice, however, the auxiliary information is often available only for the sample units. Then we can still estimate the population quantity of interest by using only categorical covariates following the suggestion of \cite{hobza2016empirical}. Suppose that they take a finite number of values, say $\bm{x}_{dj}\in \{\bm{z}_1,\dots,\bm{z}_L \}$ $\forall d$, $\forall j$ with $\bm{z}_l$ denoting the resulting classes. For these we obtain
\begin{equation}\label{eq:estimates_unit_level_bin}
\bar{\mu}^U_d=\frac{\mu^U_d}{N_d},\quad \mu^U_d=\sum_{j=1}^{N_d}p_{dj}=\sum_{l=1}^{L}N_{dl}r_{dl},\quad \mbox{ with } r_{dl}=\frac{\exp(\bm{z}_l\bm{\beta} +\delta u_d)}{1+\exp(\bm{z}_l\bm{\beta} +\delta u_d)},
\end{equation}
where the $N_{dl}=\#\{l\in \mathcal{P}_d:\bm{x}_{dj}=\bm{z}_l \}$ are known. \cite{hobza2016empirical} derive BP $\tilde{\mu}^U_d(\bm{\theta})$ and EBP $\hat{\mu}^U_d(\hat{\bm{\theta}})$ for $\mu^U_d$ as well as for other quantities in \eqref{eq:estimates_unit_level_bin}. Due to the computational burden of the analytical estimator, we use bootstrap for obtaining the MSE. 
\section{Simultaneous Intervals and Multiple Testing}\label{sec:SCI_EBP}
To construct confidence intervals for ${\zeta}_d$ that account for the effect of estimates from other areas, we need to find a region $\mathcal{I}_{1-\alpha}$ such that $P(\zeta_d \in \mathcal{I}_{1-\alpha}\; \forall d\in[D])=1-\alpha$. 
Define 
\begin{eqnarray}\label{eq:s}
	S_{0}=\max_{d=1,\dots,D}\left\lvert  S_{0d}\right\rvert , \quad 
	\mbox{ with } \ S_{0d}=\frac{\hat{\zeta}_d-\zeta_d}{\hat{\sigma}(\hat{\zeta}_d)},\quad \forall d\in [D], 
	&& \\ \label{eq:def_quant}    \qquad  \mbox{ and }\
	q_{S_0}^{(1-\alpha)}\coloneqq \inf\{t\in \mathbb{R}: P(S_0\leqslant t)\geqslant 1 -\alpha\} , 
\end{eqnarray}
with $\hat{\sigma}(\hat{\zeta}_d)$ being an estimate of the variability of EBP. We then consider
\begin{equation}\label{eq:S_proba}
	{ } \hspace{-0.4cm}
	\alpha=P\left(\left\lvert \hat{\zeta}_d-\zeta_d \right\rvert > q_{S_0}^{(1-\alpha)} \hat{\sigma}(\hat{\zeta}_d)\; \text{ for some } d\in[D]\right)
	=P\left(\max_{d=1,\dots, D}\left\lvert  \frac{\hat{\zeta}_d-
		\zeta_d}{\hat{\sigma}(\hat{\zeta}_d)} \right\rvert >q_{S_0}^{(1-\alpha)} \right) .
\end{equation}
Constructing SCI boils down to the estimation of $q_{S_0}^{(1-\alpha)}$, as one can define then 
\begin{equation}\label{eq:unif_band_S}
	\mathcal{I}^{S}_{1-\alpha}=
	\bigtimes_{d=1}^D\mathcal{I}^{S}_{d,1-\alpha}, \quad\text{with}\quad\mathcal{I}^{S}_{d,1-\alpha}= \left\{\hat{\zeta}_d \pm q_{S_0}^{(1-\alpha)}\times \hat{\sigma}(\hat{\zeta}_d)\right\},
\end{equation}
where $\bigtimes$ denotes a generalized Cartesian product. $\mathcal{I}^{S}_{1-\alpha}$ covers all $\zeta_d$ with probability $1-\alpha$, i.e., its joint confidence level is $1-\alpha$. In contrast, for each $q_{S_{0d}}^{(1-\alpha)}$ defined analogously to $q_{S_{0}}^{(1-\alpha)}$, with $S_0$ replaced by $|S_{0d}|$, individual area CI (iCI) are given by 
\begin{equation} \label{iCI}
\mathcal{I}^{iCI}_{d,1-\alpha} =  \left\{\hat{\zeta}_d \pm q_{S_{0d}}^{(1-\alpha)} \times \hat{\sigma}(\hat{\zeta}_d)\right\} \quad \forall d\in [D]   . 
\end{equation}
By construction, iCI does not contain $\zeta_d$ for at least $100\alpha\%$ of all areas.

\begin{remark} \label{remark:simultaneity}
$\mathcal{I}^{iCI}_{d,1-\alpha}$ is designed to cover $\zeta_d$ at individual confidence level. Consequently, 
the joint coverage probability of iCIs decreases in a cumulative way for increasing $D$. This 
highlights the need to construct SCI.  
Nevertheless, maintaining $1-\alpha$ simultaneous confidence level of SCI $\mathcal{I}^S_{1-\alpha}$ makes its constituents $\mathcal{I}^{S}_{d,1-\alpha}$ wider than corresponding iCIs $\mathcal{I}^{iCI}_{d,1-\alpha}$. 
This is not surprising because  $\mathcal{I}^{iCI}_{d,1-\alpha}$ and $\mathcal{I}^{S}_{d,1-\alpha}$ were constructed to cover different sets which serve distinct inferential purposes. 
It is worth mentioning that the length of $\mathcal{I}^{S}_{d,1-\alpha}$ stabilizes as for growing $D$ we observe two opposite trends: the increase of area parameters to cover and the decrease of MSE (cf., Table \ref{tab:EPI_RW_Poisson} and Table \ref{tab:EPI_RW_bin}). 
\end{remark} 

The SCI defined in equation \eqref{eq:unif_band_S} is not operational as the distribution of $S_0$ is unknown. The problem can be circumvented by bootstrap approximation: for $b_1=1,\dots,B_1$ set
\begin{equation}\label{eq:S_boot}
S^{(b_1)}_{B}=\max_{d=1,\dots, D}\left\lvert S^{(b_1)}_{Bd} \right\rvert,\quad S^{(b_1)}_{Bd}=\frac{\hat{\zeta}^{*(b_1)}_d-	\zeta^{*(b_1)}_d}{\hat{\sigma}^{*(b_1)}(\hat{\zeta}^*_d)},
\end{equation}
and approximate the critical value $ 
q_{S_B}^{(1-\alpha)}\coloneqq\inf\{t\in \mathbb{R}: P(S_B\leqslant t|(\bm{y},\bm{X}))\geqslant 1 -\alpha\} , 
$
by a $[(1-\alpha)B_1+1]^{th}$ order statistic of the $S^{(b_1)}_{B}$. Then the bootstrap equivalent of \eqref{eq:unif_band_S} is
\begin{equation}\label{eq:SCI_boot_bin}
\mathcal{I}^{B}_{1-\alpha}=
\bigtimes_{d=1}^D\mathcal{I}^{B}_{d,1-\alpha}, \quad\text{where}\quad\mathcal{I}^{B}_{d,1-\alpha}=\left\{\hat{\zeta}_d \pm q_{S_B}^{(1-\alpha)}\times \hat{\sigma}(\hat{\zeta}_d)\right\}.
\end{equation}
An alternative approach to (\ref{eq:s}) could be to take computationally simpler non-studentized statistics. Yet, already \cite{diciccio1996} pointed out that the lack of studentization results in slower convergence rates. Since the application of non-studentized SCI did not yield satisfactory results, we decided not to include them. 

Our methodology is also applicable for hypothesis testing.
Consider the test problem  
\begin{equation}\label{eq:test_proc}
	H_0:\bm{B\zeta}=\bm{b}\quad vs.\quad H_1:\bm{B\zeta}\neq\bm{b}, 
\end{equation}
where $\bm{B}\in \mathbb{R}^{D'\times D}$, $D'\leqslant D$, $\bm{b}\in \mathbb{R}^{D'}$. We are interested in max-type statistic $t_{H}$ such that
\begin{equation}\label{eq:mult_test_quant}
	t_{H}\coloneqq \max_{d=1,\dots, D'}\left \lvert t_{H_d} \right\rvert,\;
	t_{H_d} =	\frac{\hat{\zeta}^H_d-b_d}{\hat{\sigma}(\hat{\zeta}^H_d)},\;
	S_{H_0}\coloneqq	\max_{d=1,\dots,D'}\left\lvert S_{H_0d}  \right\rvert,\;
	S_{H_0d}=\frac{\hat{\zeta}^H_d -
		\zeta^H_d}{\hat{\sigma}(\hat{\zeta}^H_d)},
\end{equation}
where $\bm{\zeta}^H=(\zeta^H_1,\dots,\zeta^H_{D'})^t\coloneqq\bm{B}\bm{\zeta}\in \mathbb{R}^{D'}$ with $\hat{\bm{\zeta}}^H$ being its estimator. One
rejects $H_0$ at the $\alpha$-level if $t_{H}\geqslant q_{H_0}^{(1-\alpha)}$ with $q_{H_0}^{(1-\alpha)}\coloneqq \inf\{t\in \mathbb{R}:P(S_{H_0}\leqslant t)\geqslant 1-\alpha \}$.
In practice we might use such a test to examine differences between area characteristics. Similarly as for SCI, we approximate $q_{H_0}^{(1-\alpha)}$ applying bootstrap to a modified version of statistic $S_B$, namely 
\begin{equation}\label{eq:q_BH}
	q_{BH_0}^{(1-\alpha)}\coloneqq \inf \{ t \in  \mathbb{R}: P (S_{BH_0}\leqslant t |(\bm{y},\bm{X}))\geqslant 1 -\alpha\}
\end{equation}
where $S_{BH_0}$ in the $b_1^{th}$ bootstrap sample is
\begin{equation}\label{eq:S_boot_H}
	S^{(b_1)}_{BH_0}=\max_{d=1,\dots, D'}\left\lvert S^{(b_1)}_{BH_0d} \right\rvert,\quad S^{(b_1)}_{BH_0d}=\frac{\hat{\zeta}^{*H(b_1)}_d-
		\zeta^{*H(b_1)}_d}{\hat{\sigma}^{*}(\hat{\zeta}^{*H(b_1)}_d)},
\end{equation}
with $\bm{\zeta}^{*H(b_1)}=(\zeta_1^{*H(b_1)},\dots,\zeta_D'^{*H(b_1)})   \coloneqq \bm{B}\bm{\zeta}^{*(b_1)}\in \mathbb{R}^{D'}$ and $\hat{\bm{\zeta}}^{*H(b_1)}=(\hat\zeta_1^{*H(b_1)},\dots,\hat\zeta_D'^{*H(b_1)})$ its corresponding estimated version.

We provide the consistency of our bootstrap based confidence intervals and tests, as well as asymptotic convergence and coverage probability. Proofs are deferred to Appendix \ref{sec:App_prop2} and \ref{sec:App_SCI}. Suppose that $\hat{\bm{\theta}}$ is consistent such that $||\hat{\bm{\theta}}-\bm{\theta}||=O_P(n^{-c})$, $c>0$. Since for the GLMM with clustered random effects the log-likelihood can be expressed as the sum of independent random components, the consistency of $\hat{\bm{\theta}}$ estimated by ML follows assuming a classical theory. The consistency under a general GLMM had been an open problem for many years until it was solved by \cite{jiang2013subset}. \cite{bianconcini2014asymptotic} and \cite{huber2004estimation} investigated the consistency of ML once we compute it using AGQ and LA respectively. For our purpose we need to prove the bootstrap consistency:
\begin{proposition}\label{prop:consis_boot_param}
Under Assumptions 1-5 from Appendix \ref{sec:App_RC} it holds that:
\begin{equation*}
\mathbb{E}^*(y^*_{dj})-	\mathbb{E}(y_{dj})=o_{P^*}(1),
\quad
\mathbb{V}ar^*(\bm{y}^*_{d})-	\mathbb{V}ar(\bm{y}_{d})=[o_{P^*}(1)]_{n_d\times n_d}, 
\quad ||\hat{\bm{\theta}}^*-\hat{\bm{\theta}}||=O_{P^*}(n^{-c}).
\end{equation*}
\end{proposition} 
Given Proposition \ref{prop:consis_boot_param}, we can derive the consistency of $\mathcal{I}^B_{1-\alpha}$ based on results from extreme value theory and asymptotic expansions of the standardized statistics using ideas from \cite{chatterjee2008parametric}. Let us assume $\hat{\sigma}(\hat{\zeta}_d)=\sqrt{\hat{g}_{1d}(\hat{\zeta}_d)}$, 
though similar results are immediate for $\hat{\sigma}(\hat{\zeta}_d)=\sqrt{mse_{(\cdot)}(\hat{{\mu}}^{(\cdot)}_d)}$ where $(\cdot)$ stands for different types of estimators.
We use $q\coloneqq q_{S_0}^{(1-\alpha)}$ where unambiguous, and denote the cumulative distribution function (c.d.f.) of $S_{0d}$ and $S_{Bd}$ by $G_d(w)=P(S_{0d}\leqslant w)$ and $G_{Bd}(w)=P(S_{Bd}\leqslant w)$. 
In Appendix \ref{sec:App_SCI} we provide asymptotic expansions for both. Define $(S_{0(d+1)}\dots,S_{0(2D)})=(-S_{01},\dots, -S_{0D})$, and 
observe that $\max\limits_{d=1,\dots,D}\left\lvert  S_{0d}\right\rvert=
\max\limits_{d=1,\dots,2D}( S_{01},\dots, S_{0D}, -S_{01},\dots, -S_{0D})$. From \eqref{eq:S_proba} we get
\begin{equation}\label{eq:S_proba_cont}
	{ } \hspace{-0.4cm}
	\mathcal{T}_D(q)=P\left( S_0  \leqslant q\right)
	=P\left( S_{01}  \leqslant q, \dots,  S_{0D} \leqslant q,  -S_{01}  \leqslant q, \dots,  -S_{0D} \leqslant q \right)
	=\prod_{d=1}^{2D} G_d(q)   .
\end{equation}
As $D\rightarrow \infty$, unless standardized, the distribution in \eqref{eq:S_proba_cont} converges to 0 or 1. In Appendix \ref{sec:App_SCI} we show that $G_{d}(w)$ is asymptotically normal, such that $P\left( S_0  \leqslant q\right)\approx \Phi^{2D}(q)$. As the standard normal distribution is in the domain of attraction of the Gumbel law, it follows that $\lim\limits_{D\rightarrow \infty } \Phi^{2D}(q/b_D+b_D)=\exp (\exp (-q))=\mathcal{T}_0(q)$, for all $q\in \mathbb{R}$ where $b_D$ is a sequence of constants \cite[see Theorem 1.5.3 in][]{leadbetter2012extremes}.
Unfortunately, this approximation has a poor convergence rate, 
but bootstrap is again a remedy. Notice that a similar representation holds for $S_B$, substituting $P$ with $P^*$ and replacing the true parameters by their estimates. Application of Poyla's theorem that combines the convergence in distribution with a convergence in $\sup$ norm results in our next proposition.

\begin{proposition}\label{prop:conv_G_G_boot}
Define $\mathcal{T}^*_D(w)=P^*\left( S_B  \leqslant q\right)$ which is a bootstrap analogue of $\mathcal{T}_D(w)$ in
(\ref{eq:S_proba_cont}). Under Assumptions 1-5 from Appendix \ref{sec:App_RC} it holds that
\begin{equation*}
\sup_{w\in \mathbb{R}}|\mathcal{T}_D(w)-\mathcal{T}^*_D(w)|=o_P(1) \ .
\end{equation*}
\end{proposition}
	
\begin{corollary}
Proposition \ref{prop:conv_G_G_boot} implies that under the same assumptions,
\begin{equation*}
P(\zeta_d\in \mathcal{I}^B_{1-\alpha}\, \forall d \in [D])\xrightarrow [\text{}]{\text{}}1-\alpha \ .
\end{equation*}
\end{corollary}
	
Since we use almost identical max-type statistics in equations (\ref{eq:mult_test_quant}) and (\ref{eq:unif_band_S}), the construction of MTP follows almost immediately from the correspondence between tests and confidence intervals. In fact, the acceptance region of our test is $\mathcal{I}^{H_0}_{1-\alpha}=\bigtimes_{d=1}^D\mathcal{I}^{H_0}_{d,1-\alpha}$, where $\mathcal{I}^{H_0}_{d,1-\alpha}= \left\{\hat{\zeta}_d - q_{S_0}^{(1-\alpha)}\times \hat{\sigma}(\hat{\zeta}_d)\leqslant b_d \leqslant \hat{\zeta}_d + q_{S_0}^{(1-\alpha)}\times \hat{\sigma}(\hat{\zeta}_d) \right\}$, i.e., we reject $H_0$ if $\bm{b}\notin \mathcal{I}^{H_0}_{1-\alpha}$. We can write $P(h_d\in\mathcal{I}^{H_0}_{1-\alpha} \forall d \in [D] )=1-\alpha$. Since this probability statement is true for any $h_d=\zeta_d$, we obtain the confidence intervals defined in \eqref{eq:unif_band_S} by inverting the test. 

\begin{corollary}
Let $H_0$ be the null hypothesis defined in \eqref{eq:test_proc} and $\alpha\in(0,1)$. Under Proposition \ref{prop:conv_G_G_boot}, we have 
\begin{equation*}
P(t_H > q_{BH_0})\leqslant \alpha + o(1) \  .
\end{equation*}
\end{corollary}

\begin{remark}
Our single-step testing procedure in \eqref{eq:test_proc} with a bootstrap critical value in \eqref{eq:q_BH} controls weakly for the family-wise error rate (FWER), and might be limited in detecting false null hypotheses once we deal with a large $D'$. Yet, we can readily extend our test to a bootstrap based step-down procedure of \cite{romano2008control} which controls the false discovery rate with a better power to detect false $H_{0d}$ than FWER. 
\end{remark}

\section{Empirical Reliability Study}\label{sec:simul}

We performed intensive simulation studies to assess the reliability of our methods. SCI and MTP for EBP were constructed with different estimators of variability under the models presented in Sections \ref{sec:Poisson_area}-\ref{sec:binomial_unit}. First, we examined the relative bias and relative root-MSE of fixed effects $\hat{\bm{\beta}}$ and variability parameter $\hat\delta$. Then, the performance of EBP was evaluated comparing bias, average absolute bias and MSE for $D=26, 52, 78$. Since they did not show any atypical pattern, the results under Poisson area-level models and logistic unit-level model were deferred to the SM. Regarding SCIs, we calculated empirical coverage probability (ECP), average interval width (AIW) and the AIW variation (AIWV):  
\begin{eqnarray*}
\text{ECP} &=&\frac{1}{K}\sum_{k=1}^{K}\mathbbm{1}\{\zeta^{(k)}_{d}\in \mathcal{I}^{S}_{1-\alpha} \,\, \forall d \in [D] \} ,
\\
\text{AIW} &=& \frac{1}{DK} \sum_{d=1}^{D}\sum_{k=1}^{K}\omega^{(k)}_d, \quad 
\omega^{(k)}_d=2{q_{(\cdot)}^{(1-\alpha)}}^{(k)} \hat{\sigma}^{(k)}(\hat{\zeta}_d ),
\\ 
\text{AIWV} &=& \frac{1}{D(K-1)} \sum_{d=1}^{D} \sum_{k=1}^{K} 
\left(\omega^{(k)}_d- \bar{\omega}_d\right)^2,\quad  
\bar{\omega}_d=\frac{1}{K}\sum_{k=1}^{K}\omega^{(k)}_d,\quad d=1\dots,D.
\end{eqnarray*}
For each simulation run $k$ we record the widths of the SCI and check whether they cover all EBPs. ECP is then computed by averaging over $K$ simulation runs and is aimed to be close to $1-\alpha$. AIW is obtained by averaging over the simulation runs and areas. Narrower intervals are preferable if its ECP is close to the nominal level. These are standard measures to assess the quality of interval estimators  \citep{chatterjee2008parametric,ganesh2009simultaneous}. 
Lower AIWV values indicate that the length is stable and does not depend on the simulation run. 
 
\subsection{Finite sample performance under area-level models}\label{sec:pois_sim}

Under the Poisson-gamma model we set $y_d\sim Poiss(\mu^{PL}_d)$, $\mu^{PL}_d=\lambda_dw_d$. 
Covariates, parameters and sample sizes are taken from our case study in Section \ref{sec:data_example}, i.e., we set $\bm{\theta}=(\bm{\beta}^t,\delta) =(10.038,7.747,-3.136,11.317,-2.466,2.480)^t$, 
and $D=\{26,52,78\}$, $n_d=1$, $\forall d \in[D]$, $n=D$. For $D=52$ we take covariates from the original sample, for $D=26$ we randomly select the areas using simple random sampling without replacement, and for $D=78$, we take the original sample plus 26 randomly selected areas, i.e., these areas enter at most twice. Parameter of interest is the area proportion of individuals below the poverty line, $\bar{\mu}^{PL}_d=\mu^{PL}_d/N_d$. The EBP for $\mu^{PL}_d$ is given in \eqref{eq:BP_Poisson}. Since $N_d$ is usually unknown, in practice it is replaced by its estimate, see equation \eqref{eq:covariates} in Section \ref{sec:data_example}. We apply double bootstrap with $B_1=1000$ first-stage and $B_2=1$ second-stage bootstrap replicates \citep[the choice of the latter is motivated by][]{erciulescu2014parametric}. We generate $K=1000$ samples with the same areas and fixed covariates, but randomly drawn $w_d$ and $y_{d}$. SCIs and iCIs are constructed as follows:
\begin{enumerate}
	\item Fit the model to the data and obtain estimates 
	$\hat{\bm{\theta}}=(\hat{\bm{\beta}}, \hat{\delta})$.
	\item For $b_1=1,\dots,B_1$ bootstrap samples, generate $w^{*(b_1)}_d\sim Gamma(\hat{\delta},\hat{\delta})$ i.i.d. and set
	\begin{equation*}
	\mu_d^{PG*(b_1)}=\hat{\lambda}_dw_d^{*(b_1)}  \quad \text{and} \quad y^{*(b_1)}_d\sim Poisson(\mu_d^{PG*(b_1)}).
	\end{equation*}
	\item For each bootstrap sample calculate $\hat{\bm{\theta}}^{*(b_1)}$,  $\hat{\mu}^{PG*(b_1)}_d(\hat{\bm{\theta}}^{*(b_1)})$,
	$AD^{(b_1)}_{PG,d}=\left\lvert \hat{{\mu}}^{PG*(b_1)}_d - {\mu}^{PG*(b_1)}_d  \right\rvert$.
	\begin{enumerate}
		\item For $b_2=1,\dots,B_2$ generate samples $w^{**(b_1,b_2)}_d\sim Gamma(\hat{\delta}^{*(b_1)},\hat{\delta}^{*(b_1)})$ i.i.d. and  
		\begin{equation*}
		\mu_d^{PG**(b_1,b_2)}=\hat{\lambda}^{*(b_1)}_dw_d^{**(b_1,b_2)} \quad \text{and} 
		\quad y^{**(b_1,b_2)}_d\sim Poisson(\mu_d^{PG**(b_1,b_2)}).
		\end{equation*}
		\item For each bootstrap sample calculate $\hat{\bm{\theta}}^{**(b_1,b_2)}$ and  $\hat{\mu}^{PG**(b_1,b_2)}_d(\hat{\bm{\theta}}^{**(b_1,b_2)})$.
		\item Set $mse^{(b_1)}_d=\frac{1}{B_2}\sum_{b_2=1}^{B_2}\left(\hat{\mu}^{PG**(b_1,b_2)}_d-\mu^{PG**(b_1,b_2)}_d\right)^2$.
	\end{enumerate}
	\item Calculate bootstrap estimates $\hat g_{PG1d} (\hat{\bm{\theta}}^{*(b_1)})$ as in \eqref{eq:g1_Poisson} as well as 
	\begin{equation*} \label{eq:mse-boot}
	mse_B(\hat{\mu}^{PG}_d)=\frac{1}{B_1}\sum_{b_1=1}^{B_1} \left(\hat{\mu}^{PG*(b_1)}_d-\mu^{PG*(b_1)}_d \right)^2\; 
	, \; mse_{BC}(\hat{\mu}^{PG}_d)=2mse_B(\hat{\mu}^{PG}_d)-\frac{1}{B_1}\sum_{b_1=1}^{B_1}mse^{(b_1)}_d.
	\end{equation*}
	\item Calculate statistic $S_{PG,B}$ with the critical value $q_{PG,S_B}^{(1-\alpha)}$ obtained
	from the bootstrap sample $\bm{S}_{PG,B}=(S^{(1)}_{PG,B},\dots S^{(B_1)}_{PG,B})^t  $, where
	\begin{equation*}\label{crit-value}
	S^{(b_1)}_{PG,B}=\max_{d=1,\dots, D}  
	{AD_{PG,d}^{*(b_1)}} / {\hat{\sigma}^{*(b1)}\left(\hat\mu^{PG*(b1)}_{d}\right) } \quad 
	\text{and} \quad  q_{PG,S_B}^{(1-\alpha)}=Q_{1-\alpha}(\bm{S}_{PG,B})
	\end{equation*}
	as well as a variance estimate for $\hat{\bm{\theta}}$:
	\begin{equation} \label{eq:var-boot-est}
	\widehat{var}(\hat{\bm{\theta}})=\frac{1}{B_1}\sum_{b_1=1}^{B_1}(\hat{\bm{\theta}}^{*(b_1)}-\bar{\bm{\theta}})(\hat{\bm{\theta}}^{*(b_1)}-\bar{\bm{\theta}})^t \  \mbox{ with  } \
	\bar{\bm{\theta}}=\frac{1}{B_1}\sum_{b_1=1}^{B1}\hat{\bm{\theta}}^{*(b_1)}  .
	\end{equation}
\end{enumerate}
We compare the performance of SCI and MTP for different variability estimates $\hat{\sigma}(\hat\mu^{PG}_{d})$ and their bootstrap equivalents $\hat{\sigma}^*(\hat\mu^{*PG}_{d})$, namely for $\hat{\sigma}(\hat\mu^{PG}_{d})=\sqrt{\hat g_{PG1d}}$ and $\hat{\sigma}(\hat\mu^{PG}_{d})= \sqrt{mse_{(\cdot)}(\hat{{\mu}}^{PG}_d)}$. Here, $mse_{(\cdot )}$ refers to either the plug-in $mse_P$, the $mse_B$ or the $mse_{BC}$, defined in \eqref{eq:MSE_Poisson}, \eqref{eq:MSE_zeta_boot} and \eqref{eq:MSE_bc1}. Steps 3(a)-(c) refer to the second-stage bootstrap which is only necessary to obtain bias corrected $mse_{BC}$. Under the Poisson-gamma model, we are interested in the estimation of poverty rates. We thus consider $\hat{\bar{\mu}}^{PG}_d=\hat{\mu}^{PG}_d/N_d$, $\hat{\bar{g}}_{{PG}1d}=\hat{g}_{{PG}1d}/N^2_d$ and $ mse_{(\cdot)}(\hat{\bar{\mu}}^{PG}_d)=mse_{(\cdot)}(\hat{\bar{\mu}}^{PG}_d)/N^2_d$.

For the Poisson-lognormal model with $y_d\sim Poisson (\mu_d^{PL})$, $\mu_d^{PL}=\nu_d \rho_d$, the parameter of interest is $\rho$ with $N_d=\nu_d$, estimated by EBP \citep{boubeta2016empirical}.
We take the fixed parameters from Section \ref{sec:data_example}, i.e., 
$(\bm{\beta}^t,\delta) =(-2.264,3.480,-0.870,4.842,0.125,0.322)^t$. 
Covariates, sample sizes, number of simulation runs and bootstrap replicates are the same as in case of the Poisson-gamma model. The variability of the $\hat{\rho}_d$ was estimated using bootstrap MSEs, that is $mse_B$ and $mse_{BC}$. To obtain estimates of SCI and iCI one can use almost the same algorithm as above by changing the way we generate $y^{*(b_1)}_d$.

\begin{table}[hbt]
\centering
\begin{adjustbox}{max width=\textwidth}
	\begin{tabular}{ccccccccc}\hline
	&  & \multicolumn{4}{c}{Poisson-gamma}                             &  & \multicolumn{2}{c}{Poisson-lognormal}                      \\ \cline{3-6} \cline{8-9} 
	&  & B             & BC            & P             & G             &  & B                            & BC                          \\
	 \cline{3-6} \cline{8-9} 
	&  & \multicolumn{4}{c}{ECP (in \%)}                                       &  & \multicolumn{2}{c}{ECP (in \%)}    \\\hline
		$D=26$ &  & 95.7          & 95.9          & 95.8          & 95.3          &  & 95.0                         & 94.6                        \\
		$D=52$ &  & 94.7          & 93.7          & 94.9          & 94.6          &  & 94.9                         & 94.8                        \\
		$D=78$ &  & 94.7          & 94.8          & 94.9          & 94.4          &  & 94.9                         & 94.6                        \\\hline
		&  & \multicolumn{4}{c}{AIW $\times 10^3$ (AIWV $\times 10^3$)}    &  & \multicolumn{2}{c}{AIW $\times 10^3$ (AIWV $\times 10^3$)} \\\hline
		$D=26$ &  & 24.4 (0.025) & 24.5 (0.028) & 24.5 (0.024) & 23.9 (0.023) &  & 18.0 (0.003)                 & 18.0 (0.004)                \\
		$D=52$ &  & 30.3 (0.030) & 30.3 (0.034) & 30.4 (0.029) & 29.8 (0.028) &  & 20.7 (0.003)                 & 20.8 (0.004)                \\
		$D=78$ &  & 33.4 (0.027) & 33.4 (0.032) & 33.5 (0.025) & 33.0 (0.024) &  & 21.9 (0.002)                 & 22.0 (0.004)        \\\hline       
	\end{tabular}
	\end{adjustbox}  \vspace{-0.25cm}
	\caption{ECP, AIW  and AIWV of SCI under area-level models. Nominal coverage: 95\%.}	
	\label{tab:EPI_RW_Poisson}
\end{table}

Table \ref{tab:EPI_RW_Poisson} summarizes the performance of 95\% SCI for $\hat{\bar{\mu}}^{PG}_d$ constructed with $mse_B$ (B), $mse_{BC}$ (BC), plug-in $mse_P$ (P) and $\hat{g}_{PG1d}$ (G). For $\hat{\rho}_d$, they were constructed using $mse_B$ (B) and $mse_{BC}$ (BC). All methods perform very well regarding the coverage ECP, even for $D=26$.  In contrast, SCIs constructed using a Bonferroni procedure yield unacceptably low ECP. For instance, for $D=52$ and $mse_B$ it equals 78\% for the Poisson-gamma, and 88\% for the Poisson-lognormal model. Therefore we do not further report them. 

\begin{figure}[htb]
	\centering \vspace{-0.0cm}
	\includegraphics[height=0.83\textwidth, angle=-90]{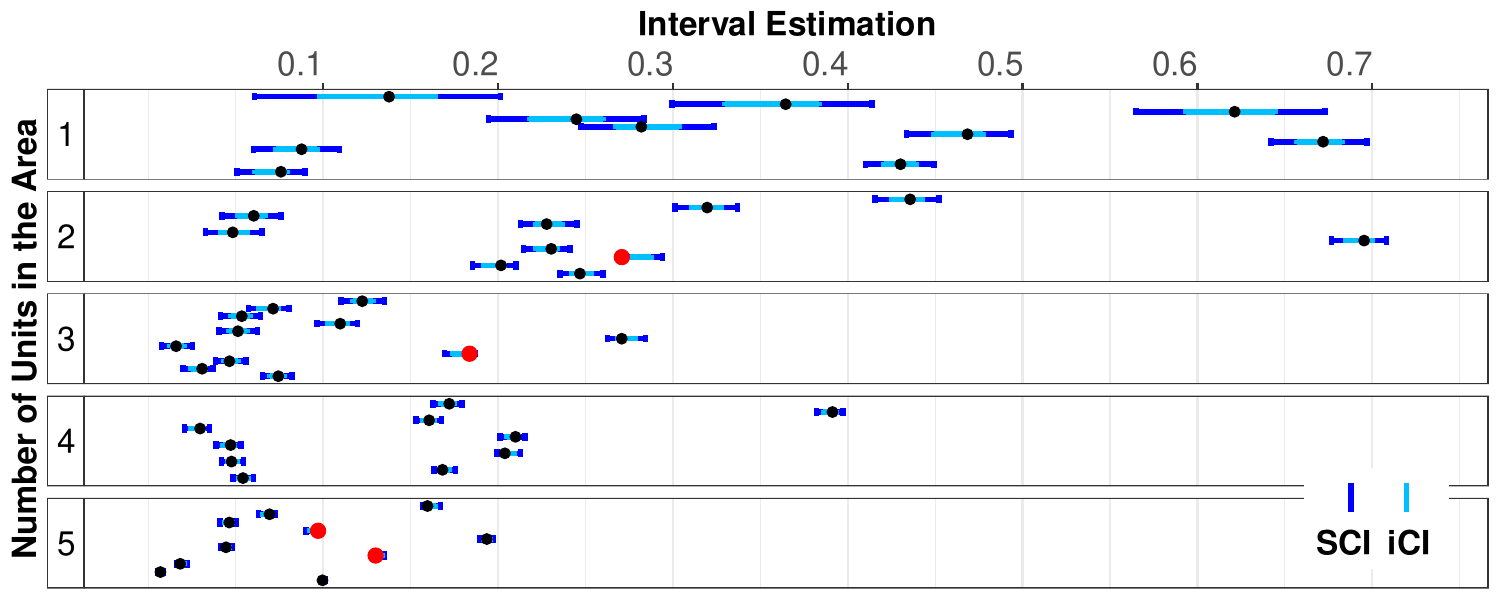}
	\vspace{-0.25cm}
		\captionof{figure}{$95\%$ iCI and SCI for proportions with $D=52$. Red dots indicate true parameters outside iCI, whereas black dots indicate true parameters inside their iCI.} 
	\label{fig:int_est_g1_95}
	\vspace{-0.2cm}
\end{figure}

Figure \ref{fig:int_est_g1_95} presents $95\%$ SCI and iCI estimates for a randomly selected simulation under the Poisson-gamma model. The plot is divided into 5 panels according to the number of units $\hat{N}^{dir}_d$ defined in \eqref{eq:covariates} with the first presenting the results for the areas with the fewest observations. The black and red dots represent the true proportions known in a simulation. The color red indicates true parameters not covered by theirs iCIs. In Figure \ref{fig:int_est_g1_95}, that holds for four of the true values ($\approx 7.7\%$). This illustrates well the difference between individual and simultaneous inference as well as a particular relevance of the latter. We obtain similar figures for the other simulations (see our SM).
\begin{figure}[htb]
\centering \vspace{-0.2cm}
\subfloat{\includegraphics[width=0.33\textwidth]{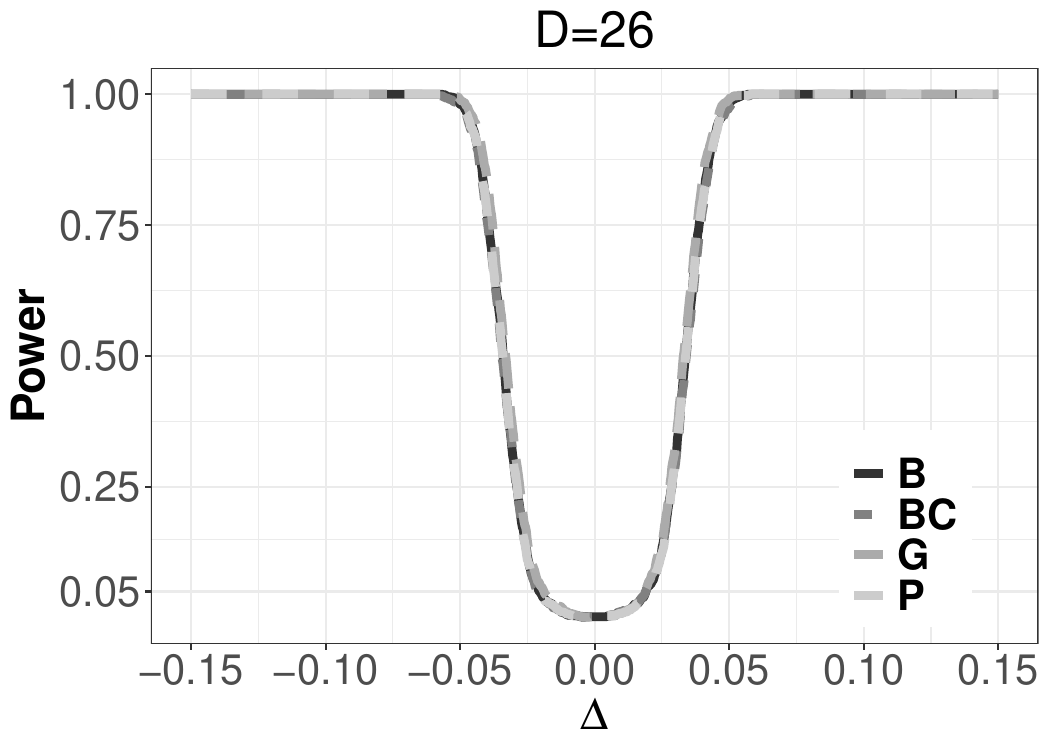}}
\subfloat{\includegraphics[width=0.33\textwidth]{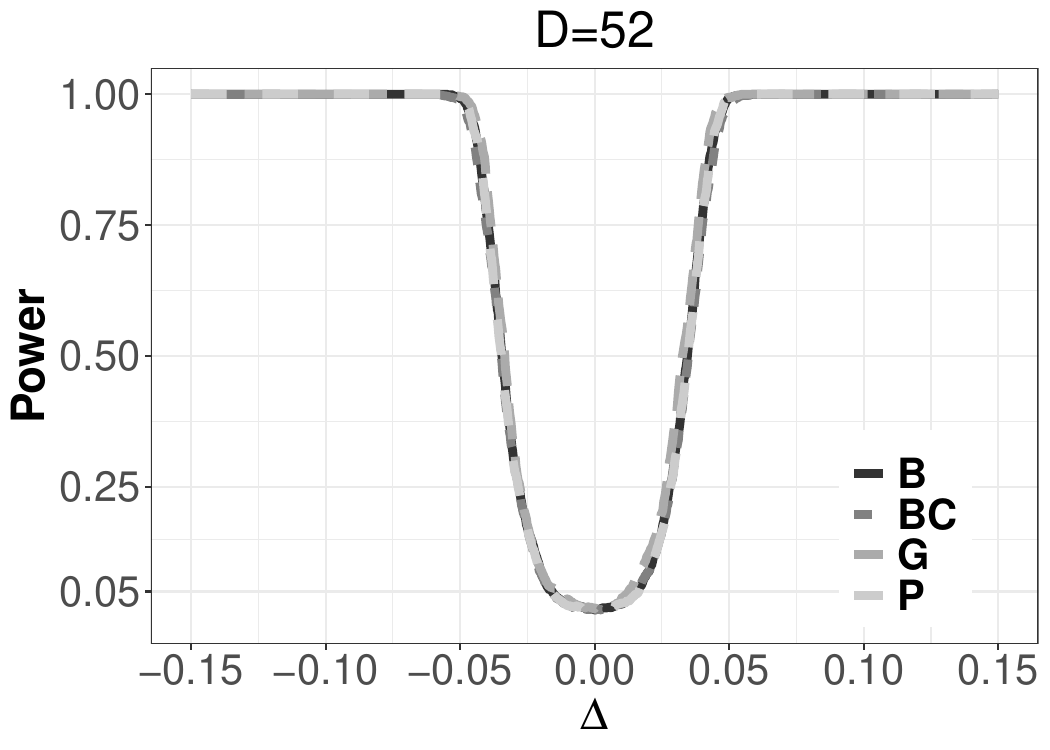}}
\subfloat{\includegraphics[width=0.33\textwidth]{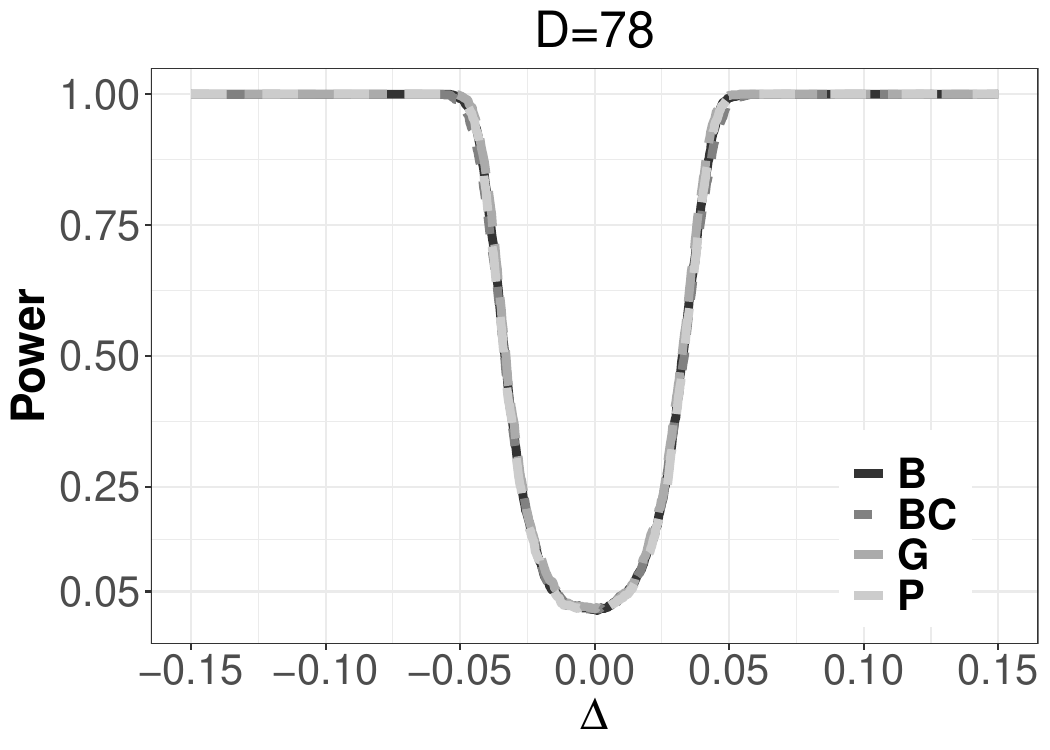}}   
\vspace{-0.25cm}
\captionof{figure}{Simulated powers for multiple test $H_0:\bar{\bm{\mu}}^{PG}=\bm{h}$ versus $H_1:\bar{\bm{\mu}}^{PG}=\bm{h}+\bm{1}_{D}\Delta$ under the area-level Poisson-gamma model; (left) $D=26$, (middle) $D=52$, (right) $D=78$.} 
\label{fig:test_res}
\vspace{-0.2cm}
\end{figure}

Finally we studied the performance of our test \eqref{eq:test_proc} under Poisson-gamma and Poisson-lognormal models. Results for the latter are in our SM as they reveal the same features. Consider $H_0:\bar{\bm{\mu}}^{PG}=\bm{b}$ vs. $H_1:\bar{\bm{\mu}}^{PG}=\bm{b}+\bm{1}_{D}\Delta$, where $\bm{b}\coloneqq\bar{\bm{\mu}}$ for the same data generating processes as before. Critical values 
are obtained from the bootstrap analogues of $S_{H_0}$ calculated similarly as in Step 5 of the algorithm above. Figure \ref{fig:test_res} shows the power functions of our test based on different variability estimates. They are visibly indistinguishable, which is not surprising given the similar ECPs and AIWs in Table \ref{tab:EPI_RW_Poisson}. For $D=52$, i.e., the sample size of the real data, the nominal level of $5\%$ is attained almost exactly under $H_0$.

\subsection{Finite sample performance of SCI under the unit-level model}

Under the unit-level model we assume $y_{dj}\sim \mathrm{Bin}(m_{dj},p_{dj})$ with $p_{dj}=\{\exp(\bm{x}_{dj}^t\bm{\beta}+\delta u_d)\}/\{1+\exp(\bm{x}_{dj}^t\bm{\beta}+\delta u_d)\}$, $m_{dj}=1$, $u_d\sim N(0,1)$. In our context, $y_{dj}$ is binary and value 1 indicates an individual below the poverty threshold defined in Section \ref{sec:data_example}. The regression parameters are taken from our case study: $\bm{\theta}= (\bm{\beta}^t,\delta)=(-2.048, 0.989, 0.172, 0.760, 0.100,0.348)^t$. Four categorical covariates result in 16 covariate classes $\bm{x}_{dj}\in\{\bm{z}_1,\dots,\bm{z}_{16}\}$ for which we need to estimate $N_{dl}$ using equation \eqref{eq:covariates}, $l=1,\dots,16$. We considered $D=\{26, 52, 78\}$  containing unit-level information with $n=\{11423,23628,35818\}$, respectively. Summary statistics for all samples are presented in Table \ref{tab:Statistics_distribution}. Furthermore, for $D=52$, $n_d$, $N_d$, $\bm{x}_{dj}$, $\bm{z}_{l}$ are the same as in our case study. The areas for $D=\{26,78\}$ were selected as in Section \ref{sec:pois_sim}. In addition, for the latter, within each of the additional area we sampled with replacement $n_d$ units (i.e., 26 newly sampled areas contained different units in comparison to the original sample). The parameter of interest is the area poverty proportion $\bar{\mu}^U_d$ defined in \eqref{eq:estimates_unit_level_bin}. Given the original sample size was $n=23628$, under the unit-level model we restrict our simulations to $K=200$, $B_1=500$ and $B_2=1$. As far as the algorithm for constructing SCI and iCI is concerned, it follows almost the same steps as  in Section \ref{sec:pois_sim}. The exact algorithm can be found in the SM. 

\begin{table}[htb]
\centering
\begin{adjustbox}{max width=\textwidth}
\begin{tabular}{cccccccccccccccccccc}\hline
	\multicolumn{6}{c}{$D=26$}                   &  & \multicolumn{6}{c}{$D=52$}                   &  & \multicolumn{6}{c}{$D=78$}                   \\ \cline{1-6} \cline{8-13} \cline{15-20} 
	Min. & $Q_1$ & $Q_2$ & Mean & $Q_3$ & Max. &  & Min. & $Q_1$ & $Q_2$ & Mean & $Q_3$ & Max. &  & Min. & $Q_1$ & $Q_2$ & Mean & $Q_3$ & Max. \\
	47   & 139   & 260   & 439  & 454   & 1714 &  & 34   & 108   & 260   & 454  & 479   & 2631 &  & 34   & 114   & 289   & 459  & 482   & 2631\\\hline
\end{tabular}
\end{adjustbox} \vspace{-0.2cm}
\caption{Summary statistics of $n_d$ under different scenarios in the simulation study.}	\label{tab:Statistics_distribution}
\end{table}

\begin{table}[htb]
	\centering
	\begin{adjustbox}{max width=\textwidth}
		\begin{tabular}{c c cc c cc}\hline
			&
			& \multicolumn{2}{c}{ECP  (in \%)} & & \multicolumn{2}{c}{AIW$\times10^3$ (AIWV$\times10^3$ )} \\\cline{3-4} \cline{6-7}
			D & & B  & BC & & B  & BC           \\\hline
			26 & &94.0  & 93.0 & &134.0 (0.193) & 135.0 (0.268) \\
			52 & &93.5 & 93.0 & &150.3 (0.139) & 150.9 (0.189) \\
			78 & &92.0  & 91.5 && 153.9 (0.125) & 155.1 (0.234) \\\hline         
		\end{tabular}
	\end{adjustbox} \vspace{-0.2cm}
\caption{ECP, AIW  and AIWV of $95\%$ SCI under the unit-level model.}	\label{tab:EPI_RW_bin}
\end{table}

Table \ref{tab:EPI_RW_bin} presents the performance of SCI constructed using $mse_B$ (B) and $mse_{BC}$ (BC). The coverage probability is somewhat lower than the nominal level. In addition, it slightly decreases with increasing $D$, whereas the AIW increase stabilizes as expected (see Remark \ref{remark:simultaneity}). The undercoverage might be related to the simulation design. Even tough the latter is popular in SAE, it is suboptimal for random effects from the asymptotic point of view ($n_d\nrightarrow\infty$, $d\in[D]$, recall Section \ref{sec:estimation_comp}). The results in Table \ref{tab:EPI_RW_bin} do not demonstrate any inconsistencies with respect to the theoretical developments, nor they exhibit unexpected findings. 
Due to their limited impact, the equivalents of Figures \ref{fig:int_est_g1_95} and  \ref{fig:test_res} for this simulation are deferred to our SM. In comparison to the area-level models, the coverage probability is worse and the average width of SCIs is much larger (it is also the case for the iCI, see our SM). Moreover, fitting unit-level models is computationally more expensive. In our case the estimation of MSE and construction of intervals took about 900 to 1000 times longer. Since the data generation processes are different, the numerical results in our simulations are not directly comparable. However, our empirical studies suggest to give some preference to the area-level modeling in the considered GLMM settings. 

Our simulations lead us to following conclusions. First, for a given sample size and data, our SCI attains the nominal coverage probability, almost independently from the choice of the estimator of  variability. In particular, the area-level models yield very accurate results even for small samples. Second, the distinction between SCI and iCI is crucial, and the latter should not be employed in comparative studies. Third, the numerical performance of our test for comparative studies is satisfying. Given the simplicity of SCI and tests based on $\sqrt{\hat g_{1d}}$, we restrict further presentations to them.

\begin{remark}\label{remark:direct_est_simul}
In our simulation study we do not analyze the performance of direct estimators for proportions, because our goal is to study the numerical performance of our MTP and SCIs, and to compare them to existing iCIs.  Since MTP and SCIs are the first tools for simultaneous inference with GLMM-based mixed parameter, we concentrate on their implementation and application to the well-known model-based estimators. These have been thoroughly examined in comparative analyses which included direct estimators \citep[see for instance][]{boubeta2016empirical,hobza2016empirical}. In our case study in Section \ref{sec:data_example}, we include direct estimators in order to have an almost model-free benchmark.
\end{remark}

\section{Predicting Poverty Rates in Galicia}\label{sec:data_example}

\begin{figure}[htb]
\centering
\vspace{-0.2cm}	
\includegraphics[width=0.85\textwidth]{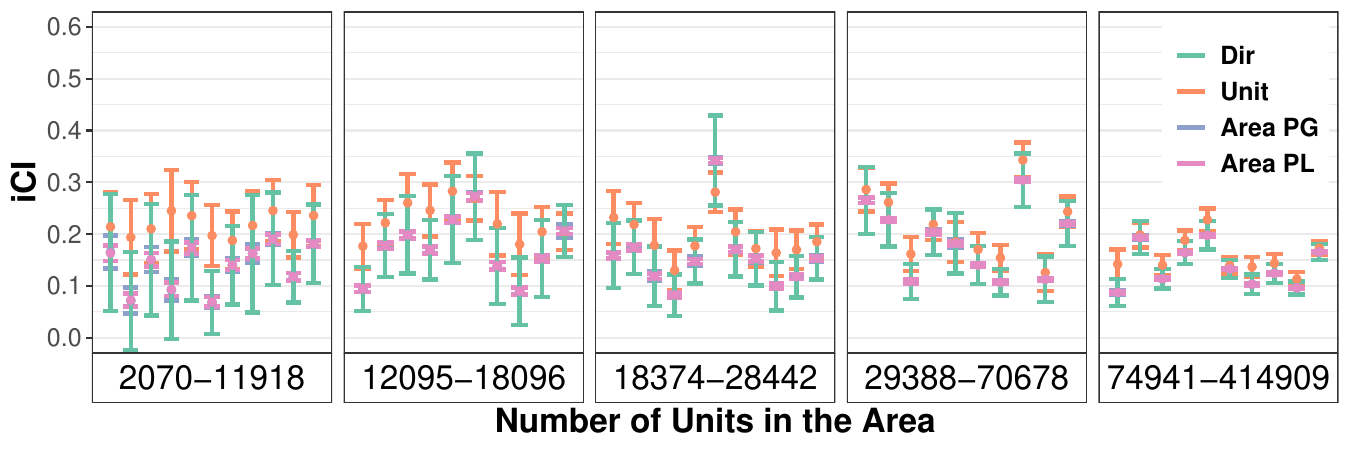}
\vspace{-0.2cm}
\captionof{figure}{Design and model-based $95\%$ iCIs. }
\label{fig:iCI}
\end{figure}

Poverty prediction is of great interest for statistical offices. It provides a basis on which local or central authorities can decide about resource allocation and related polices. The interest is not in individual, randomly chosen small areas but in the total picture. Resource distribution requires comparative statistics, and one would thus provide SCI instead of iCI. We illustrate our methodology calculating point estimates, iCIs and SCIs for the poverty rates in each county of Galicia, i.e., the proportions of inhabitants who live under a poverty line. We make use of a general part of the Structural Survey for Homes (SSH) in Galicia in 2015 with 23628 individuals within 9203 households located in 52 counties (small areas). The survey does not produce official estimates at the area level, but we managed to recover the direct estimates of the totals of people below the poverty line $(Y_{d})$, as well as the number of inhabitants $(N_d)$ for each county. For the area-level models, we need to calculate the number of units which fall into a particular category $(X_{di})$, e.g., number of employees or of graduates in each county of Galicia, $i=1,\dots, p$. The latter are used to obtain the proportions of individuals in each category $\bar{X}_{di}=X_{di}/N_d$. For the unit-level model, we need to obtain the number of units $N_{dl}$ falling into artificially created categories $\bm{z}_{dl}$, $d=1,\dots, D$, $l=1,\dots, L$, see Section \ref{sec:binomial_unit}. The explicit formulas are 
\begin{equation}\label{eq:covariates}
\begin{split}
\hat{Y}^{dir}_d&=\sum_{j\in \mathcal{R}_d} w_{dj}y_{dj},\quad
\hat{N}^{dir}_d=\sum_{j\in \mathcal{R}_d} w_{dj},\quad
\hat{N}^{dir}_{dl}=\sum_{j\in \mathcal{R}_{d}} w_{dj}\bm{x}_{dj}\mathbbm{1}_{\{\bm{x}_{dj}=\bm{z}_l\}},\\
\hat{X}^{dir}_{di}&=\sum_{j\in \mathcal{R}_d} w_{dj}x_{dji},\quad\text{and}\quad
\hat{\bar{X}}^{dir}_{di}=\hat{X}^{dir}_{di}/\hat{N}^{dir}_d,
\end{split}
\end{equation}
where $\mathcal{R}_d\in\mathcal{P}_d$ are the sample elements belonging to area $d$, $d\in [D]$, $w_{dj}$ sampling weights, and $y_{dj}$ binary variables with 1 indicating that an individual is below the poverty line. The poverty threshold is calculated from the survey. It is set to 0.6 of the median household income per capita in Galicia, i.e., we do not use county specific poverty lines. This income is calculated in each household according to scale developed by the Organisation for Economic Co-operation and Development (the same technique is used by Eurostat). The model based approach of this paper assumes that the estimates in equation \eqref{eq:covariates} are considered to be known, non-random quantities, following \cite{lopez2015small}. SSH provides many categorical, auxiliary variables. Under the unit-level model these are binary variables with 1 indicating that a person belongs to a particular category, whereas under area-level models we use the county proportions. We considered four variables for labour status: children (ls0), employed (ls1), unemployed (ls2), inactive (ls3), and four covariates for education: less than primary (ed0), primary (ed1), first and second level secondary (ed2), higher education (ed3). Furthermore, we analyzed three variables for the size of municipality: less than 10 000 (sm1), 10 000-50 000 (sm2), more than 50 000 (sm3). We have also investigated the effect of two variables indicating the nationality, that is, Spanish (n1), not Spanish (n2). Finally, we examined five age variables: $<15$ (age1),  $15-24$ (age2), $25-49$ (age3), $50-64$ (age4), $>=65$ (age5). 
We are interested in $\bar{\mu}^{(\cdot)}_d \coloneqq \mu^{(\cdot)}_d/N_d$ with $(\cdot)$ standing for $PG$ or $U$ in case of Poisson-gamma and binomial model, respectively, and in $\rho_d$ in case of the Poisson-lognormal model. We first compute estimates of proportions and their variances using the same formulas as \cite{boubeta2016empirical}:
\begin{equation}\label{eq:direct_prop}
\hat{p}^{dir}_d=\hat{\bar{Y}}^{dir}_d=\frac{\bar{Y}^{dir}_d}{\hat{N}^{dir}_d}, \quad  \hat{var}(\hat{p}^{dir}_d)=\frac{1}{(\hat{N}^{dir}_d)^2} \sum_{j\in \mathcal{R}_d} w_{dj} (1- w_{dj})\left(y_{dj}-\hat{p}^{dir}_d\right)^2.
\end{equation}
We used estimates in \eqref{eq:direct_prop} to construct design-based iCI intervals (Dir) displayed in Figure \ref{fig:iCI}. Following \cite{lopez2015small}, we then proceed with a variable selection inspired by the simulation results. More specifically, under the Poisson-gamma model we check if any of the levels of categorical variables for labor status, education and age are significant at the $\alpha=0.05$ level.  We examined these covariates in the first place, because they turned out to be important in earlier studies on poverty rates \citep[see, for instance,][]{boubeta2016empirical}. In this way we selected ls2, ed2 and age2. Afterwards, we tested the levels of variables nationality and the size of the municipality and we additionally retained sm1 which was significant after the selection of ls2, ed2 and age2. The same categories were then used to other models, see Table \ref{tab:coef_Galicia}. As we do not carry out a causality analysis, we refrain ourselves from a discussion of the magnitude or signs of estimates. We only notice that under the Poisson-gamma model, the signs are consistent with our expectations; unemployment and young age are associated with higher poverty rates, whereas higher level of studies or living in a small municipality is associated with lower poverty rates.

\begin{table}[h]
\centering     \vspace{-0.1cm}
\begin{adjustbox}{max width=\textwidth}
\begin{tabular}{cccccccccccccccc}\hline
	&  & \multicolumn{4}{c}{Poisson-gamma}             &  & \multicolumn{4}{c}{Poisson-lognormal}                        &  & \multicolumn{4}{c}{Unit-level   model}        \\ \cline{3-6} \cline{8-11} \cline{13-16} \hline
	&  & $\hat{\beta}$ & SE    & $z$-value & $P(>|z|)$ &  & $\hat{\beta}$ & SE & z value & Pr(\textgreater{}|z|) &  & $\hat{\beta}$ & SE    & $z$-value & $P(>|z|)$ \\\hline
	Int  &  & 10.038        & 0.669 & 15.005    & 0.000     &  & -2.264        & 0.341      & -6.633  & 0.000                 &  & -2.048        & 0.067 & -30.415   & 0.000     \\
	ls2  &  & 7.747         & 3.091 & 2.506     & 0.012     &  & 3.480         & 1.577      & 2.207   & 0.027                 &  & 0.989         & 0.052 & 19.160    & 0.000     \\
	ed2  &  & -3.136        & 1.201 & -2.611    & 0.009     &  & -0.8703       & 0.612      & -1.422  & 0.155                 &  & 0.172         & 0.039 & 4.442     & 0.000     \\
	age2 &  & 11.317        & 4.023 & 2.813     & 0.005     &  & 4.842         & 2.057      & 2.354   & 0.019                 &  & 0.760         & 0.058 & 13.033    & 0.000     \\
	sm1  &  & -2.466        & 0.267 & -9.224    & 0.000     &  & 0.125         & 0.136      & 0.918   & 0.358                 &  & 0.100         & 0.050 & 1.993     & 0.046    \\\hline
\end{tabular}
\end{adjustbox}  \vspace{-0.2cm}
\caption{Estimates of regression parameters under the area- and the unit-level models with $\hat{\delta}^{PG}=2.48$, $ \hat{\delta}^{PL}=0.32$ and $\hat{\delta}^U=0.35$, respectively.}
\label{tab:coef_Galicia}
\end{table}


Figure \ref{fig:iCI} shows point and iCI estimates of proportions under Poisson-gamma (PG), Poisson-lognormal (PL) and binomial (Unit) models together with direct estimates (Dir). In this plot we compare point estimates 
within four modeling frameworks; we do not compare them across different areas within the same model. First, the variability reflected by the width of iCIs decreases with the number of units in each area $\hat{N}^{dir}_d$ defined in \eqref{eq:covariates}. Second, even though the area sample sizes $n_d$, $d\in[D]$ are not that small, the iCI of direct estimates are wider than model based estimates, which is in accordance with the literature. The width difference is especially pronounced when comparing area-level-based with design-based direct estimates -- the latter entirely cover the former. Unit-level model-based point and interval estimates are different with much wider iCIs than under area-level models, but still overlapping with the direct estimates. Only in one case (6$^{th}$ area in the third panel), the iCIs under area-level models do not overlap with the iCI under the unit-level model which indicates a possible bias in one of the approaches. In contrast, both area-level models produce almost identical estimates. 

\begin{figure}[htb]
\centering
\vspace{-0.05cm}
\includegraphics[width=0.82\textwidth]{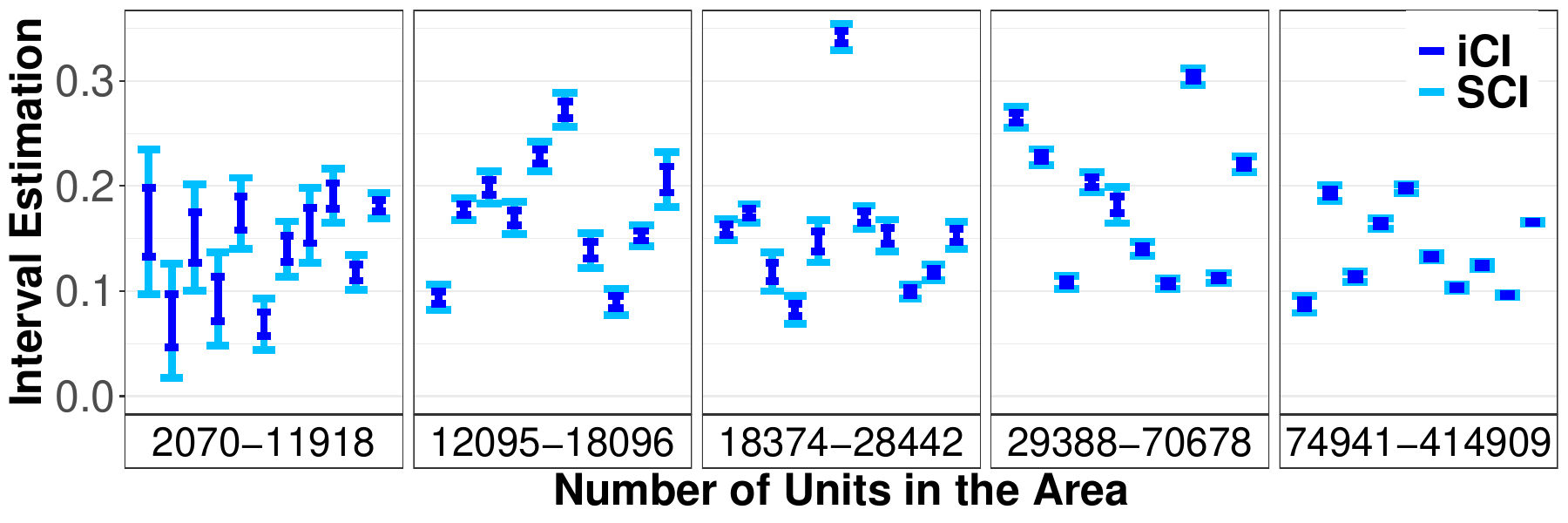}
\vspace*{-1em}
\captionof{figure}{$95\%$ iCI and bootstrap SCI estimates for EBP poverty rates in counties of Galicia.}
\label{fig:plot_order}
\vspace{-0.05cm}
\end{figure}

\begin{figure}[htb]
\vspace*{-0.05cm}
\centering
\includegraphics[width=0.99\textwidth]{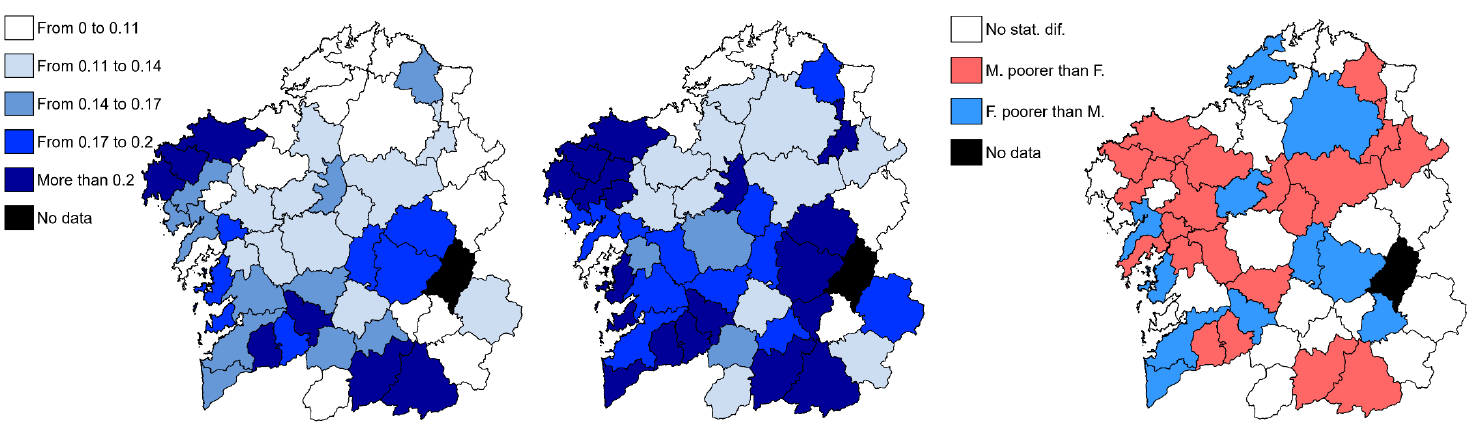}
\vspace*{-0.8em}
\captionof{figure}{SCIs of EBP poverty proportions: (left) lower boundary, (middle) upper boundary; (right) significant differences in poverty rates between women (F) and men (M).} 
\label{fig:galicia_map}
\end{figure}

Figure \ref{fig:plot_order} presents bootstrap iCI and SCI for $\hat{\bar{\mu}}^{PG}_d$, $d=1,\dots, D$ constructed with $\hat \sigma(\hat{\zeta}_d) = \sqrt{\hat g_{PG1d}}$ as defined in \eqref{eq:g1_Poisson}. The plot is divided into five panels according to the numbers of units in each area obtained by direct estimates of county inhabitants $\hat{N}^{dir}_d$ in \eqref{eq:covariates}. Figure \ref{fig:plot_order} serves as an illustration of the differences between individual and simultaneous inference. When comparing iCIs and SCIs, in many cases (e.g., first and  second county of the first panel in Figure \ref{fig:plot_order}) iCI would insinuate statistically different poverty rates, whereas SCI does not confirm this claim. Such multiple comparisons are valid only if we use SCIs. In addition, at least 5\% of the true poverty rates are not contained in their iCIs. Analogous figures under the Poisson-lognormal and binomial models lead to the same conclusions. They are thus deferred to the SM. Further model selection and specification testing might be interesting but they are beyond the scope of this article. 
Since we do not know which model is closer to the real data generating process, we proceed with the Poisson-gamma area-level model, as it is reliable and the least computer intensive. Left and middle panel of Figure \ref{fig:galicia_map} depict the resulting maps of the counties with the corresponding lower and upper bounds of our SCI. We observe a higher rate of poverty in the interior and a south-western part of the region whereas a lower level is typical for the northern part. These conclusions are similar to those drawn by \cite{boubeta2017poisson}. 

Finally, we investigate whether men and women are equally affected by poverty. 
We wish to test for equality on the county level across Galicia. 
Testing for each county individually at $\alpha=5\%$ error level results in rejection of at least 5\% of the hypotheses of no significant difference. 
We thus use our MPT and consider clusters created from the cross-section of sex and county such that ${\bm{\zeta}}\in \mathbb{R}^{104}$. We test $H_0:\bm{B} \bm{\zeta}=\bm{0}_{52}$ versus $H_1:\bm{B} \bm{\zeta}\neq\bm{0}_{52}$ where $\bm{B}\in \mathbb{R}^{52\times104}$ with rows being vectors with $1$ on the $2d-1$ place, $-1$ on $2d$ place, and $0$ elsewhere. The max-type test statistic yields $t_{H} =\max_{d=1,\dots, D}\frac{|\bm{B}\hat{\bm{\zeta}}|}{\hat{\sigma}(\hat{\bm{\zeta}})}\approx 20.489$ while the bootstrap critical value under $H_0$ is $q_{BH_0}^{(1-\alpha)}\approx2.999$. Thus, we strongly reject $H_0$. 
However, our test does not support the hypothesis that women are more affected than men, or vice versa, see the right panel of Figure \ref{fig:galicia_map}. Additional results  are deferred to our SM. 


\begin{remark}
	Imagine that Galician counties were considered as part of a macro region, e.g., Spain with $D_S$ counties, 
	and consider two inferential problems: (a) the calculation of SCIs for the poverty rates in all $D_S$ Spanish counties, (b) the calculation of SCIs only for $D$ Galician counties, but using all data. Following Remark \ref{remark:simultaneity}, we expect that the widths of our SCIs in Figure \ref{fig:plot_order} would increase in case (a) to maintain the joint coverage probability of $95\%$ for all $D_S>D$ counties. In contrast, they would most likely slightly decrease in case (b). In fact, the simultaneous coverage probability of $95\%$ would be requested for the set of $D$ counties, but SCI would be constructed using a more precise estimate of MSE computed using a larger data set with $D_S$ counties.	  
\end{remark}

\section{Conclusions}\label{sec:conc}

We developed a methodology that allows for statistically valid simultaneous inference for EBP under GLMM. We constructed SCI and MTP applying a combination of max-type statistics and consistent bootstrap estimation of its distribution. These tools enable practitioners to make comparisons between areas. In contrast, 
the iCIs are not suitable for such comparative analyses because they are constructed at individual confidence level and disregard an additional variation which arises in joint studies. We do not claim that SCIs and MTP are better than iCIs or t-tests. The former simply complete the toolbox for statistical inference for mixed parameter $\zeta_d$. Similarly, the simultaneous inference completes the individual inference for fixed parameters. 
We introduced various versions of statistics to construct SCI and MTP. Within our framework, all of them exhibited similar performances without indicating a clear winner. 

Our methodology can be extended to more complicated data structures such as GLMM with spatial or temporal correlation \citep[see, e.g.,][]{hobza2018small,chandra2019small}. One could also consider spatio-temporal or nonparametric models to build SCI by adjusting the statistic $S_0$ and choosing a bootstrap procedure accordingly. Apart from a mathematical challenge to develop a valid asymptotic theory, these extensions would require a construction of an appropriate bootstrap scheme and its computationally efficient implementation. 

\appendix
\section{Appendix}\label{sec:App}

\subsection{Regularity conditions}\label{sec:App_RC}
In this section we state the regularity conditions used in our derivations.
\begin{enumerate}
	\item $\hat{u}_d=\argmaxA_{u_d\in \mathbb{R}}\{\log g_d(\bm{y_d}|u_d,\bm{\theta})+\log h (u_d) \} $.
	\item $l(\bm{\theta})$ exists and is well-defined if: (a) $l(\bm{\theta})$ is continuous, uniquely maximized and $\bm{\theta}_0\in\Theta$, where $\bm{\theta}_0$ is a true parameter value; 
	(b) $l(\bm{\theta})$ and $\hat{l}(\bm{\theta})$ are concave; (c) $\bm{\theta_0}$ is an interior point of the parameter space and the estimator $\hat{\bm{\theta}}$ is an interior point of the neighborhood of $\bm{\theta}_0$; 
	(d) $\hat{l}(\bm{\theta})$ converges uniformly in probability to $l(\bm{\theta})$.
	\item $\bm{x}_{dj}$ are bounded and $\mathbb{E}(y^m_{dj})<\infty$ for all $d\in [D]$, $j\in [n_d]$, where $m$ is suitable large. 
	\item For each fixed $\bm{y}$, a score equation is continuously differentiable and $\mathbb{E}\{R(\bm{\theta}_0)\}=0$. 
	\item $\liminf\limits_{n}\lambda[n^{-1}\mathbb{V}\mathrm{ar}\{R(\bm{\theta})\}]>0$ and  $\liminf\limits_{n}\lambda[-n^{-1}\mathbb{E} \{\nabla R(\bm{\theta})\}]>0$ where $\nabla R(\bm{\theta}) =\frac{\partial R(\bm{\theta})}{\partial \bm{\theta}}$ and $\lambda[A]$ indicates the smallest eigenvalue of matrix $A$.
\end{enumerate}
The first two conditions refer to the log-likelihood function 
\citep[see, for example, ][]{bianconcini2014asymptotic}, whereas conditions 3-5 are needed for the derivation of the $\mathrm{MSE}$ estimators.

\subsection{Proof of Proposition 2}\label{sec:App_prop2}
Let $y^*_{dj}\sim Exp. Family (\bm{\theta})$. If $u^*_d$ is sampled from a suitable distribution, then we have $\gamma_{dj}^*=M(\mathbb{E}(y^*_{dj}|u_d))=x^t_{dj}\hat{\bm{\beta}}+u_d^*$. Furthermore, $\mathbb{V}ar^*(\bm{y}^*_d)=\mathbb{V}ar^*(\mathbb{E}^*(\bm{y}^*_d|u^*_d))+\mathbb{E}^*(\mathbb{V}ar^*(\bm{y}^*_d|u_d^*))$. The first part of the Proposition follows from the way we generate the random effects as well as results on the consistency of $\hat{\bm{\theta}}$. To show the second part we consider a general score equation. Replace $\bm{y}$ by $\bm{y}^{*}$ and set $\bm{\theta}=\hat{\bm{\theta}}$, i.e., $R^*(\bm{\theta})=\frac{\partial l^*(\hat{\bm{\theta}})}{\partial \hat{\bm{\theta}}}=\sum_{d=1}^{D}\frac{\partial \log f_d(\bm{y}^*_d|\hat{\bm{\theta}}) }{\partial \hat{\bm{\theta}}}=0$. Then $\mathbb{E}^*\{R^*(\bm{\theta})\}=0$ at $\bm{\theta}=\hat{\bm{\theta}}$ which yields consistency of $\hat{\bm{\theta}}^*$. \hfill $\square$
 
\subsection{Proof of Proposition 3}
\label{sec:App_SCI}

Let $\zeta_d$ be a general EBP, $g_d\coloneqq g_d(\bm{\theta})$ and $\hat{g}_d\coloneqq \hat{g}_d(\hat{\bm{\theta}})$. Assume that $||\hat{g}_d-g_d||=O_P(n^{-c})$. The proof uses ideas of \cite{chatterjee2008parametric}. We investigate the properties of $G_d(a)$.  
\begin{eqnarray*}
G_{d}(a)&=&P\left(\frac{\hat{\zeta}_d-\zeta_d}{\sqrt{\hat{g}}_d}\leqslant a\right)=\mathbb{E}\left(P\left[\frac{\tilde{\zeta}_d-\zeta_d}{\sqrt{g}_d}\leqslant a + 
\left\{ \frac{ a(\sqrt{\hat{g}}_d-\sqrt{g}_d) + \tilde{\zeta}_d-\hat{\zeta_d}}{\sqrt{g}_d}    \right\}
\right]  \Bigg\rvert \bm{y}_d \right)\\
&=&\mathbb{E}\left[\Phi\{a+Q(a,\bm{y}_d)\}\right]=\Phi(a)+\phi(a)\mathbb{E}\left\{Q(a,\bm{y}_d) \right\}-
2^{-1}a\phi(a)\mathbb{E}\left\{Q^2(a,\bm{y}_d) \right\}\\
&&+2^{-1}\mathbb{E}\left[ \int_{a}^{a+Q(a,\bm{y}_d)}   
\{a+Q(a,\bm{y}_d)-x\}^2(x^2-1)\phi(x)\mathrm{d} x  \right].
\end{eqnarray*}
Applying some classical results and a triangle inequality, it follows 	that the last term is bounded by $\mathbb{E}|Q|^3$, and is of smaller order than the first three terms. Therefore, the first step towards consistency of SCIs is to quantify the asymptotic expansions of $\mathbb{E}\{Q(a,\bm{y}_d)\}$ and $\mathbb{E}\{Q^2(a,\bm{y}_d)\}$. We decompose $Q(a,\bm{y}_d)$ into 
\begin{equation*}
Q(a,\bm{y}_d)= g_d^{-1/2}(\tilde{\zeta}_d-\hat{\zeta_d}) + ag_d^{-1/2}(\hat{g}^{1/2}_d-g^{1/2}_d) = Q_1+Q_2.
\end{equation*}
Let $\psi$ be a twice differentiable function with respect to $\bm{\theta}$, $y_{d.}=y_{d1}+\dots +y_{dn_d}$, $\forall d\in[D]$. Observe that $y_{d.}=y_d$ under an area-level model. The specific form of $\psi$ depends on the choice of the GLMM (for instance, under the Poisson-gamma model we spelled it out in~\eqref{eq:BP_Poisson}). Function $\psi$ satisfies the decomposition
\begin{equation}\label{eq:cons_zeta}
\begin{split}
\hat{\zeta}_d-\tilde{\zeta}_d&=\psi_d(y_{d.},\hat{\bm{\theta}})-\psi_d(y_{d.},\bm{\theta})=\left\{\frac{\partial}{\partial \bm{\theta}} \psi_d(y_{d.},\bm{\theta}) \right\}^t\left( \hat{\bm{\theta}} -\bm{\theta}\right)\\
&+ \frac{1}{2}\left( \hat{\bm{\theta}} -\bm{\theta}\right)^t  \left\{\frac{\partial^2}{\partial^2 \bm{\theta}} \psi_d(y_{d.},\bm{\theta}) \right\}  \left( \hat{\bm{\theta}} -\bm{\theta}\right)+o_P(||  \hat{\bm{\theta}} -\bm{\theta} ||^2).
\end{split}
\end{equation}

Let $C=2c$ where $c>0$. Since we assume $||\hat{\bm{\theta}}-\bm{\theta}||=O_P(n^{-c})$, we have 
\begin{equation}\label{eq:cons_zeta2}
\mathbb{E}\left[ \{\hat{\zeta}_d(\hat{\bm{\theta}})-\tilde{\zeta}_d(\bm{\theta}) \}^2 \right]=\frac{1}{n^{C}}\mathbb{E}\left( \left[ \left\{\frac{\partial}{\partial \bm{\theta}} \psi_d(y_{d.},\bm{\theta}) \right\}^t n^c\left( \hat{\bm{\theta}} -\bm{\theta}\right)  	\right]^2 \right)+o(n^{-C}).
\end{equation}
As for $Q_1$, it has been found in equation \eqref{eq:cons_zeta} that
\begin{equation*}
\mathbb{E}(\hat{\zeta}_d-\tilde{\zeta}_d) =\frac{1}{n^c}\mathbb{E}
\left[ \left\{\frac{\partial}{\partial \bm{\theta}} \psi_d(y_{d.},\bm{\theta})\right\}^t 
n^c\left( \hat{\bm{\theta}} -\bm{\theta}\right)  \right] +o\left(n^{-c}\right),
\end{equation*}
and $\mathbb{E}\left\{(\hat{\zeta}_d-\tilde{\zeta}_d)^2\right\} =O(n^{-C}) $, thanks to the result in equation \eqref{eq:cons_zeta2}. Furthermore, observe that $g_d$ is of order $O(1)$ which leads to $\mathbb{E}(Q_1)=O(D^{-c})$ as well as $\mathbb{E}(Q^2_1)=O(D^{-C})$.	When we turn to $Q_2$, we have an immediate simplification 
$Q_2=ag_d^{-1/2}(\hat{g}^{1/2}_d-g^{1/2}_d) =a\left\{(\hat{g}_d/g_d)^{1/2} -1\right\}$.
Let $g_d$ be twice differentiable with respect to $\bm{\theta}$. Similarly to the computations above, we have the expansion
\begin{equation*}
\hat{g}_d(\hat{\bm{\theta}})=g_d(\bm{\theta})+
\left(\frac{\partial}{\partial \bm{\theta}} g_d(\bm{\theta})\right)^t 
\left( \hat{\bm{\theta}} -\bm{\theta}\right)+
\frac{1}{2}\left( \hat{\bm{\theta}} -\bm{\theta}\right)^t  \left(\frac{\partial^2}{\partial^2 \bm{\theta}} g_d(\bm{\theta}) \right)  \left( \hat{\bm{\theta}} -\bm{\theta}\right)+o_P(||  \hat{\bm{\theta}} -\bm{\theta} ||^2).
\end{equation*} 
Therefore we obtain
\begin{equation*}
\mathbb{E}\{\hat{g}_d(\hat{\bm{\theta}})\}=g_d(\bm{\theta})+
\frac{1}{n^C}\mathbb{E}\left[\left\{\frac{\partial}{\partial \bm{\theta}} g_d(\bm{\theta})\right\}^t n^{C} \left( \hat{\bm{\theta}} -\bm{\theta}\right)\right]+ O(n^{-C}).
\end{equation*} 
It follows that $\mathbb{E}\left\{(\hat{g}/g)^{1/2}\right\}=O(n^{-c})$,  $\mathbb{E}(Q_2)=O(n^{-C})$ and $\mathbb{E}(Q^2_2)=O(n^{-C})$. We can deduce that $G_d(a)$ attains the asymptotic expansion 
$G_d(a)=\Phi(a)+n^{-c}\gamma (a,\bm{\theta})+O(n^{-C})$.
A similar expansion can be established for $G_d^*(a)$ if we replace $\bm{\theta}$ with $\hat{\bm{\theta}}$ and $P$ with $P^*$.

\newpage
\section{Supplementary material}\label{sec:App2}
\subsection{Introduction} \label{sec:intro2}

This document contains additional information to the paper {\sl Simultaneous Inference for Empirical Best Predictors with a Poverty Study in Small Areas} which introduces uniform inference for mixed parameters in Generalized Linear Mixed Models (GLMM). Specifically, we consider empirical best predictor (EBP) and construct simultaneous confidence interval (SCI) and multiple testing procedure (MTP). Our main motivation stems from the gap in the statsitical literature --  existing methods for mixed parameter do not allow for valid comparative analyses. For example, in a great majority of literature on small area estimation (SAE) only individual confidence intervals (iCI) are considered. 

The rest of this document is organized as follows. In Section \ref{sec:MSE_EBP} we provide a decomposition of the Mean Squared Error (MSE) of the EBP which was used in the main article. Sections \ref{sec:App_Poiss}, \ref{sec:App_Poisson_Lognormal} and \ref{sec:App_Bin} present more details on the estimation assuming the area-level Poisson-gamma, the area-level Poisson-lognormal and the unit-level binomial models, respectively. In particular, Section \ref{sec:MSE_Poisson} contains the derivation of a new MSE estimator, whereas in other sections we present results of various simulation studies assessing and comparing the finite sample performances of estimators for fixed effects, EBPs and their variability. Finally, in Section \ref{sec:data_example2}, we provide additional information on the case study of poverty across counties in Galicia, Spain. This includes comparisons of confidence intervals and diagnostic residual plots.

\subsection{MSE of EBP under GLMM}\label{sec:MSE_EBP}

We start with a decomposition of MSE of EBP under GLMM. Consider a general EBP $\hat{\zeta}_d$
\begin{equation}\label{eq:MSE_dec}
	\begin{split}
		\mathrm{MSE}(\hat{\zeta}_d)&=\mathbb{E}[(\hat{\zeta}_d -\zeta_d)^2]=\mathbb{E}[ (\hat{\zeta}_d -\tilde{\zeta}_d+\tilde{\zeta}_d-\zeta_d)^2  ]\\
		&=\mathbb{E}[ (\hat{\zeta}_d -\tilde{\zeta}_d)^2 +(\tilde{\zeta}_d -\zeta_d )^2+2\{(\hat{\zeta}_d -\tilde{\zeta}_d ) (\tilde{\zeta}_d -\zeta_d )\}]\\
		&=\mathbb{E}[(\hat{\zeta}_d-\tilde{\zeta}_d)^2]
		+\mathbb{E}[(\tilde{\zeta}_d-\zeta_d)^2]\eqqcolon g_{2d}+g_{1d},
	\end{split}
\end{equation}
where the fourth equality follows by the law of iterated expectation, that is 
\begin{equation*}
	\begin{split}
		&\mathbb{E}[  (\hat{\zeta}_d -\tilde{\zeta}_d )   (\tilde{\zeta}_d -\zeta_d )]=
		\mathbb{E}\{  (\hat{\zeta}_d -\tilde{\zeta}_d ) \mathbb{E}[  (\tilde{\zeta}_d -\zeta_d ) |\bm{y}_d] \}=0,
	\end{split}
\end{equation*}
by the definition of $\tilde{\zeta}_d(\bm{\theta})=\mathbb{E}[\zeta_d|\bm{y}_d]$.
Furthermore, we can decompose $g_{1d}$ to obtain
\begin{equation}\label{eq:M_1_Poisson}
	g_{1d}=\mathbb{E}[(\tilde{\zeta}_d -\zeta_d )^2] =\mathbb{E}(\tilde{\zeta}^2_d)+\mathbb{E}(\zeta_d^2)-2\mathbb{E}[\tilde{\zeta}_d\mathbb{E}(\zeta_d|\bm{y}_d)]
	=\mathbb{E}(\zeta_d^2) -\mathbb{E}(\tilde{\zeta}_d^2)
\end{equation}
once again by the definition of $\tilde{\zeta}_d$.

\subsection{Additional details on the area-level Poisson-gamma model}\label{sec:App_Poiss}
\subsubsection{Estimation of parameters}\label{sec:App_Poiss_param}
The derivation of the likelihood for the Poisson-gamma model proceeds as follows:
\begin{equation}\label{eq:likelihood_Poisson2}
	\begin{split}
		&  \mathcal{L}^{PG}(\bm{\theta}) \coloneqq 
		f^{PG}(\bm{y}|\bm{\theta})=\prod_{d=1}^{D}\int_{0}^{\infty}\frac{\exp\left(-\mu^{PG}_d\right)\mu_d^{PGy_d}}{y_d!}f(w_d)dw_d\\
		&=\prod_{d=1}^{D} \frac{\lambda_d^{y_d}\delta^{\delta} \Gamma (y_d+\delta)}{y_d!\Gamma(\delta)(\delta+\lambda_d)^{y_d+\delta}}
		\int_{0}^{\infty}\frac{(\delta+\lambda_d)^{y_d+\delta}\exp(-(\delta+\lambda_d)w_d)w_d^{y_d+\delta-1}}{\Gamma(y_d+\delta)}d w_d   
		\\ & = \prod_{d=1}^{D} \frac{\Gamma(y_d+\delta)}{\Gamma(y_d+1) \Gamma(\delta)}\left(\frac{\delta}{\delta+\lambda_d}\right)^{\delta}
		\left(\frac{\lambda_d}{\delta+\lambda_d} \right)^{y_d} .
	\end{split}
\end{equation}
The log-likelihood is proportional to 
\begin{equation}\label{eq:loglike_Poisson}
	l^{PG}(\bm{\theta})=\sum_{d=1}^{D}\left( \sum_{j=1}^{y_d-1} 
	\log(1+\delta^{-1} j) +y_d\log \lambda_d -(y_d+\delta)\log(1+\delta^{-1}\lambda_d) \right),
\end{equation}
where $\Gamma(b+c)/\Gamma(c)=c(c+1)\dots(c+b-1)$ if $b\geqslant1$ and $\sum_{j=1}^{y_d-1}\log(1+\delta^{-1} j) =0$ if $y_d-1<0$.
A similar expression as in \eqref{eq:loglike_Poisson} has been used by \cite{lawless1987negative} who derived the estimating equations, and implemented a scoring algorithm to obtain an estimator for ${\bm{\theta}}$.

As far as the estimation method of $\bm{\theta}$ is concerned, we followed the suggestions of \cite{lawless1987negative} and implemented Newton-Raphson and Fisher scoring algorithms. Application of both methods leads to a following iterative scheme
\begin{equation}
	\bm{\theta}^{(i+1)}=\bm{\theta}^{(i)}-\bm{H}_{PG}^{-1}(\bm{\theta}^{(i)})\bm{R}_{PG}(\bm{\theta}^{(i)}),\end{equation}
where $\bm{R}_{PG}(\bm{\theta})$ is a score vector composed of the first derivative of the likelihood. Furthermore, $\bm{H}_{PG}(\bm{\theta})$ stands for the observed information matrix $\bm{J}_{PG}(\bm{\theta})=-\bm{H}_{PG}(\bm{\theta})$ under a Newton-Raphson algorithm, and the information matrix $\bm{I}_{PG}(\bm{\theta})=\mathbb{E}\{\bm{J}_{PG}(\bm{\theta})\}$ under Fisher scoring. Let the score vector $\bm{R}_{PG}(\bm{\theta})$ be defined as
\begin{equation*}
	\bm{R}_{PG}(\bm{\theta})=(R_{PG1}(\bm{\theta}),R_{PG2}(\bm{\theta}),\dots,R_{PGp+1}(\bm{\theta}))^t.
\end{equation*}
Its components are given as follows 
\begin{equation*}
	R_{PGk}=\frac{\partial l^{PG}}{\partial \beta_k}=\sum_{d=1}^{D}\frac{x_{dk}(y_d-\lambda_d)}{1+\alpha\lambda_d}, \quad k=1,2,\dots,p,
\end{equation*}
\begin{equation*}
	R_{PGk+1}=\frac{\partial l^{PG}}{\partial \alpha}=\sum_{d=1}^{D}
	\left[\sum_{j=1}^{y_d-1} \left( \frac{j}{1+\alpha j}\right) +\alpha^{-2}\log (1+\alpha \lambda_d ) 
	-\frac{(y_d+\alpha^{-1})\lambda_d}{1+\alpha \lambda_d}  \right], 
\end{equation*}
with $\alpha=\delta^{-1}$. The observed information matrix is composed of the negative second derivatives of the likelihood, that is 
\begin{equation*}
	J_{PGkr}=-\frac{\partial^2 l^{PG}}{\partial \beta_k\partial \beta_r}=\sum_{d=1}^{D} \frac{(1+\alpha y_d)\lambda_d x_{dr} x_{dk}}{(1+\alpha \lambda_d)^2},\quad k,r=1,\dots,p,
\end{equation*}  
\begin{equation*}
	J_{PGk(r+1)}=-\frac{\partial^2 l^{PG}}{\partial \beta_k\partial \alpha}=\sum_{d=1}^{D} \frac{\lambda_d(y_d-\lambda_d)x_{dr}}{(1+\alpha \lambda_d)^2}, \quad k=1,\dots,p,
\end{equation*}
\begin{equation*}
	J_{PG(k+1)(r+1)}=-\frac{\partial^2 l^{PG}}{\partial^2 \alpha}=\sum_{d=1}^{D}
	\left[ \sum_{j=1}^{y_d-1} \left( \frac{j}{1+\alpha j}  \right)^2+\frac{2\log(1+\alpha\lambda_d)}{\alpha^3}-\frac{2\alpha^{-2} \lambda_d}{1+\alpha\lambda_d}    -\frac{(y_d+\alpha^{-1})\lambda^2_{d}}{(1+\alpha \lambda_d)^2}  \right].
\end{equation*}
On the other hand, the Fisher information matrix, which is assured to be positive definite, is composed of    
\begin{equation*}
	I_{PGkr}=\sum_{d=1}^{D} \frac{\lambda_d x_{dr} x_{dk}}{(1+\alpha \lambda_d)},\quad k,r=1,\dots,p,
\end{equation*}  
\begin{equation*}
	I_{PGk(r+1)}=0, \quad k=1,\dots,p,
\end{equation*}
\begin{equation*}
	I_{PG(k+1)(r+1)}=\alpha^{-4}\sum_{d=1}^{D} \left[ \mathbb{E}\sum_{j=1}^{y_d-1} \left( \alpha^{-1}+ j \right)^{-2} -\frac{\alpha \lambda_d}{\lambda_d+\alpha^{-1}} \right].
\end{equation*}
The details of this derivation and some suggestions regarding the calculation are in \cite{lawless1987negative}. It is necessary to point out that we followed one of them and first we maximized $l(\bm{\theta})$ with respect to fixed parameters for selected values of $\alpha$ using Newton-Raphson or Fisher scoring algorithms.  In this way one obtains a profile likelihood $l(\tilde{\bm{\theta}}(\alpha),\alpha)$. Since the Fisher scoring algorithm turned out to be numerically more stable, Section \ref{sec:poisson_sim} and Section 4 and 5 of the main article contain the results obtained using this scheme. We follow the suggestion of \cite{boubeta2016empirical} when it comes to the choice of the starting values, namely we set $\bm{\beta}^{(0)}=\tilde{\bm{\beta}}$, where $\tilde{\bm{\beta}}$ is the maximum likelihood estimator under the model without the random effects. When it comes to $\alpha$, we use the properties of a random variable which is distributed according to negative binomial, that is $\sigma^2_d\coloneqq\mathbb{V}ar(y_d)=\lambda_d+\alpha \lambda_d^2$. Therefore $\sigma^2_d$ can be estimated applying 
\begin{equation*}
	\tilde{\sigma}^2_d=\frac{1}{D}\sum_{d=1}^{D}\{y_d- \exp(\bm{x}_d^t\tilde{\bm{\beta}})\}^2\quad \text{ setting } \quad \alpha^{(0)}=\frac{\tilde{\sigma}^2_d-\exp(\bm{x}_d^t\tilde{\bm{\beta}})}{\exp(2\bm{x}_d^t\tilde{\bm{\beta}})} .
\end{equation*}

\subsubsection{Estimation of the MSE of the EBP}\label{sec:MSE_Poisson}

Let us turn to the derivation of the MSE for  EBP $\hat{\mu}^{PG}$ for which the best predictor (BP) is defined by
\begin{equation}\label{eq:BP_Poisson2}
	\mathbb{E}(\mu^{PG}_d|y_d)=\frac{\int_{0}^{\infty} \lambda_d w_d g(y_d|w_d)h(w_d) dw_d}
	{\int_{0}^{\infty}g(y_d|w_d)h(w_d) dw_d} =\frac{A_d^{PG}(y_d, \bm{\theta})}{C_d^{PG}(y_d,\bm{\theta})}=\frac{\lambda_d(y_d+\delta)}{(\lambda_d+\delta)}\eqqcolon\psi_d^{PG}(y_d,\bm{\theta}) . 
\end{equation}
Firstly, we focus on $\mathbb{E}(\zeta_d^2)$ from \eqref{eq:M_1_Poisson}:
\begin{equation*}
	\begin{split}
		\kappa_{1d}\coloneqq\mathbb{E}(\mu_d^{PG2})&=\int_{0}^{\infty}\lambda_d^2w_d^2f(w_d)\mathrm{d}w_d=\int_{0}^{\infty}\lambda_d^2w_d^2\frac{\delta^{\delta}\exp(-w_d\delta)w_d^{\delta-1}}{\Gamma(\delta)}\mathrm{d}w_d\\
		&=\frac{\lambda_d^2\delta^{\delta}\Gamma(\delta+2)}{\Gamma(\delta)\delta^{\delta+2}}\int_{0}^{\infty}\frac{\delta^{\delta+2}\exp(-w_d\delta)w_d^{\delta+2-1}}{\Gamma(\delta+2)}\mathrm{d}w_d
		=\frac{\lambda_d^2 (\delta+1) }{\delta}   . 
	\end{split}
\end{equation*}
Then, the $\mathbb{E}(\tilde{\zeta}_d^2)$ under Poisson-gamma area-level model is 
\begin{equation*}
	\begin{split}
		\kappa_{2d}\coloneqq\mathbb{E}(\tilde{\mu}^{PG2}_d)&=\mathbb{E}\{\mathbb{E}^2(\mu_d^{PG}|y_d)\}
		=\mathbb{E}\{\psi_d^{PG2}(y_d,\bm{\theta})\} \\ &
		=\mathbb{E}\left\{\frac{\lambda^2_d(y_d+\delta)^2}{(\lambda_d+\delta)^2}\right\}
		=\sum_{j=0}^{\infty}\frac{\lambda^2_d(j+\delta)^2}{(\lambda_d+\delta)^2}P(y_d=j)  ,
	\end{split}
\end{equation*}
where we used \eqref{eq:BP_Poisson2} to obtain the final expression.
When it comes to the estimation of $g_{2d}$ in \eqref{eq:MSE_dec} under our Poisson-gamma area-level model, we used the Taylor expansion applied in the main document in equation (28) and the result from equation (29). This gives
\begin{equation}\label{eq:the_last_term_MSE}
	\mathbb{E}\left[ \{\hat{\mu}^{PG}_d(\hat{\bm{\theta}})-\tilde{\mu}^{PG}_d(\bm{\theta}) \}^2 \right]=\frac{1}{D}\mathbb{E}\left( \left[ \left\{\frac{\partial}{\partial \bm{\theta}} \psi^{PG}_d(y_d,\bm{\theta}) \right\}^t \sqrt{D}\left( \hat{\bm{\theta}} -\bm{\theta}\right)  
	\right]^2 \right)+o(1/D)  .
\end{equation}
Afterwards, we employed the construction of \cite{jiang2001empirical} to ML estimators. We define an estimator $\hat{\bm{\theta}}_{d-}$
based on $\bm{y_{d-}}=(y_{1},\dots y_{d-1}, y_{d}, \dots y_{D} )$ and $\hat{\mu}^{PG}_{d-}=\psi^{PG}_d(y_d,\hat{\bm{\theta}}_{d-})$, and we replace $\hat{\bm{\theta}}$ \eqref{eq:the_last_term_MSE} with $\hat{\bm{\theta}}_{d-}$. As a result,
\begin{equation}
	\begin{split}
		c_{-d}(\bm{\theta})&=\mathbb{E}\left( \left[ \left\{\frac{\partial}{\partial \bm{\theta}} \psi^{PG}_d(y_d,\bm{\theta}) \right\}^t \sqrt{D}\left( \hat{\bm{\theta}}_{d-} -\bm{\theta}\right)  \right]^2 \right)\\
		&=\sum_{j=1}^{\infty}\mathbb{E}\left( \left[ \left\{\frac{\partial}{\partial \bm{\theta}} \psi^{PG}_d(y_d,\bm{\theta}) \right\}^t \sqrt{D}\left( \hat{\bm{\theta}}_{d-} -\bm{\theta}\right)   \right]^2 \lvert_{y_d=j}\right)P(y_d=j)\\
		&=\sum_{j=1}^{\infty} \left\{\frac{\partial}{\partial \bm{\theta}} \psi^{PG}_d(y_d,\bm{\theta}) \right\}^t\mathbb{V}ar_{-d}(\bm{\theta})\left\{\frac{\partial}{\partial \bm{\theta}} \psi^{PG}_d(y_d,\bm{\theta}) \right\}P(y_d=j),
	\end{split}
\end{equation}
where $\mathbb{V}ar_{-d}(\bm{\theta})=D\mathbb{E}
\left[ \left( \hat{\bm{\theta}}_{d-} -\bm{\theta}\right) \left( \hat{\bm{\theta}}_{d-} -\bm{\theta}\right) ^t \right]$ which does not depend on the value of $y_d$. Therefore
\begin{equation*}
	\mathrm{MSE}(\hat{\mu}^{PG}_{-d})=g_{1d}(\bm{\theta})+\frac{1}{D}c_{-d}(\bm{\theta})+o(1/D).
\end{equation*}
Furthermore, if we suppose that conditions 1-2 from Appendix of the main document hold, the plug-in MSE is given by 
\begin{equation}\label{eq:MSE_Poiss}
	\mathrm{MSE}(\hat{\mu}^{PG}_d)=\kappa_{1d}(\bm{\theta})-\kappa_{2d}(\bm{\theta})+\frac{1}{D}c_d(\bm{\theta})+o(1/D),
\end{equation}
with $c_d$ defined as above replacing $\hat{\bm{\theta}}_{d-}$ and $\mathbb{V}ar_{-d}$ by
$\hat{\bm{\theta}}_{d}$ and $\mathbb{V}ar_{d}$.

\subsubsection{Finite sample performance}\label{sec:poisson_sim}

As we have mentioned in the main document, for the performance study of fixed effects $\bm{\beta}$ and variability parameter $\delta$ we used the relative bias (RBIAS) and the relative root MSE (RRMSE) which are defined as follows
\begin{equation*}
	RBIAS(\hat{\theta}_j)= \frac{1}{K}\sum_{k=1}^{K}(\hat{\theta}^{(k)}_j -\theta_j) / |\theta_j|  
	, \quad 
	RRMSE(\hat{\theta}_j) = \left( \sqrt{\frac{1}{K}\sum_{k=1}^{K}(\hat{\theta}^{(k)}_j -\theta_j)^2} \right) / |\theta_j|,
\end{equation*}
where $K$ stands for the number of simulations and $\theta_j\in \bm{\theta}=(\bm{\beta}, \delta)$, $j=1,\dots,6$. 

In our simulation study, we used EBP to estimate proportions, in preparation for the case study where we estimate poverty rates for different small areas. Under the area-level Poisson-gamma model we thus considered $\zeta_d:=\bar\mu^{PG}_d=\mu^{PG}_d/N_d$. The empirical performance of the EBP $\hat{\zeta}_d$ is evaluated using bias ($B_d$), average absolute bias (B), MSE ($E_d$) and average MSE ($E$), calculated as follows:
\begin{equation*}
	B_d=\frac{1}{K}\sum_{k=1}^{K}(\hat{\zeta}_d^{(k)}-\zeta^{(k)}_d), \quad B=\sum_{d=1}^{D}|B_d|/D,\quad
	E_d=\frac{1}{K}\sum_{k=1}^{K}(\hat{\zeta}_d^{(k)}-\zeta^{(k)}_d)^2, \quad E=\sum_{d=1}^{D}|E_d|/D.
\end{equation*}
We also calculated relative root MSE and the relative bias of the EBP using slightly modified formulas (compared to the fixed parameters), namely 
\begin{equation*}
	RBIAS(\hat{\zeta}^{(k)}_d)=B_d/\bar{\zeta}^{(k)}_d, \quad RRMSE(\hat{\zeta}^{(k)}_d)=
	\sqrt{E_d}/\bar{\zeta}^{(k)}_d\quad , \text{ where}\quad \bar{\zeta}^{(k)}_d=\frac{1}{K}\sum_{k=1}^{K}\zeta^{(k)}_d.
\end{equation*}
A detailed setting of the simulation study under the area-level Poisson-gamma model is given in the main document. Table \ref{tab:rb_rm_Poisson} outlines RBIAS and RRMSE of $\hat{\bm{\theta}}$. It is apparent that both criteria decrease with an increasing number of areas. 
\begin{table}[htb]
	\centering
	\begin{adjustbox}{max width=\textwidth}
		\begin{tabular}{c cccccc}\hline
			D  & $\hat{\beta}_0$  & $\hat{\beta}_1$  & $\hat{\beta}_2$  & $\hat{\beta}_3$  & $\hat{\beta}_4$  & $\hat{\delta}$  \\\hline
			26 & 0.0018 (0.1007)  & -0.1006 (0.7333) & -0.0235 (0.5965) & -0.0519 (0.5710) & -0.0187 (0.1596) & 0.5364 (0.6934) \\
			52 & -0.0026 (0.0691) & -0.0201 (0.4043) & -0.0339 (0.4110) & -0.0337 (0.3561) & 0.0001 (0.1053)  & 0.1540 (0.2758) \\
			78 & -0.0043 (0.0551) & -0.0004 (0.3330) & -0.0193 (0.3440) & 0.0035 (0.2897)  & 0.0018 (0.0902)  & 0.0887 (0.2043)\\\hline
		\end{tabular}
	\end{adjustbox}
	\caption{RBIAS and RRMSE (in parenthesis) for ML estimators under area-level Poisson-gamma model.}
	\label{tab:rb_rm_Poisson}
\end{table}

Furthermore, Table \ref{tab:B_E_p_d_poisson} summarizes bias and MSE of  EBP $\hat{\bar{\mu}}^{PG}_d$, $d=1,\dots, D$ for quantiles of set $\{1,\dots,D\}$ where the domains are sorted by the sample size. The first column shows the average absolute biases and the average MSE. The results confirm a very good performance of this estimator.   

\begin{table}[htb]
	\centering
	\begin{adjustbox}{max width=\textwidth}
		\begin{tabular}{ccccccccc}\hline
			$D$    & $d$  & $B_d(E_d)$       & $D$    & $d$  & $B_d(E_d)$       & $D$    & $d$  & $B_d(E_d)$       \\\hline
			26   & 6  & -0.0001 (0.0000) & 52   & 11 & -0.0002 (0.0000) & 78   & 16 & -0.0003 (0.0000) \\
			& 11 & -0.0001 (0.0000) &      & 21 & -0.0001 (0.0000) &      & 32 & -0.0001 (0.0001) \\
			& 16 & 0.0001 (0.0000)  &      & 32 & 0.0000 (0.0001)  &      & 47 & 0.0000 (0.0001)  \\
			& 21 & 0.0001 (0.0001)  &      & 42 & 0.0002 (0.0002)  &      & 63 & 0.0002 (0.0003)  \\
			$B(E)$ &    & 0.0002 (0.0001)  & $B(E)$ &    & 0.0003 (0.0001)  & $B(E)$ &    & 0.0003 (0.0002) \\\hline
		\end{tabular}
	\end{adjustbox}
	\caption{$B_d$, $E_d$ and their averages $B$ and $E$ for the estimators of ${\bar{\mu}}^{PG}_d$ under the area-level Poisson-gamma model.}
	\label{tab:B_E_p_d_poisson}
\end{table}

\begin{figure}[htb]
	\centering
	\subfloat{\includegraphics[width=0.45\textwidth]{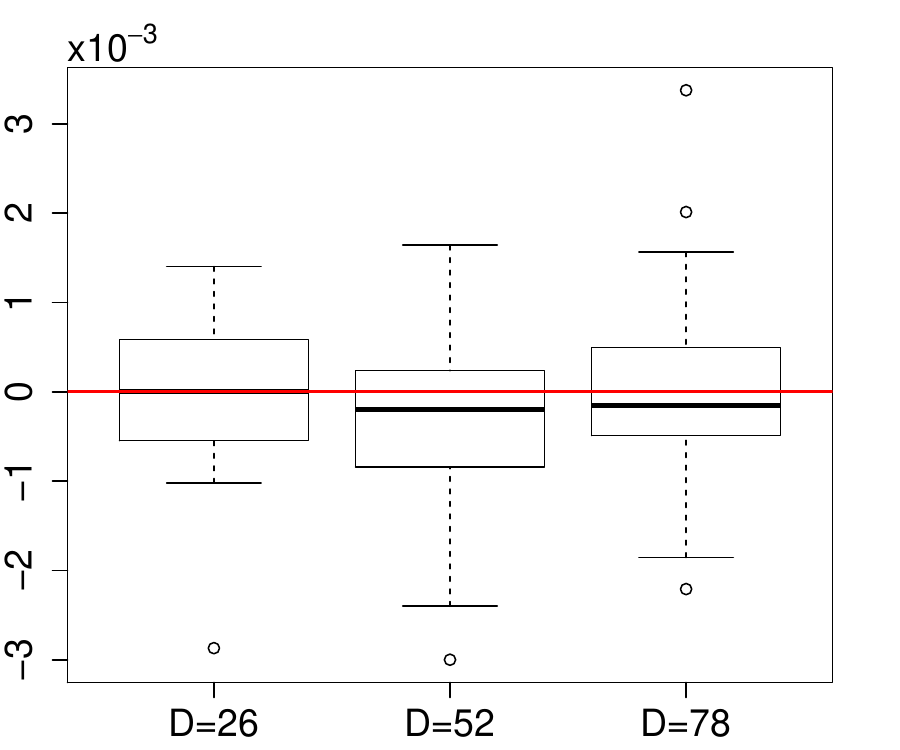}}
	\subfloat{\includegraphics[width=0.45\textwidth]{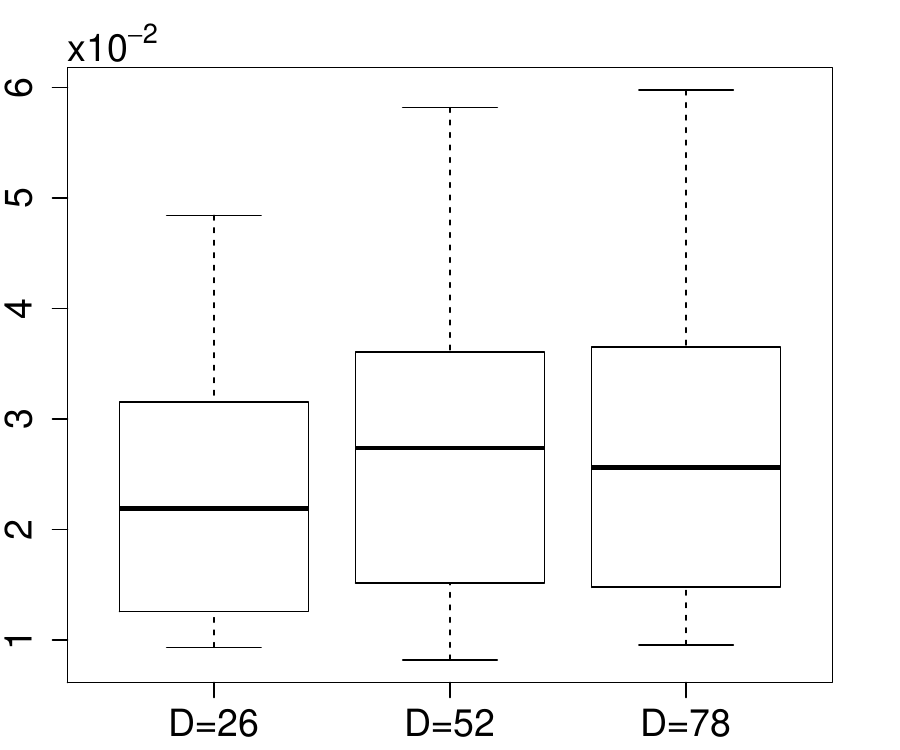}}
	\captionof{figure}{Box plots with RBIAS (left) and RRMSE (right) for $\hat{\zeta}^{(k)}_d$ under the area-level Poisson-gamma model.} 
	\label{fig:bias_mse}
\end{figure}

\begin{figure}[htb]
	\centering          \vspace{-0.7cm}
	\subfloat{\includegraphics[width=0.45\textwidth]{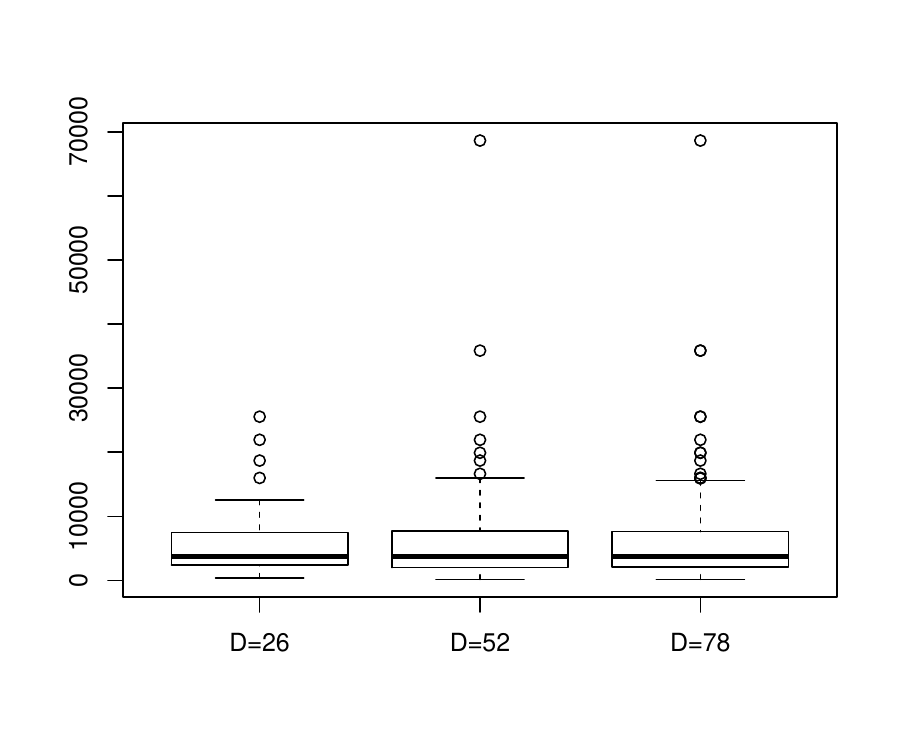}}
	\subfloat{\includegraphics[width=0.45\textwidth]{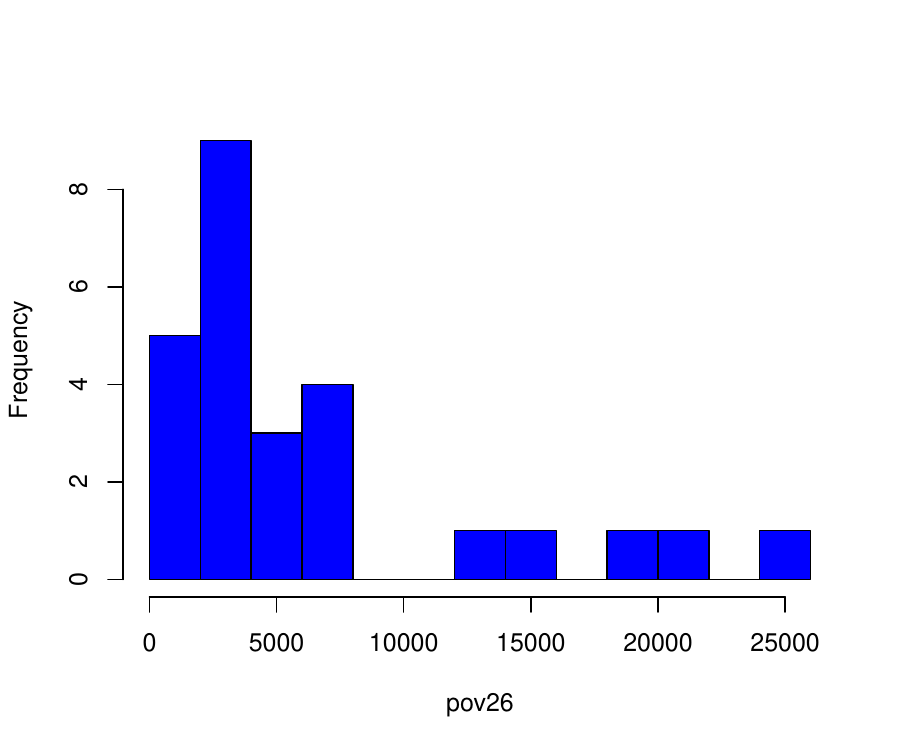}}\\
	\vspace{-5mm}
	\subfloat{\includegraphics[width=0.45\textwidth]{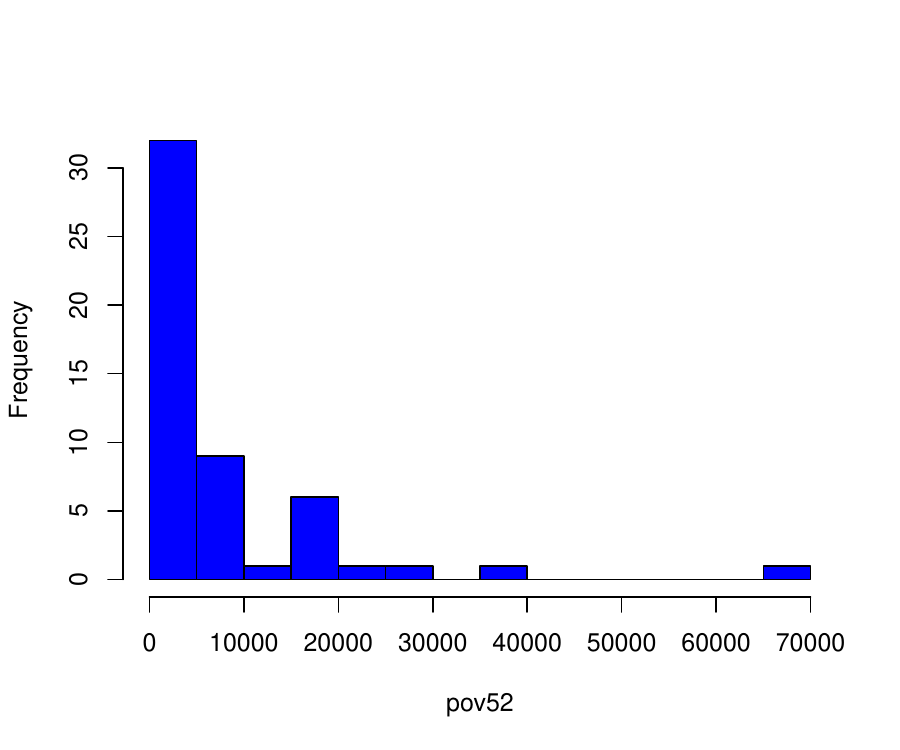}}
	\subfloat{\includegraphics[width=0.45\textwidth]{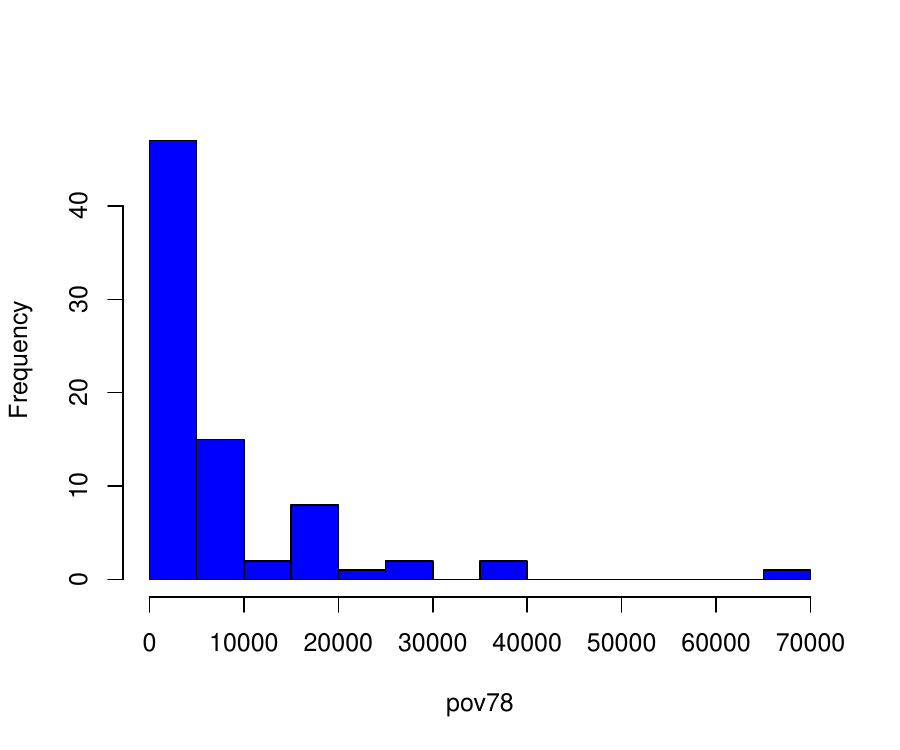}}
	\captionof{figure}{Box plot and histograms of response variable for $D=26$, $D=52$ and $D=78$.} 
	\label{fig:resp_hist_box}
\end{figure}

On the other hand, Figure \ref{fig:bias_mse} depicts RBIAS and RRMSE of $\hat{\zeta}^{(k)}_d$ under different sample sizes. The most striking feature, already mentioned in the main document, is an outstanding performance for $D=26$ and poorer results for $D=52$ and $D=78$. Recall that the smallest sample was obtained by a simple random sampling from the pool of true areas. Relatively better performance can be thus explained by the fact that a couple of problematic areas were not selected. Extreme values are the most challenging to estimate -- we found out that a sample with $D=26$ was randomly truncated, because it did not include 3 lowest and 2 highest counts of people below poverty level (see histograms and box plot in Figure \ref{fig:resp_hist_box}). As a consequence, the model was easier to fit and the EBP was estimated more accurately. Regarding $D=52$ and $D=72$, their performances are similar with lower median, yet more outliers. 

\begin{figure}[htb]
	\centering   \vspace{-0.7cm}
	\subfloat{\includegraphics[width=0.33\textwidth]{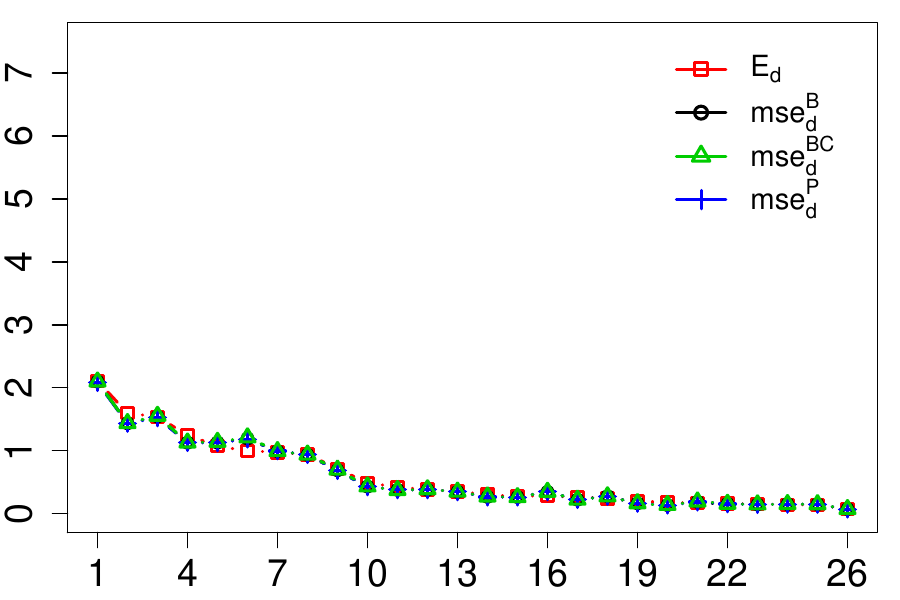}}
	\subfloat{\includegraphics[width=0.33\textwidth]{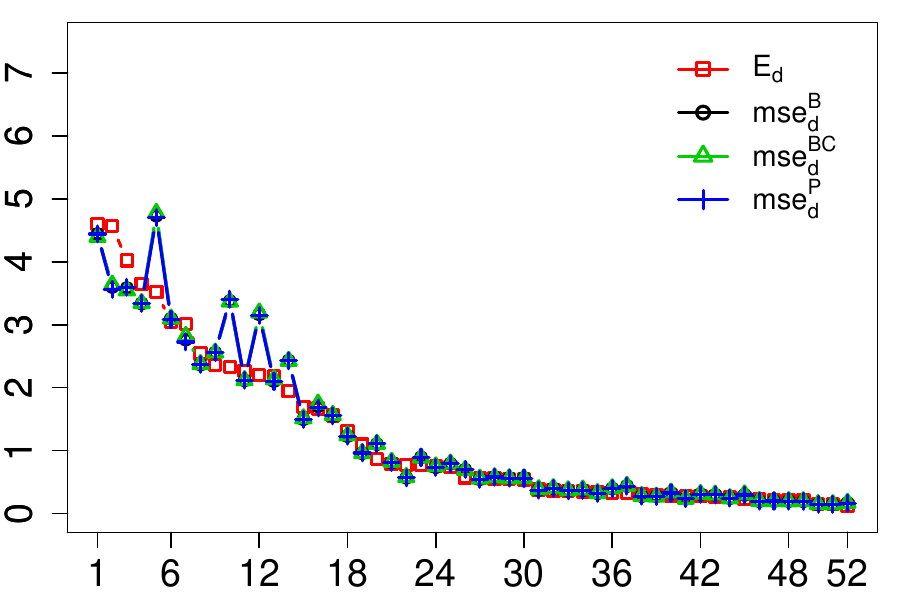}}
	\subfloat{\includegraphics[width=0.33\textwidth]{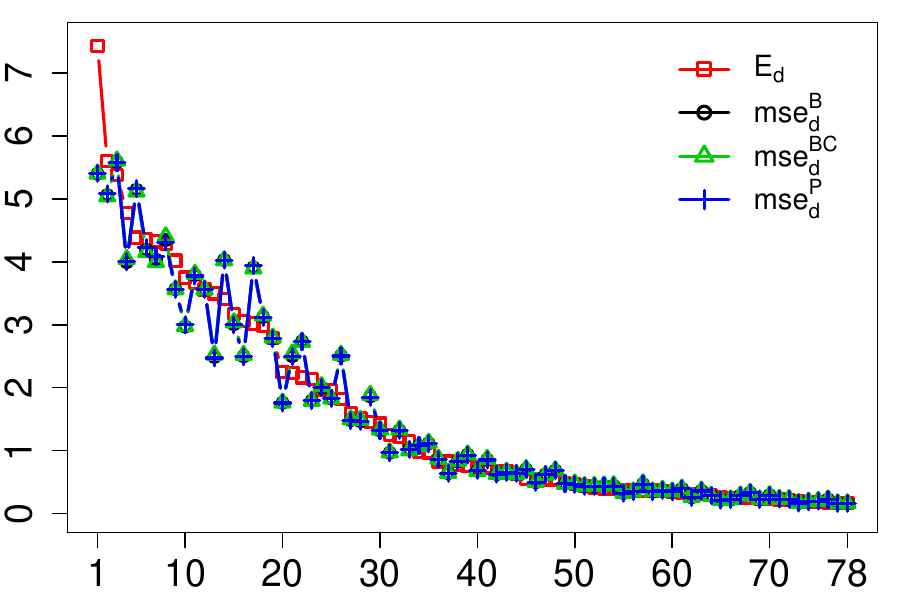}}   
	\captionof{figure}{MSE estimators ($\times10^4$): (left) D = 26, (middle) D = 52 and (right) D = 78 under the area-level Poisson-gamma model.} 
	\label{fig:MSE_est_p}
\end{figure}

\begin{figure}[htb]
	\centering   \vspace{-0.7cm}
	\subfloat{\includegraphics[width=0.33\textwidth]{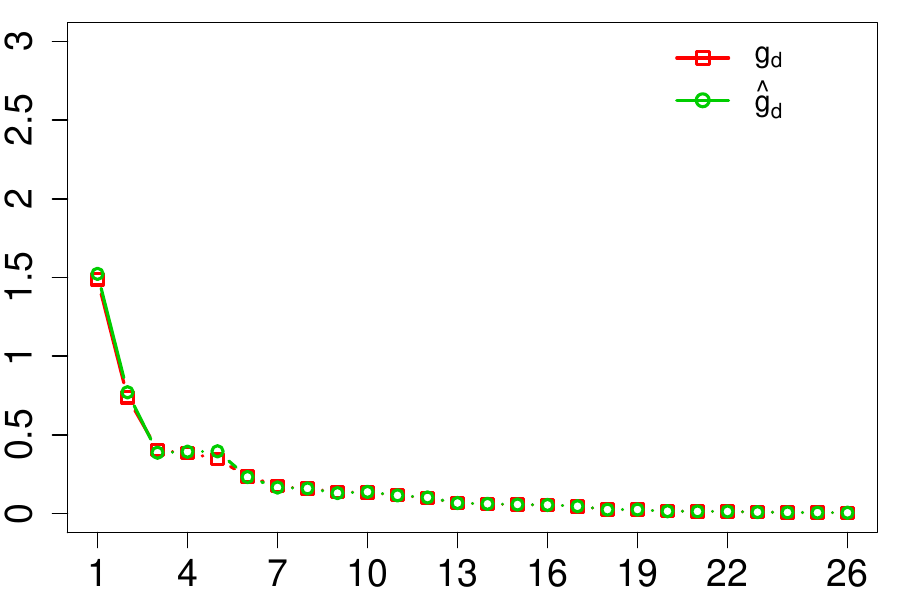}}
	\subfloat{\includegraphics[width=0.33\textwidth]{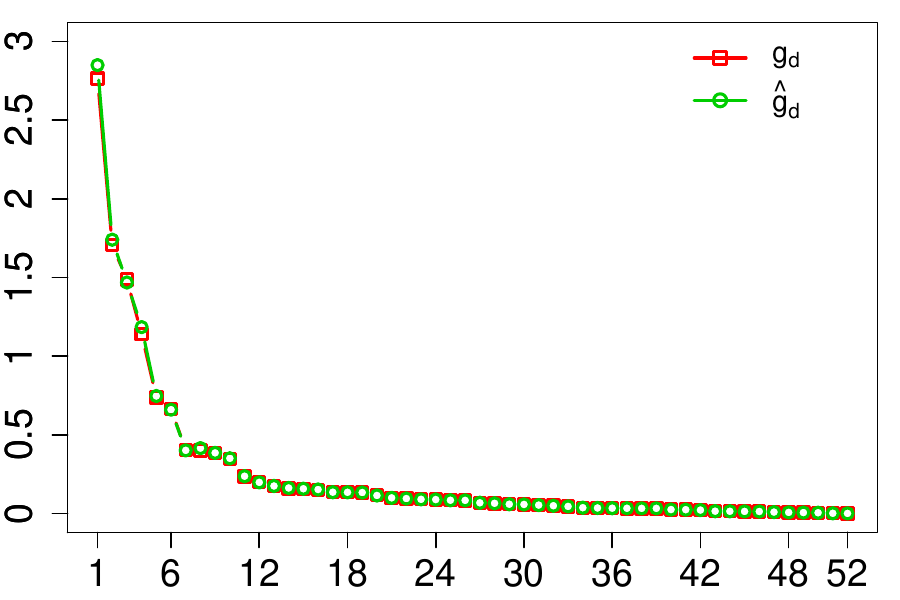}}
	\subfloat{\includegraphics[width=0.33\textwidth]{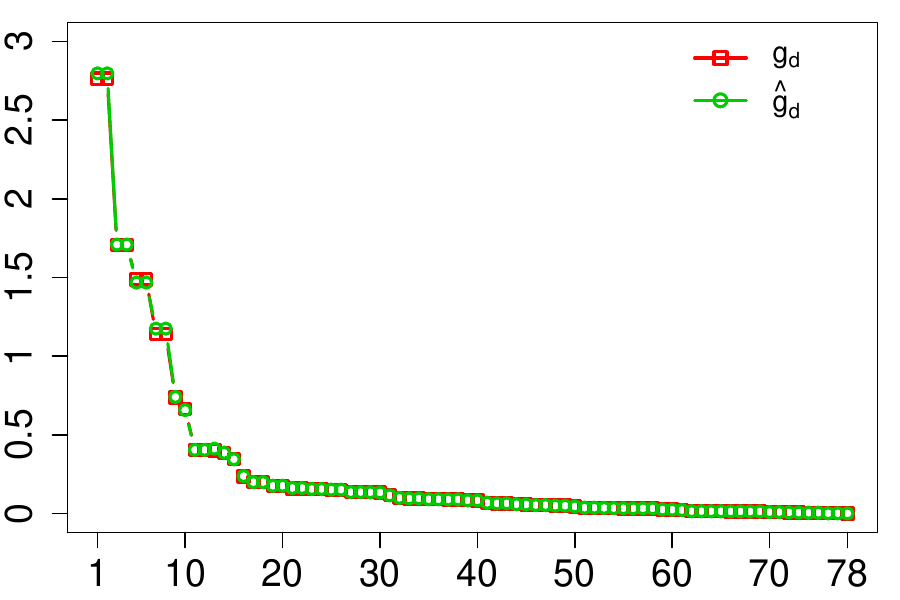}} 
	\captionof{figure}{$g_{1d}$ estimators ($\times10^4$): (left) D = 26, (middle) D = 52 and (right) D = 78 under area-level Poisson-gamma model.} 
	\label{fig:g1_est_p}
\end{figure}

Figure \ref{fig:MSE_est_p} and Figure \ref{fig:g1_est_p} illustrate the performance of MSE and $\hat{g}_{1d}$ respectively for each area $d=1,\dots,D$. They are sorted in decreasing order. As far as MSE estimation is concerned, the estimators approximate well empirical MSE ($E_d$). The results are similar for bootstrap, bias corrected bootstrap and plug-in estimators. For $D=26$, they are almost identical with the true value of MSE while for $D=52$ and $D=78$ they deviate slightly from truth for the highest values and become almost indistinguishable for the lowest values which is motivated by the same reasoning as results in Figure \ref{fig:bias_mse}. The results for estimating $g_{1d}$ in Figure \ref{fig:g1_est_p} are even more accurate.

\subsection{Additional details on the area-level Poisson-lognormal model}\label{sec:App_Poisson_Lognormal}

We provide more detailed information about the finite sample performance of EBP estimators and SCI under the Poisson-lognormal area-level model. We can, however, keep this section much shorter than the above because this model was examined thoroughly in the literature, see for example, \cite{cameron2013regression}, \cite{franco2015borrowing} and \cite{boubeta2016empirical}. The detailed simulation setting for the Poisson-lognormal model is provided in the main document.

\begin{table}[htb]
	\centering
	\begin{adjustbox}{max width=\textwidth}
		\begin{tabular}{ccccccc}\hline
			& $\hat{\beta}_0$ & $\hat{\beta}_1$ & $\hat{\beta}_2$ & $\hat{\beta}_3$ & $\hat{\beta}_4$ & $\hat{\delta}$ \\\hline
			$D=26$ & 0.009 (0.219)   & 0.010 (0.764)   & -0.009 (0.986)  & 0.028 (0.656)   & -0.032 (1.582)  & -0.111 (0.179) \\
			$D=52$ & 0.001 (0.150)   & -0.006 (0.451)  & -0.012 (0.705)  & -0.005 (0.410)  & -0.026 (1.079)  & -0.052 (0.111) \\
			$D=78$ & 0.000 (0.122)   & -0.009 (0.349)  & -0.009 (0.613)  & -0.002 (0.343)  & -0.017 (0.875)  & -0.030 (0.083)\\\hline
		\end{tabular}
	\end{adjustbox}
	\caption{RBIAS and RRMSE (in parenthesis) for ML estimators under area-level Poisson-lognormal model.}
	\label{tab:rb_rm_Poisson_l}
\end{table}

\begin{figure}[htb]
	\centering           \vspace{-0.7cm}
	\subfloat{\includegraphics[width=0.33\textwidth]{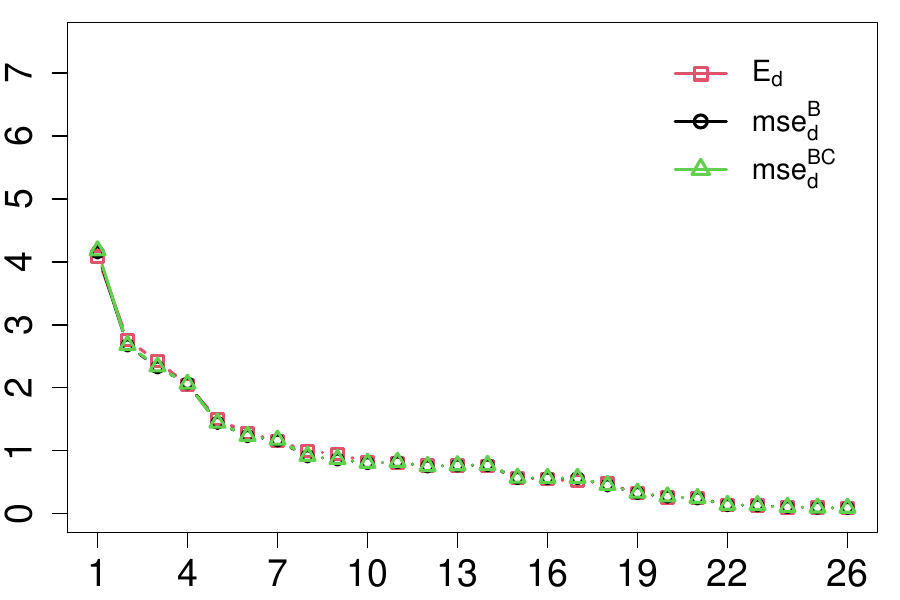}}
	\subfloat{\includegraphics[width=0.33\textwidth]{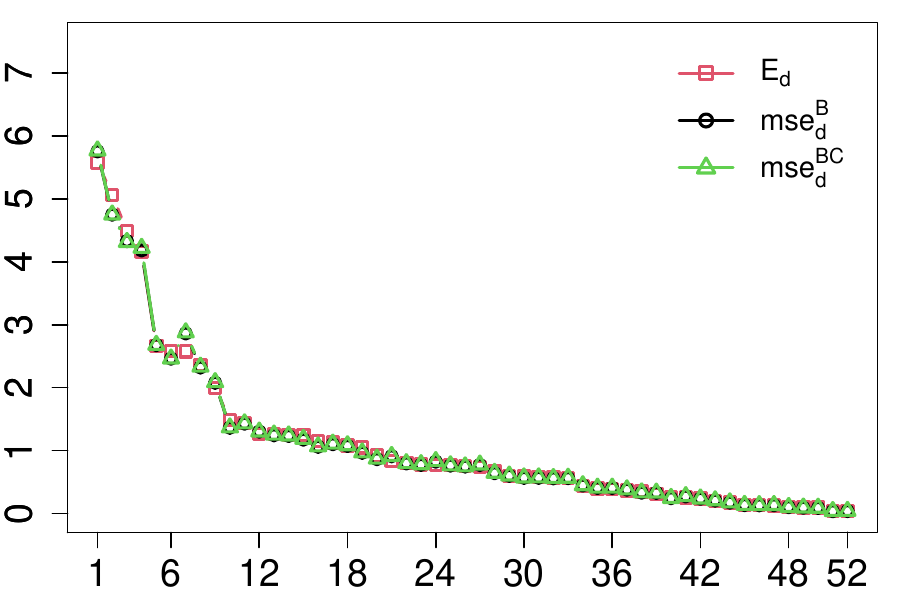}}
	\subfloat{\includegraphics[width=0.33\textwidth]{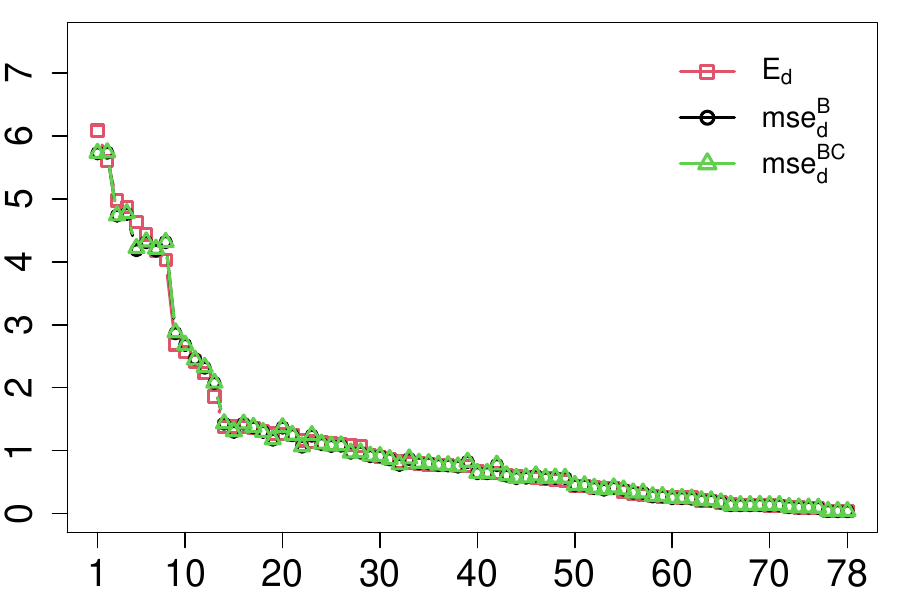}}   
	\captionof{figure}{MSE estimators ($\times10^4$): (left) D = 26, (middle) D = 52 and (right) D = 78 under area-level Poisson-lognormal model.} 
	\label{fig:MSE_est_Poisson_l}
\end{figure}

\begin{table}[htb]
	\centering
	\begin{adjustbox}{max width=\textwidth}
		\begin{tabular}{ccccccccc}\hline
			$D$    & $d$ & $B_d(E_d)$         & $D$    & $d$ & $B_d(E_d)$         & $D$    & $d$ & $B_d(E_d)$ \\\hline
			$26$     & $6$   & -0.00005 (0.00000) & $52$     & $11$  & -0.00008 (0.00000) & $78$     & 16  & -0.00006 (0.00000)  \\
			& 11  & -0.00001 (0.00001) &        & 21  & -0.00002 (0.00001) &        & 32  & -0.00001 (0.00001) \\
			& 16  & 0.00002 (0.00001)  &        & 32  & 0.00002 (0.00001)  &        & 47  & 0.00002 (0.00001)  \\
			& 21  & 0.00003 (0.00001)  &        & 42  & 0.00007 (0.00001)  &        & 63  & 0.00005 (0.00001)  \\
			$B(E)$ &    & 0.00007 (0.00001)  & $B(E)$ &    & 0.00008 (0.00001)  & $B(E)$ &    & 0.00008 (0.00001) \\\hline
		\end{tabular}
	\end{adjustbox}
	\caption{$B_d$, $E_d$ and their averages $B$ and $E$ for the estimators of ${\rho}_d$ under area-level Poisson-lognormal model.}
	\label{tab:B_E_p_d_poisson_l}
\end{table}
We study first the estimators of the fixed parameters and the variance parameters. Table \ref{tab:rb_rm_Poisson_l} displays RBIAS and RRMSE of $\bm{\theta}$. Similarly to the results in Section \ref{sec:poisson_sim}, they are decreasing with a growing number of areas. Recall that in the main document we examined the estimation of proportion for each area. Assuming the Poisson-lognormal area-level model we considered  $\zeta_d = \rho_d$ with $\nu_d = N_d$. Table \ref{tab:B_E_p_d_poisson_l} confirms a very good performance of the EBP $\hat{\rho}_d$. In Figure \ref{fig:MSE_est_Poisson_l} we can observe that also the MSE bootstrap estimators for the EBP perform well (under this model, we did not implement an analytical MSE nor an estimator for $g_{1d}$). 

\begin{figure}[htb]
	\centering       
	\includegraphics[height=0.99\textwidth, angle=-90]{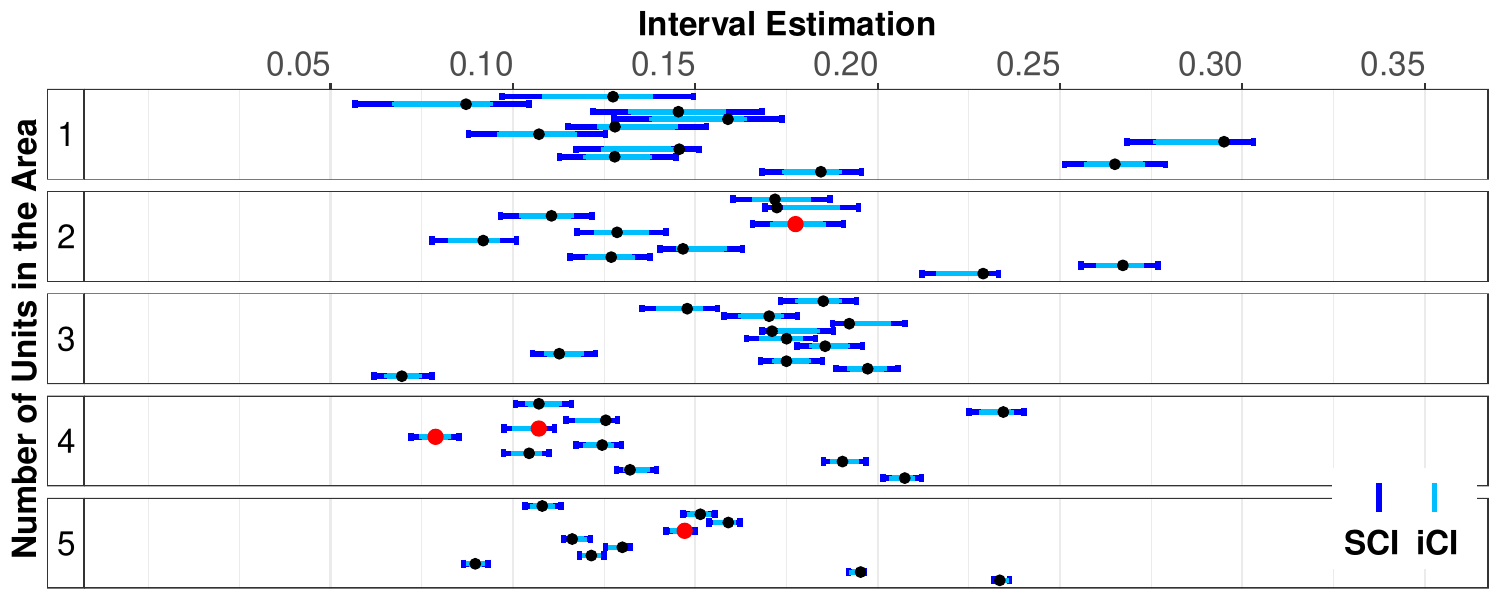}
	\captionof{figure}{iCI and bootstrap SCI for proportions under the area-level Poisson-lognormal model, $D=52$. Red dots indicate true parameters outside iCI, whereas black dots indicate true parameters inside their iCI.} 
	\label{fig:int_est_c52area2}
\end{figure}

Figure \ref{fig:int_est_c52area2} presents iCI and our SCI for proportions constructed under the area-level Poisson-lognormal model. Their behavior is similar to the figures for the Poisson-gamma model in the main article. Observe that four out of true parameters (i.e., $7.7\%$) are not contained in their iCIs whereas our SCIs cover all of them.

\begin{figure}[htb]
	\centering
	\subfloat{\includegraphics[width=0.33\textwidth]{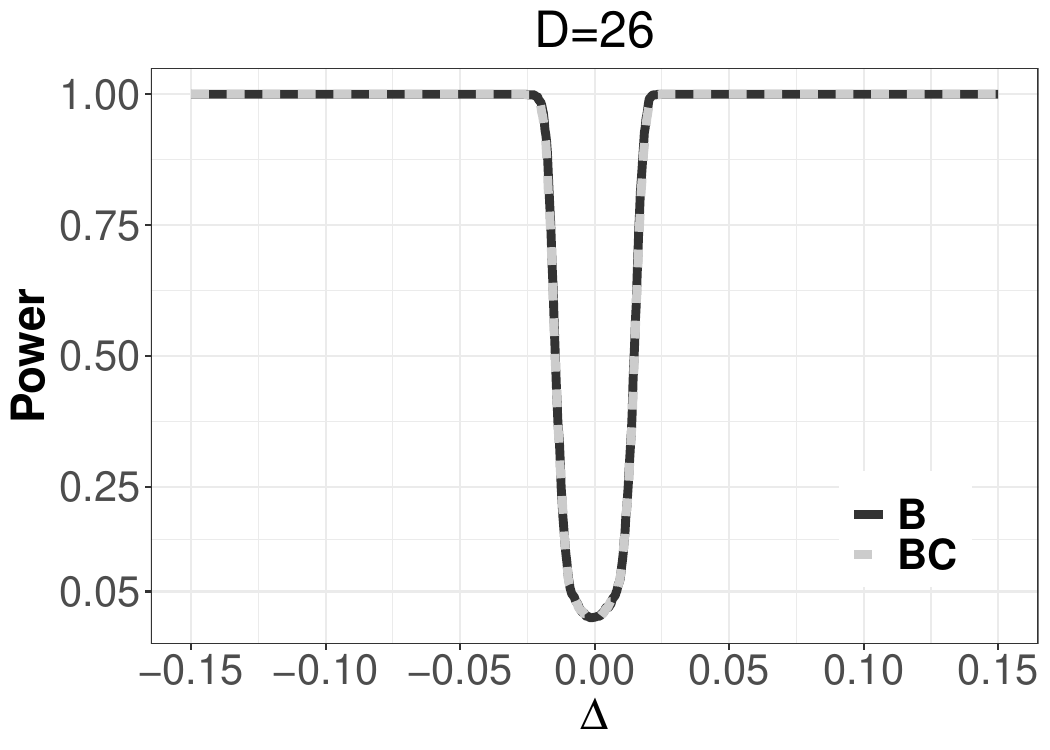}}
	\subfloat{\includegraphics[width=0.33\textwidth]{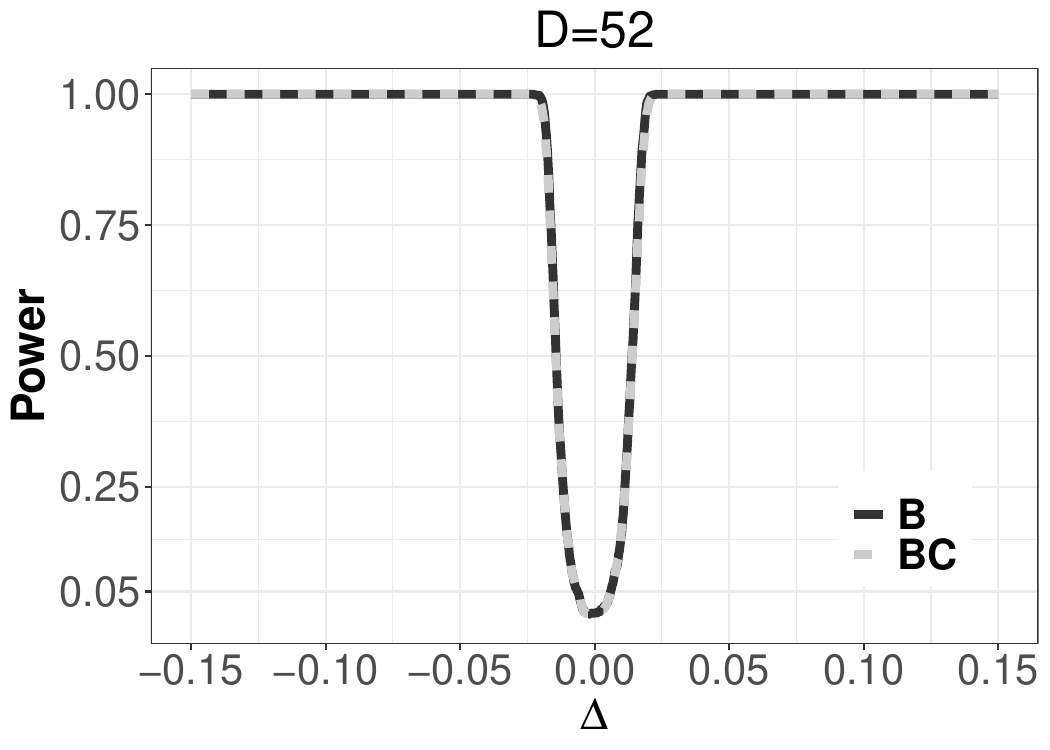}}
	\subfloat{\includegraphics[width=0.33\textwidth]{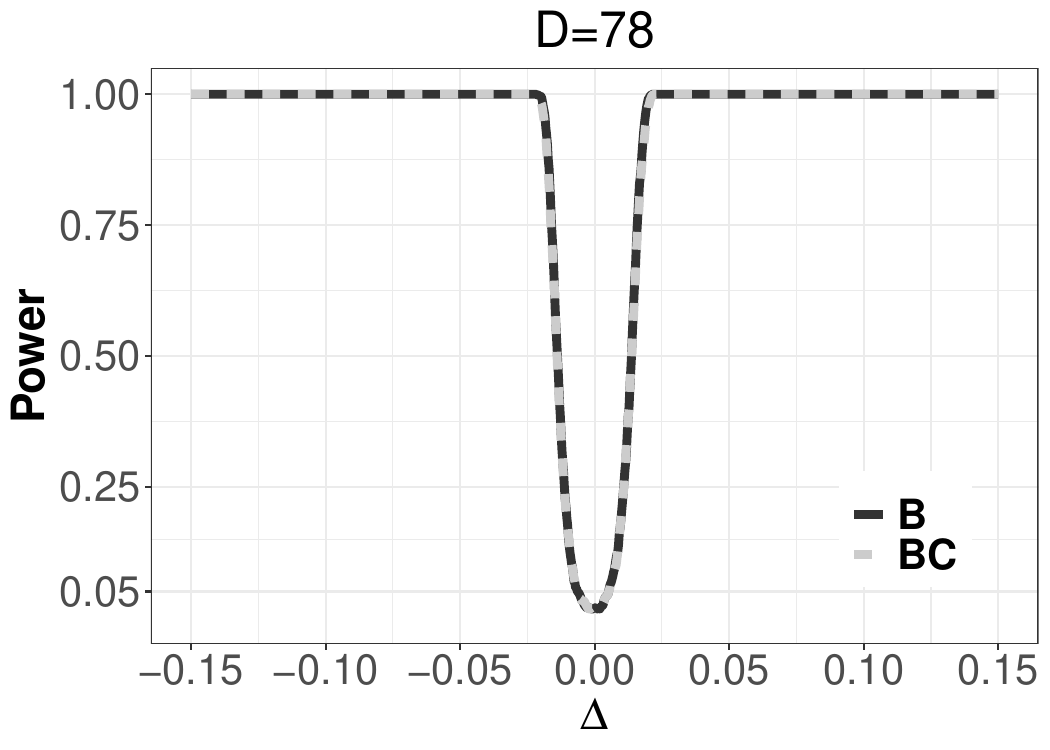}}   
	\captionof{figure}{Simulated power for testing $H_0:\bar{\bm{\mu}}=\bm{b}$ versus $H_1:\bar{\bm{\mu}}=\bm{b}+\bm{1}_{D}\Delta$ under the area-level Poisson-lognormal model; (left) $D=26$, (middle) $D=52$, (right) $D=78$.} 
	\label{fig:test_area2}
\end{figure}

Finally, Figure \ref{fig:test_area2} shows simulated powers for our simultaneous testing procedure. Their behaviour is similar to the analogous figures assuming the Poisson-gamma model in the main article.

\subsection{Additional details on the unit-level logit model}\label{sec:App_Bin}

\subsubsection{Estimation of the parameters}\label{sec:App_Bin_Est}

The unit-level binomial model is widely applied for counts or binary responses, and it has been comprehensively discussed by, e.g., \cite{hobza2016empirical}. We propose a different estimation method. However, since almost all theoretical assumptions are more or less identical to those proposed by the authors we are going to try to be as concise as possible not to hamper the presentation of the results. 

There exist many methods to estimate the vector of model parameters $\bm{\theta}$. A direct maximization of the log-likelihood cannot be easily performed due to the presence of an integral in the log-likelihood function. Nevertheless, one can approximate the integral using adaptive Gaussian quadrature (AGQ) or approximate the integrand applying Laplace approximation. We proceed with the former as it can be seen as a higher order version of the latter, i.e., it provides smaller approximation error \citep{bianconcini2014asymptotic}. AGQ is popular and has been implemented by, e.g., \cite{rabe2005maximum} and \cite{joe2008accuracy}. Another idea is to use the method of simulated moments suggested by \cite{jiang1998consistent} which was employed explicitly to estimate model parameters under unit-level logit and area-level Poisson-lognormal models by \cite{hobza2016empirical} and \cite{boubeta2016empirical} respectively. In what follows, we  proceed with the likelihood based methods which might be considered as a comparative study to the two aforementioned. Since our derivation follows tightly the theory developed by \cite{bianconcini2014asymptotic} under generalized linear latent variable model (GLLVM), with some notational changes, but otherwise almost identical, we provide only the most important points.  

Consider a GLMM with one dimensional normally distributed random effect. In this case, the structure of the exponential family allows us to write the likelihood contribution from each area $d$ in a simplified form defining functions $i_d$
\begin{equation*}
	i_d(\bm{\theta},u_d)=\sum_{j=1}^{n_d}\left\{\frac{y_{dj}\gamma_{dj}-b(\gamma_{dj})}{\varphi}+c(y_{dj},\varphi) \right\}-\frac{1}{2}u_d^2  .
\end{equation*}
Contributions from each area are conditionally independent. Therefore, the marginal distribution of the whole data vector under the exponential family is given by 
\begin{equation*}\label{eq:whole_like}
	\mathcal{L}^{G}(\bm{\theta})\coloneqq f^G(\bm{y}|\bm{\theta})=\prod_{d=1}^{D}f^G_d(\bm{y}_d|\bm{\theta})
	=(2\pi)^{-D/2}\prod_{d=1}^{D}\int_{\mathbb{R}}\exp\left\{ i_d(\bm{\theta},u_d)\right\}\mathrm{d}u_d  .
\end{equation*}
The log-likelihood is then \begin{equation}\label{eq:loglikelihood}
	l^G(\bm{\theta})=-\frac{D}{2}\log(2\pi)+
	\sum_{d=1}^{D} \log \int_{\mathbb{R}} \exp\left\{ i_d(\bm{\theta},u_d)\right\} \mathrm{d}u_d  .
\end{equation}
The asymptotic properties of AGQ were studied by \cite{bianconcini2014asymptotic} under GLLVM. Due to the close similarity between both models, we can use their results with a slightly different notation. Under GLMM a score equation is given by
\begin{equation}\label{eq:score}
	\begin{split}
		\bm{R}^G(\bm{\theta})&=\sum_{d=1}^{D}\frac{\partial \log f^G_d(\bm{y}_d|\bm{\theta}) }{\partial \bm{\theta}}=\sum_{d=1}^{D}
		\frac{1}{f^G_d(\bm{y}_d|\bm{\theta})}\int_{\mathbb{R}} \frac{\partial }{\partial \bm{\theta}}\left\{g_d(\bm{y}_d|u_d,\bm{\theta})  h(u_d)  \right\} \mathrm{d}u_d\\
		&
		=\sum_{d=1}^{D}\int_{\mathbb{R}} \bm{R}^G_d(\bm{\theta};u_d)h(u_d|\bm{y}_d;\bm{\theta})\mathrm{d}u_d=\sum_{d=1}^{D}\mathbb{E}_{\bm{u}|\bm{y}}\left\{ \bm{R}^G_d(\bm{\theta};u_d)\right\},
	\end{split}
\end{equation}
where $ \bm{R}^G_d(\bm{\theta};u_d)$ denotes $\partial \log f^G_d(\bm{y}_d|\bm{\theta}) /\partial\bm{\theta}=  \partial \{ \log g_d(\bm{y}_d|u_d,\bm{\theta})+\log h(u_d)  \}/\partial \bm{\theta}$. The application of the AGQ requires rewriting the equation for a conditional likelihood under a specific form, namely 
\begin{equation}\label{eq:f_d_AGQ}
	f^G_d(\bm{y}_d|\bm{\theta})=\int_{\mathbb{R}}\frac{g_d(\bm{y}_d|u_d,\bm{\theta})  h(u_d) h_1(u_d|\hat{u}_d,\hat{\delta}) }{h_1(u_d|\hat{u}_d,\hat{\delta}) }\text{d}u_d,
\end{equation}
where $h_1(\cdot|\hat{u}_d,\hat{\delta})$ is a normal p.d.f.\ with the following first and second moments
\begin{equation*}
	\begin{split}
		\hat{u}_d&=\argmaxA_{u_d\in \mathbb{R}} \left\{\log g_d(\bm{y}_d|u_d,\bm{\theta})+\log h(u_d) \right\},\\ \hat{\delta}^2&= \left[\frac{-\partial^2 \left\{ \log g_d(\bm{y}_d|u_d,\bm{\theta})+\log h(u_d)  \right\}}{\partial^2 u_d} \right]\Bigg\lvert_{u_d=\hat{u}_d}^{-1}  .
	\end{split}
\end{equation*}
For multidimensional random area effects, one would need to find a Cholesky decomposition of the variance covariance matrix \citep[see][]{tuerlinckx2006statistical,bianconcini2014asymptotic}. In case of one dimensional random effect we can just take a square root of $\hat{\delta}^2$. Furthermore, equation \eqref{eq:f_d_AGQ} can be approximated by
\begin{equation}\label{eq:AGH}
	f^G_d(\bm{y}_d|\bm{\theta},\varphi)\approx\hat{f}^G_d(\bm{y}_d|\bm{\theta}, \varphi)=2^{1/2}\hat{\delta} \sum_{r=1}^{m}g_d(\bm{y}_d|t^A_{dr},\bm{\theta}^e )h_d(t^A_{dr})w^A_r,
\end{equation}
where  $t^A_{dr}=\hat{u}_d+\hat{\delta}\sqrt{2}t_{r}$ and $w^A_{r}=\exp(t^2_{r})w_{r}$ denote AGQ nodes and weights, with $t_{dr}$ and $w_r$ being the classical Gaussian quadrature nodes and weights respectively \citep{abramowitz1966handbook}. Using \eqref{eq:AGH} we can obtain the approximation of the whole likelihood \eqref{eq:loglikelihood}
\begin{equation}\label{eq:loglikelihood_hat}
	\hat{l}^G(\bm{\theta})=-\frac{D}{2}\log(2\pi)+\sum_{d=1}^{D} \log \left[ 2^{1/2} \hat{\delta} \sum_{r=1}^{m} \exp\left\{\frac{ y_{dj}\gamma^A_{dj}-b(\gamma^A_{dj})}{\varphi } +c(y_{dj},\varphi)  
	-\frac{t_{dr}^{A2}}{2}\right\} w^A_r\right]   .
\end{equation}
To estimate the model parameters, we use the score equation based on \eqref{eq:loglikelihood_hat} 
\begin{equation}\label{eq:score_AGH}
	\hat{\bm{R}}^G(\bm{\theta})=\frac{\partial \hat{l}^G(\bm{\theta})}{\partial \bm{\theta}}=\sum_{d=1}^{D}\frac{ \sum_{r=1}^{m}\bm{R}^G_d(\bm{\theta},t_{dr}^A) g(\bm{y}_d|\bm{\theta},t^A_{dr} )h(t^A_{dr})w^A_r}{ \sum_{r=1}^{m}g(\bm{y}_d|\bm{\theta},t^A_{dr} )h(t^A_{dr})w^A_r }=\sum_{d=1}^{D}\hat{\mathbb{E}}_{\bm{u}|\bm{y}}\{\mathcal{S}_d(\bm{\theta},u_d)\}  .
\end{equation}
Further details for the general derivatives under GLVMM are given by \cite{bianconcini2014asymptotic}.

As far as the statistical properties of AGQ based estimators are taken into account, \cite{bianconcini2014asymptotic} proved their consistency 
which follows from the relation to the Laplace approximation. Moreover, the estimators are asymptotically normally distributed since they belong to the class of M-estimators \citep{huber1964robust} which are implicitly defined by a function $\Psi$ using equation $\sum_{d=1}^{D}\Psi(\bm{y}_d,\bm{\theta})=0$.
Note that the AGQ $\Psi$ function is given in equation \eqref{eq:score_AGH}. As pointed out by \cite{huber2004estimation}, one needs to verify if the regularity conditions on the log-likelihood function for the consistency and the asymptotic normality (see condition 2 in Appendix A.1 of the main article) are satisfied for each p.d.f. of $y_{dj}$. In case of the binomial model these were already checked by \cite{bianconcini2014asymptotic} under GLVMM and the proof for GLMM would be identical with a slight change of notation. 
As far the unit-level model is considered, we can use equation \eqref{eq:loglikelihood_hat} to obtain the approximated likelihood
\begin{equation*}
	\begin{split}
		\hat{l}^G(\bm{\theta})&=\sum_{d=1}^{D} \left( -\frac{1}{2}\log\pi +\log\hat{\sigma}+\log \left[\sum_{r=1}^{m} \exp\left\{ \sum_{j=1}^{n_d}\log{m_{dj}\choose y_{dj}} +\sum_{j=1}^{n_d} y_{dj}(  \bm{x}_{dj}^t\bm{\beta}+\delta t^A_{dk} )\right .\right .\right. \\   &\left .\left .\left.- \sum_{j=1}^{n_d}m_{dj}\log[1+\exp\{ \bm{x}_{dj}^t\bm{\beta}+\delta t^A_{dk}   \}   ] -\frac{t^{A2}_{dk}}{2} \right\} w^A_{dk}  \right]\right),
	\end{split}
\end{equation*}
and the resulting score equations for each parameter
\begin{equation*}
	\begin{split}
		\hat{\bm{R}}_d(\beta_i;t^{A}_{dk})&=\sum_{j=1}^{n_d}\left[y_{dj}x_{dji}+\frac{m_{dj}\exp(x_{dji}\beta_i-\delta t^A_{dk})x_{dji}}{1+\exp(x_{dji}\beta_i+\delta t^A_{dk})}\right],\quad i=1,\dots,p\\\hat{\bm{R}}_d(\delta;t^{A}_{dk})&=\sum_{j=1}^{n_d}\left[y_{dj}t^A_{dk}+\frac{m_{dj}\exp(x_{dji}\beta_i-\delta t^A_{dk})t^A_{dk}}{1+\exp(x_{dji}\beta_i+\delta t^A_{dk})}\right] .
	\end{split}
\end{equation*}
The equations might be solved using, for example, quasi Newton-Raphson \citep{bianconcini2014asymptotic} or a different numerical method suitable to solve nonlinear equations. Notice that under this class of models fast implementations of AGQ are available in some packages of the statistical software language \textbf{\textsf{R}}.

\begin{remark}
	AGQ might be inappropriate for the area-level model due to a potential lack of consistency, which can be seen from the rate of convergence in \cite{bianconcini2014asymptotic}. 	To circumvent this problem, \cite{jiang1998asymptotic} suggests using the method of simulated moments which exhibits an asymptotic order $O(D^{-1})$ and does not depend on $n_d$. This approach was applied,  e.g., in \cite{hobza2016empirical}. 
\end{remark}

\begin{remark}
	The numerical results for our different simulation settings are not directly comparable, because the data generation processes are different to guarantee that each method is studied with reference to the assumed model. A comparative study based on a common model-design would be interesting but has a different focus and it might be a topic for future research; \cite{namazi2015level} and \cite{hidiroglou2016comparison} have conducted interesting studies in this direction.
	Nevertheless, our findings are consistent with a general statistical theory: a more complex modeling with random effects and artificially created covariate classes inflates the variability and leads to noisier estimators.
	Moreover, fitting unit-level models is computationally more expensive. For our sample sizes the estimation of MSE and construction of intervals took about 900 to 1000 times longer. To be more specific, the computation performed using a laptop with 32 GB RAM and an Intel$^{\textsuperscript{\textregistered}}$ Core$^{\text{\tiny{TM}}}$ i9-9980HK processor with CPU 2.40 GHz, takes up to 2 days, whereas for the area-level model just about 3 minutes. In addition, EBP under the unit-level model is estimated using 16 different covariate classes, which requires the estimation of the size of 16 times more artificially created clusters. Moreover, GLMM with random effects and numerical approximations inflates the variance. 
\end{remark}

\subsubsection{Estimation of EBP }\label{sec:EBP_Bin}

Under the unit-level binomial model, BP $\tilde{\bar{\mu}}^U_d$ of $\bar{\mu}^U_d$ defined in the main article can be obtained using similar derivations as in \cite{hobza2016empirical}, that is
\begin{eqnarray*} 
	&&\tilde{r}_{dl}(\bm{\theta})\coloneqq\mathbb{E}(r_{dl}|\bm{y}_d)=\frac{\int_{\mathbb{R}} \frac{\exp(\bm{z}^t_{l}\bm{\beta} +\delta u_d )}{1 +\exp{(\bm{z}^t_{l}\bm{\beta} +\delta u_d )}} g^U(\bm{y}_d|u_d,\bm{\theta})h(u_d) du_d  }{\int_{\mathbb{R}} g^U(\bm{y}_d|u_d,\bm{\theta})h(u_d) du_d  }=\frac{A^{U}_{dl}(y_{dj},\bm{\theta})}{C^U_d(\bm{y}_{d},\bm{\theta})},\\
	&&\tilde{\mu}^U_d(\bm{\theta})\coloneqq\mathbb{E}(\mu^U_{d}|\bm{y}_d)=\sum_{l=1}^{L}N_{dl}\tilde{r}_{dl}(\bm{\theta} )\eqqcolon\psi^U_d(\bm{d},\bm{\theta}) \quad \text{and} \quad \tilde{\bar{\mu}}^U_{d}(\bm{\theta})=\frac{\tilde{\mu}^U_d}{N_d},\\
	&&\tilde{u}_d(\bm{\theta})\coloneqq\mathbb{E}(u_{d}|\bm{y}_d)=\frac{\int_{\mathbb{R}} u_d g^U(\bm{y}_d|u_d,\bm{\theta})h(u_d) du_d  }{\int_{\mathbb{R}} g^U(\bm{y}_d|u_d,\bm{\theta})h(u_d) du_d  }=\frac{A^{uU}_{d}(\bm{y}_{d},\bm{\theta})}{C^U_d(\bm{y}_{d},\bm{\theta})} , 
\end{eqnarray*}
where 
\begin{eqnarray*}
	A^{U}_{dl}&=&\int_{\mathbb{R}}\frac{\exp(\bm{z}^t_{l}\bm{\beta}+\delta u_d)}{1+\exp(\bm{z}_{l}^t\bm{\beta}+\delta u_d )} 
	\exp\left[   \delta y_{d\cdot}u_d-\sum_{j=1}^{n_d} m_{dj}\log \{1+\exp(\bm{x}^t_{dj}\bm{\beta} +\delta u_d  )   \}     \right]  h(u_d) \mathrm{d}u_d,\\
	A^{uU}_{d}&=&\int_{\mathbb{R}} u_d 
	\exp\left[    \delta y_{d\cdot}u_d-\sum_{j=1}^{n_d} m_{dj}\log \{1+\exp(\bm{x}^t_{dj}\bm{\beta} +\delta u_d  )   \}\right]  h(u_d) \mathrm{d}u_d,   \\  
	C^U_d&=&\int_{\mathbb{R}} \exp\left[    \delta y_{d\cdot}u_d-\sum_{j=1}^{n_d} m_{dj}\log \{1+\exp(\bm{x}^t_{dj}\bm{\beta} +\delta u_d  )   \}  \right]  h(u_d) \mathrm{d}u_d  .
\end{eqnarray*}
The EBP equivalents of $\tilde{r}_{dl}$, $\tilde{\mu}^U_{d}$, $\tilde{\bar{\mu}}^U_{d}$ and $\tilde{u}_d$ are $\hat{p}^c_{dl}=\tilde{p}^c_{dl}(\hat{\bm{\theta}})$, $\hat{\mu}^U_{d}=\tilde{\mu}^U_{d}(\hat{\bm{\theta}})$, $\hat{\bar{\mu}}^U_{d}=\tilde{\bar{\mu}}^U_{d}(\hat{\bm{\theta}})$ and $\hat{u}_d=\tilde{u}_d(\hat{\bm{\theta}})$, where $\hat{\bm{\theta}}$ denotes a consistent estimator of $\bm{\theta}$. They can be calculated using a Monte Carlo simulation, as described in \cite{hobza2016empirical}.

\subsubsection{Finite sample performance}\label{sec:binomial_NR}
A detailed setting of the simulation under the unit-level model is described in the main document. Employing the simulation steps of \cite{hobza2016empirical}, p.671, we provide below a shortened version of the algorithm to obtain SCIs under unit-level model. 
\begin{enumerate}
	\item Follow the steps of the parametric bootstrap procedure from \cite{hobza2016empirical}, p.671.
	\item Calculate $AD^{(b_1)}_{U,d}=\left\lvert \hat{\bar{\mu}}^{U*(b_1)}_d -\bar{\mu}^{U*(b_1)}_d  \right\rvert$.
	\item Compute statistic $S_{U,B}$ and the critical value $q_{U,S_B}^{(1-\alpha)}$
	\begin{equation*}
		S^{(b_1)}_{U,B}=\max_{d=1,\dots, D}\frac{AD_{U,d}^{*(b_1)}}{\hat{\sigma}^{*} }\quad 
		\text{and}\quad  q_{U,S_U}^{(1-\alpha)}=Q_{1-\alpha}(\bm{S}_{U,B})
		\quad \text{where}\quad \bm{S}_{U,B}=(S^{(1)}_{U,B},\dots S^{(B_1)}_{U,B})^t, 
	\end{equation*}
	as well as the variance of $\bm{\theta}$, setting $\bar{\bm{\theta}}=\frac{1}{B_1}\sum_{b_1=1}^{B1}\hat{\bm{\theta}}^{*(b_1)}$, and
	\begin{equation*}
		\widehat{var}(\hat{\bm{\theta}})=\frac{1}{B_1}\sum_{b_1=1}^{B_1}(\hat{\bm{\theta}}^{*(b_1)}-\bar{\bm{\theta}})(\hat{\bm{\theta}}^{*(b_1)}-\bar{\bm{\theta}})^t   . 
	\end{equation*}
\end{enumerate}

\begin{table}[htb]
	\centering
	\begin{adjustbox}{max width=\textwidth}
		\begin{tabular}{ccccccc}
			\hline
			D  & $\hat{\beta}_0$  & $\hat{\beta}_1$   & $\hat{\beta}_2$ & $\hat{\beta}_3$  & $\hat{\beta}_4$  & $\hat{\delta}$   \\\hline
			26 & 0.0027 (0.0451)  & -0.0050 (0.0682)  & 0.0058 (0.2990) & -0.0083 (0.1017) & -0.0485 (0.6725) & -0.0572 (0.1903) \\
			52 & -0.0042 (0.0360) & -0.0013 (0.0537 ) & 0.0176 (0.2415) & 0.0088 (0.0726)  & 0.0118 (0.4895)  & -0.0022 (0.1367) \\
			78 & -0.0027 (0.0263)                & 0.0037 (0.0418)                & -0.0046 (0.1773)               & 0.0043 (0.0645)                &  -0.0081 (0.4163)            & -0.0170 (0.1035)             \\\hline
		\end{tabular}
	\end{adjustbox}
	\caption{RBIAS and RRMSE (in parenthesis) for AGQ estimates under unit-level model.}
	\label{tab:rb_rm_log_bin}
\end{table}

\begin{table}[htb]
	\centering
	\begin{adjustbox}{max width=\textwidth}
		\begin{tabular}{ccccccccc}\hline
			D    & d  & $B_d(E_d)$       & D    & d  & $B_d(E_d)$       & D    & d  & $B_d(E_d)$       \\\hline
			26   & 6  & -0.0008 (0.0002) & 52   & 11 & -0.0007 (0.0002) & 78   & 16 & -0.0013 (0.0002) \\
			& 11 & -0.0004 (0.0003) &      & 21 & -0.0001 (0.0003) &      & 32 & -0.0003 (0.0003) \\
			& 16 & -0.0002 (0.0005) &      & 32 & 0.0005 (0.0005)  &      & 47 & 0.0003 (0.0005)  \\
			& 21 & 0.0002 (0.0008)  &      & 42 & 0.0015 (0.0008)  &      & 63 & 0.0013 (0.0008)  \\
			B(E) &    & 0.0007 (0.0005)  & B(E) &    & 0.0010 (0.0005)  & B(E) &    & 0.0012 (0.0005)\\\hline 
		\end{tabular}
	\end{adjustbox}
	\caption{$B_d$, $E_d$ and their averages $B$ and $E$ for $\hat{\bar{\mu}}^U_d$ under unit-level binomial model.}
	\label{tab:B_E_p_d}
\end{table}

\begin{figure}[htb]
	\centering  \vspace{-0.7cm}
	\subfloat{\includegraphics[width=0.45\textwidth]{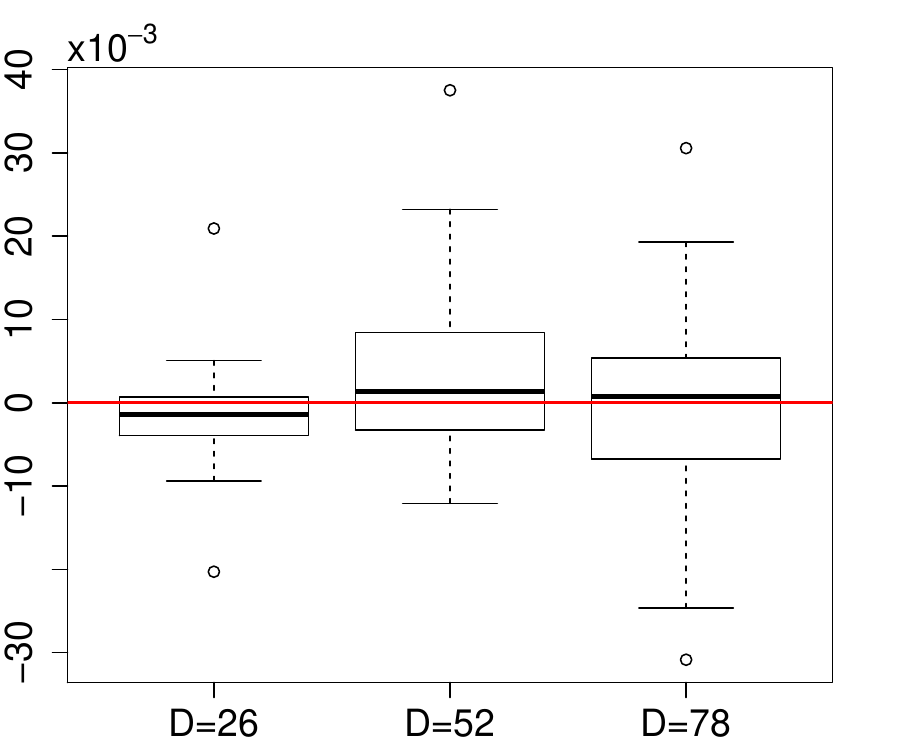}}
	\subfloat{\includegraphics[width=0.45\textwidth]{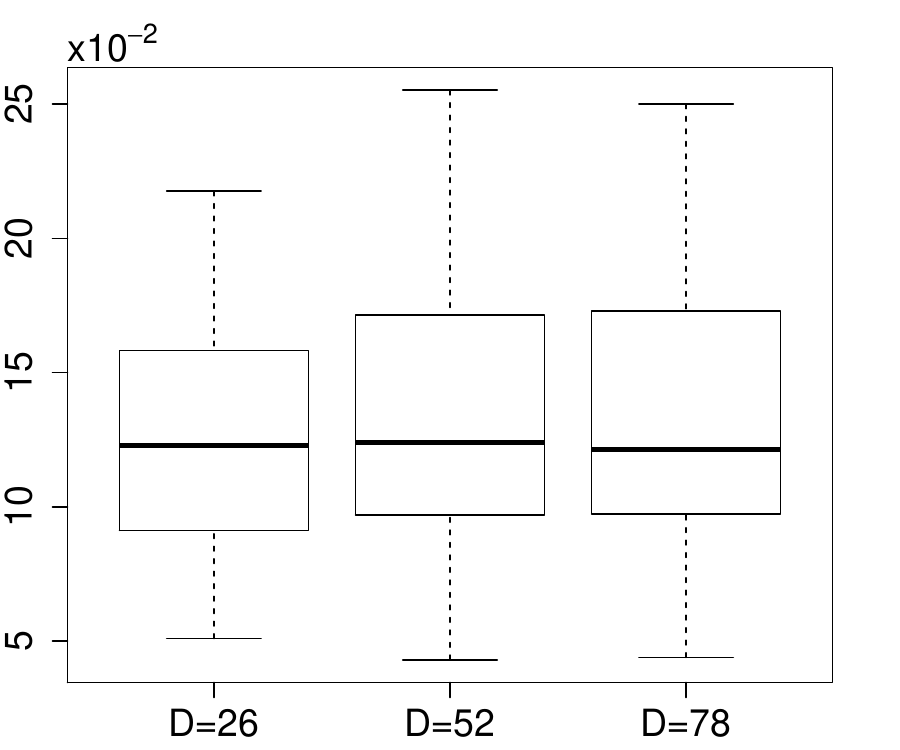}}
	\captionof{figure}{Box plots with (left) RBIAS and (right) RRMSE for $\hat{\bar{\mu}}^U_d$ under the unit-level binomial model.} 
	\label{fig:bias_mse_unit}
\end{figure}
Table \ref{tab:rb_rm_log_bin} outlines RBIAS and RRMSE of $\bm{\theta}$. Similarly to the area-level Poisson model, the quality of estimates improves with a growing sample size. On the other hand, a very good performance of EBP $\hat{\bar{\mu}}^U_d$ is shown in Table \ref{tab:B_E_p_d}. When it comes to the relative bias and the relative root MSE in Figure \ref{fig:bias_mse_unit}, their behaviour is similar to the equivalents under the area-level model, nevertheless with a higher magnitude. 
\begin{figure}[htb]
	\centering    \vspace{-0.7cm}
	\subfloat{\includegraphics[width=0.33\textwidth]{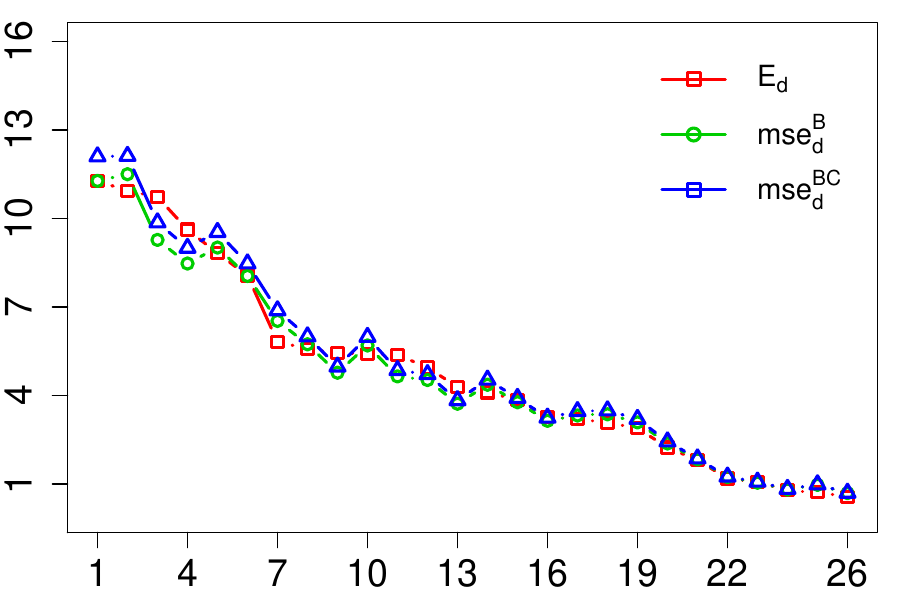}}
	\subfloat{\includegraphics[width=0.33\textwidth]{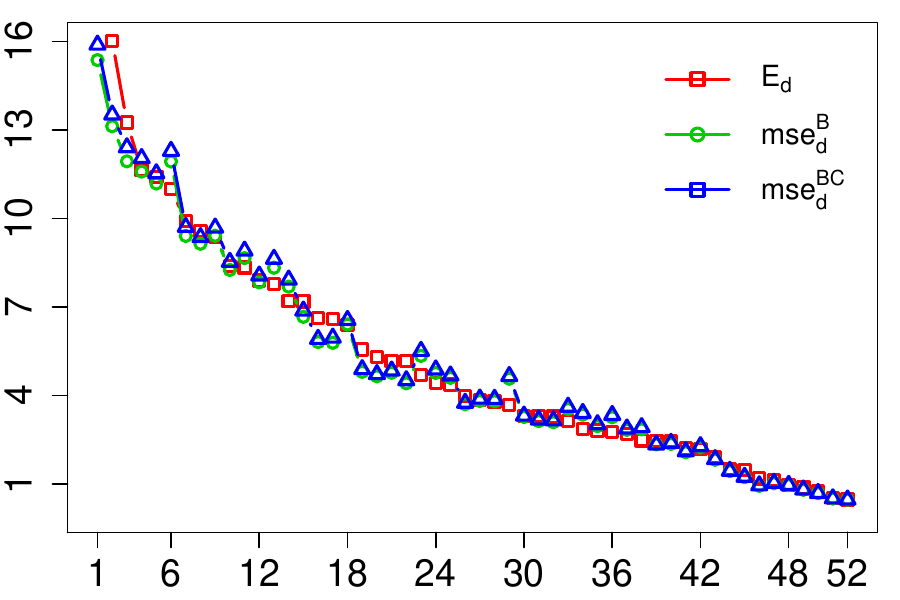}}
	\subfloat{\includegraphics[width=0.33\textwidth]{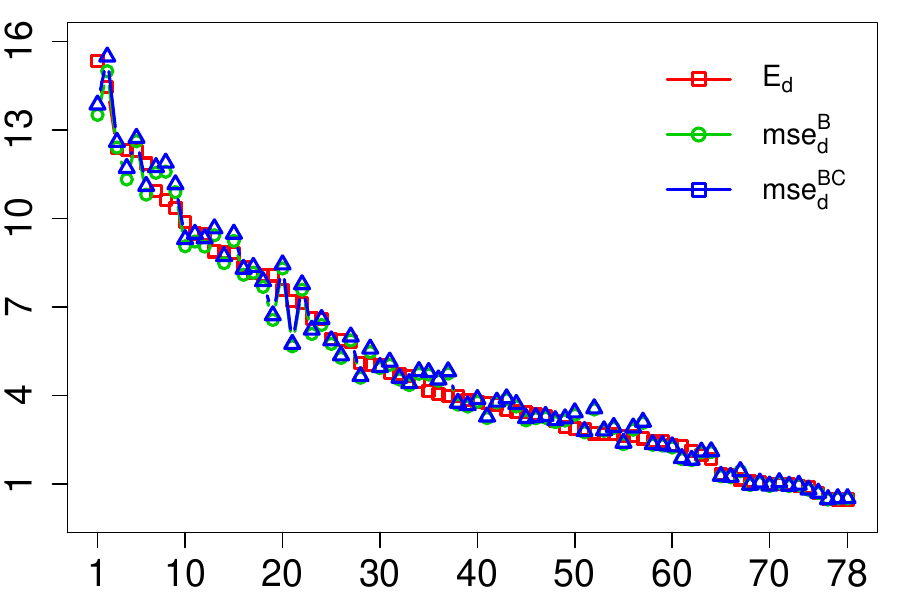}}   
	\captionof{figure}{MSE estimators ($\times10^4$): (left) D = 26, (centre) D = 52 and (right) D = 78 under unit-level binomial model.} 
	\label{fig:MSE_est_unit}
\end{figure}
The estimates of MSE behave similarly as in \cite{hobza2016empirical} as can be seen in Figure \ref{fig:MSE_est_unit}. 
\begin{figure}[htb]
	\centering  
	\includegraphics[height=0.99\textwidth, angle=-90]{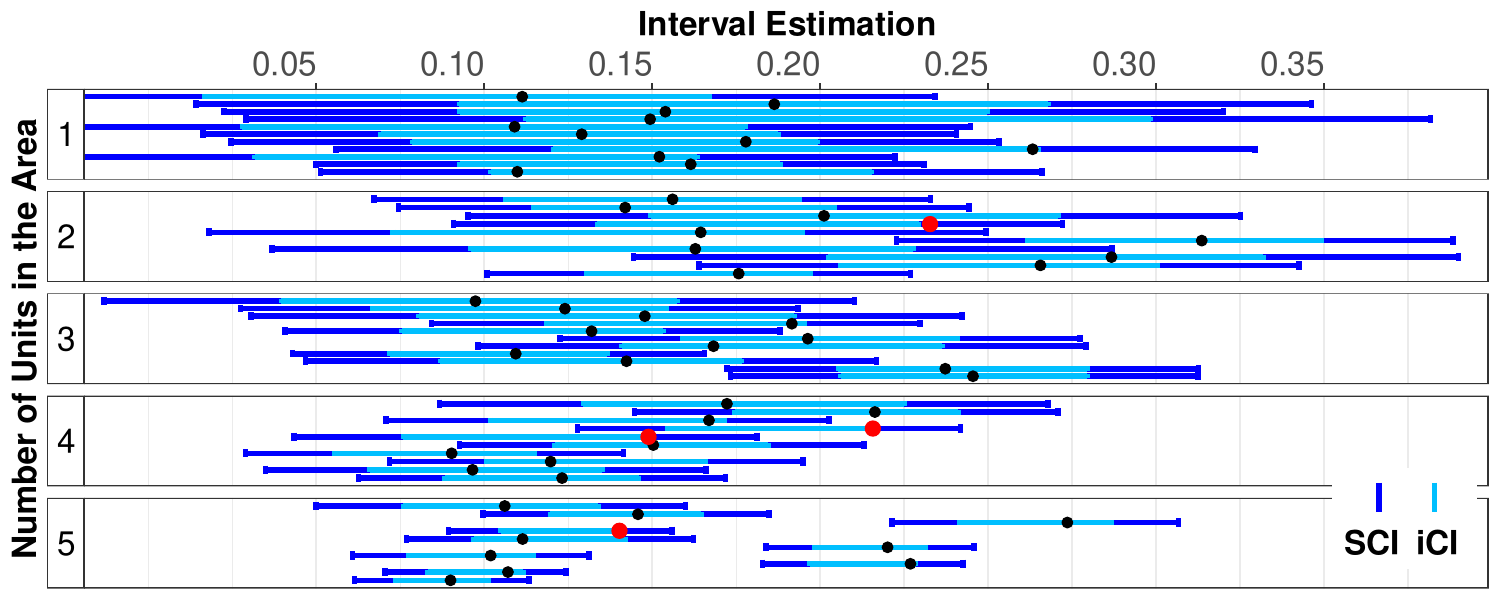}
	\captionof{figure}{iCI and bootstrap SCI for proportions under the unit-level binomial model, $D=52$. Red dots indicate true parameters outside iCI, whereas black dots indicate true parameters inside their iCI.} 
	\label{fig:int_est_95_u}
\end{figure}

Figure \ref{fig:int_est_95_u} displays iCI and SCI under the unit-level binomial model for a randomly selected simulation. Red dots indicate true parameters outside iCI, whereas black dots indicate true parameters inside their iCI. We can draw the same conclusions as under the area-level models.
\begin{figure}[h]
	\centering
	\subfloat{\includegraphics[width=0.33\textwidth]{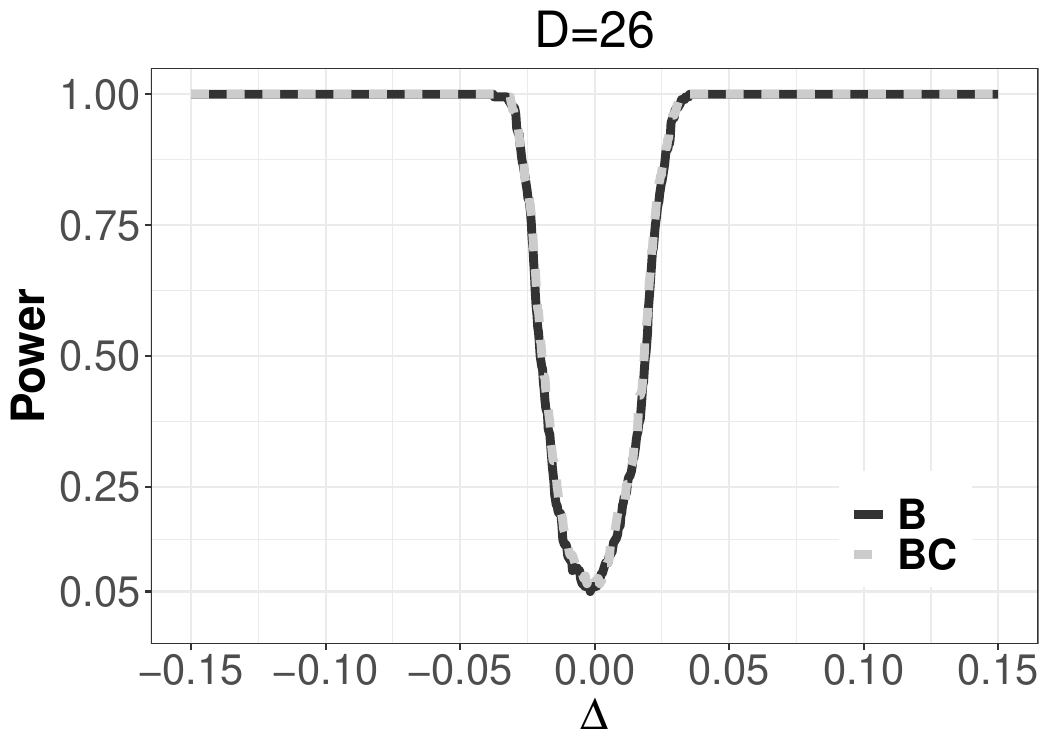}}
	\subfloat{\includegraphics[width=0.33\textwidth]{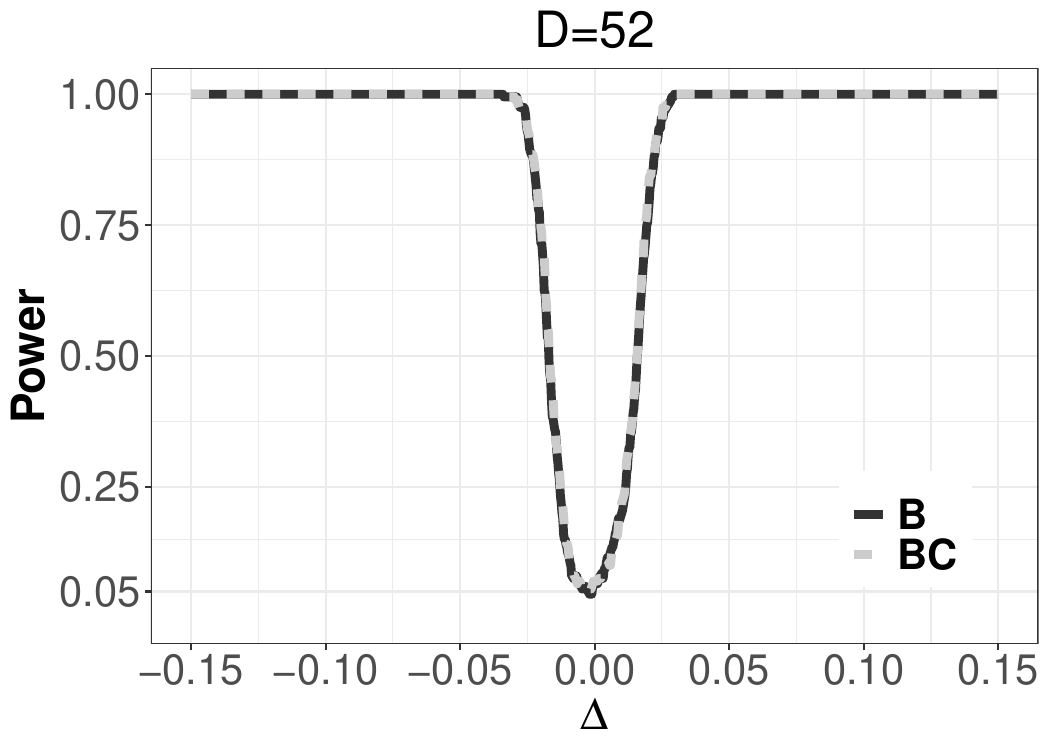}}
	\subfloat{\includegraphics[width=0.33\textwidth]{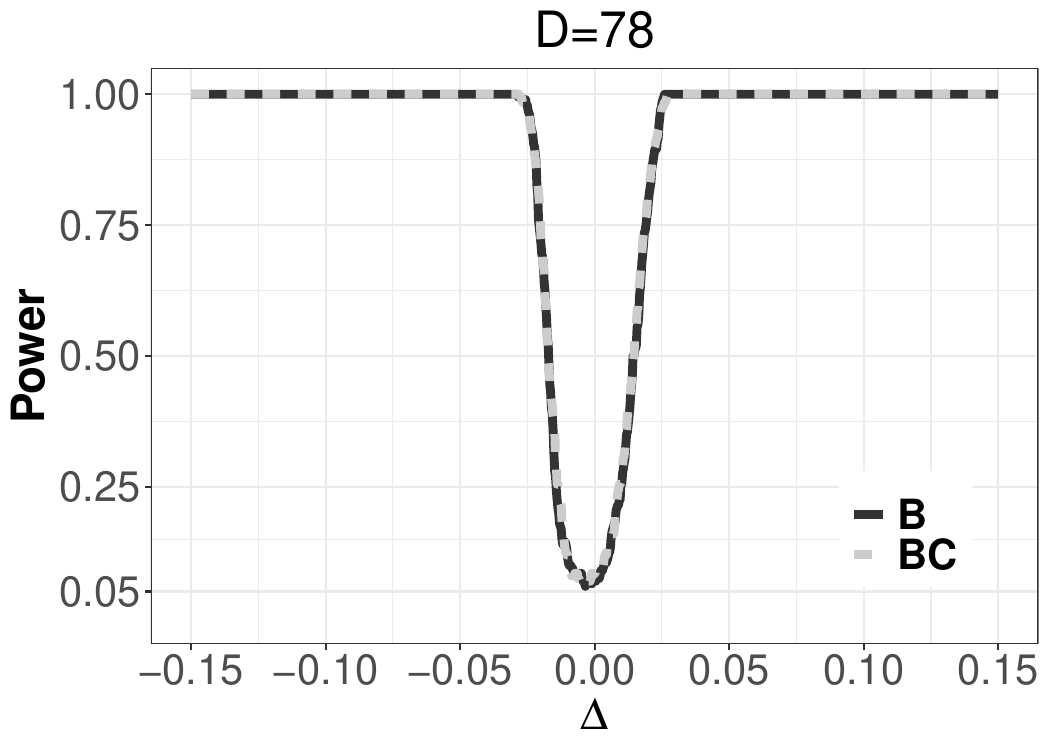}}   
	\captionof{figure}{Simulated power for testing $H_0:\bar{\bm{\mu}}=\bm{h}$ versus $H_1:\bar{\bm{\mu}}=\bm{h}+\bm{1}_{D}\Delta$ under the unit-level binomial model; (left) $D=26$, (middle) $D=52$, (right) $D=78$.} 
	\label{fig:test_res_u}
\end{figure}
Finally, Figure \ref{fig:test_res_u} presents results of our multiple testing procedure. 
Notice that already for $D=26$, the power of our test attains the nominal level of $\alpha=0.05$.

\subsection{Additional details on Predicting Poverty Rates in Galicia}\label{sec:data_example2}
In this section we provide some additional results to complete the case study on the predictions of poverty rates in Galicia in the northern part of Spain. We start with diagnostic plots that are commonly used in small area estimation. 

\begin{figure}[htb]
	\centering
	\subfloat{\includegraphics[width=0.33\textwidth]{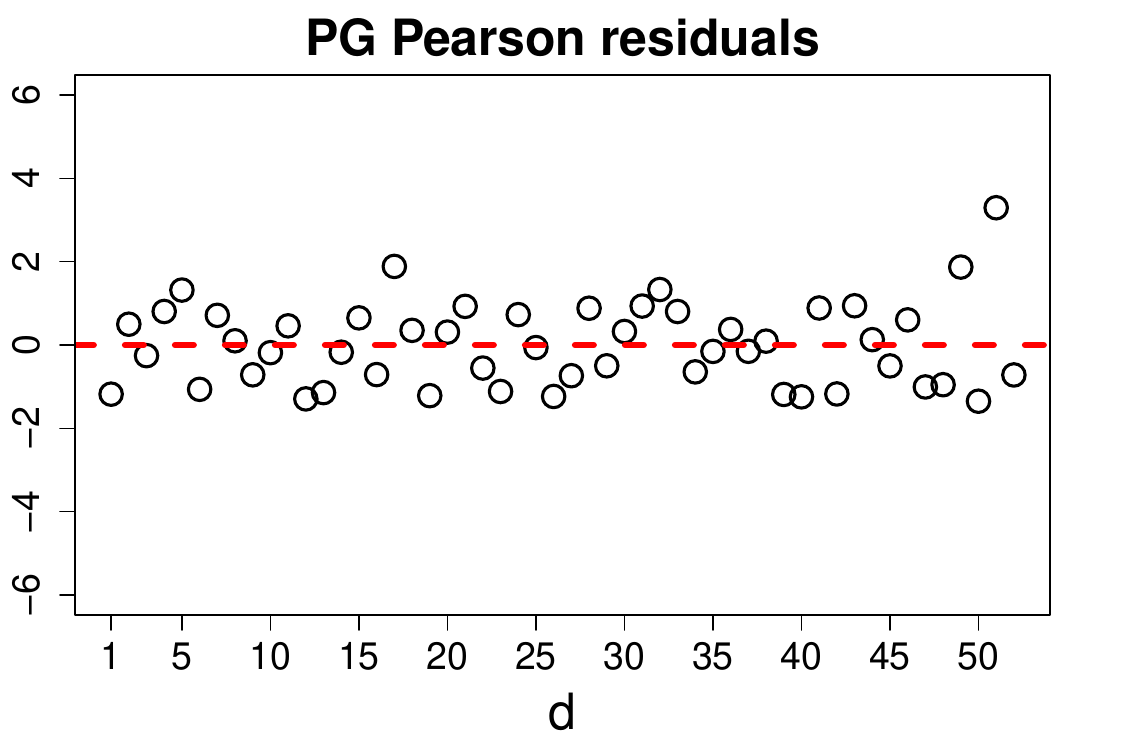}}
	\subfloat{\includegraphics[width=0.33\textwidth]{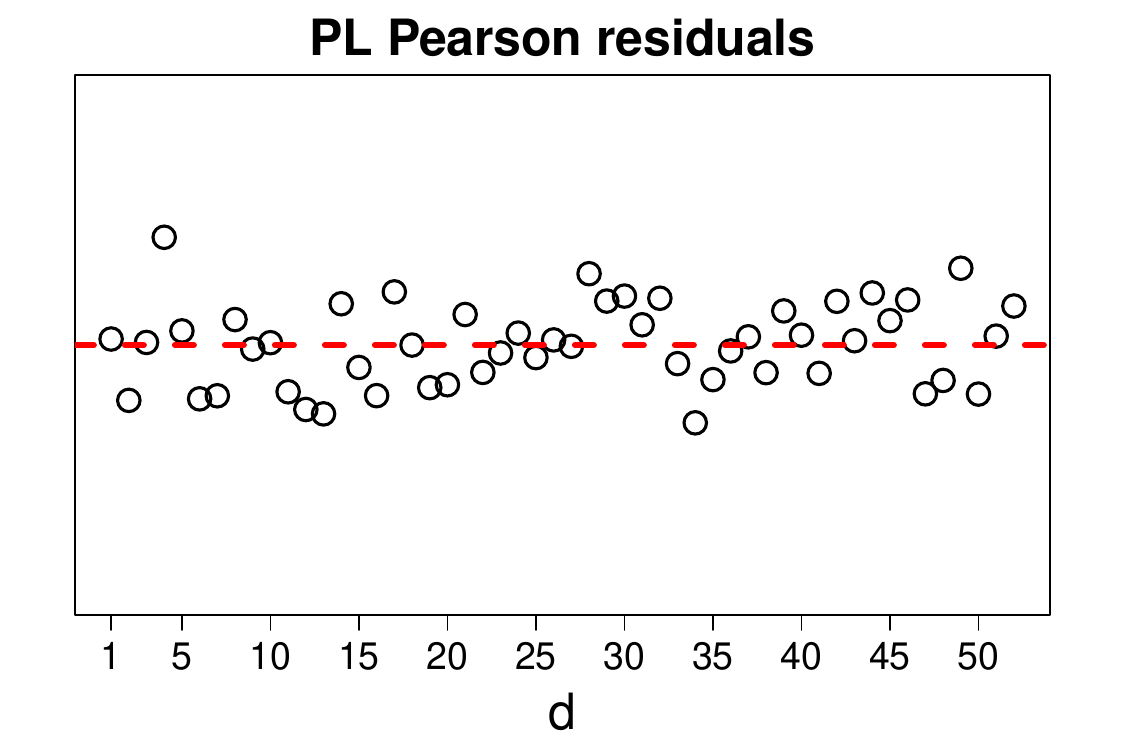}}
	\subfloat{\includegraphics[width=0.33\textwidth]{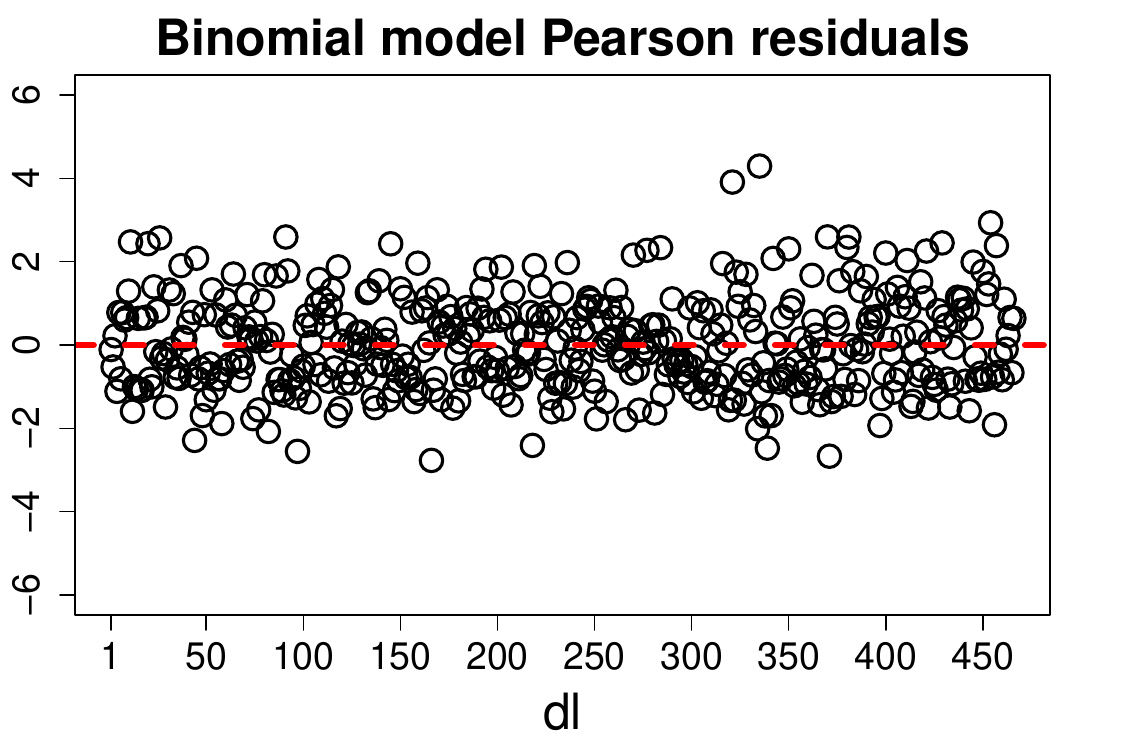}}
	\vspace*{-1em}\\
	\captionof{figure}{Diagnostic plots with Pearson residuals: (left) Poisson-gamma model, (middle) Poisson-lognormal model, (right) unit-level model.} 
	\label{fig:pearson_res}
\end{figure}

Figure \ref{fig:pearson_res} illustrates diagnostic plots with Pearson residuals for the Poisson-gamma model in the left panel, Poisson-lognormal model in the middle panel, and the binomial unit-level in the right panel. In the left and middle panel, conditionally on the random effects, $y_d$ is supposed to be distributed according to the Poisson distribution (see main document) and the Pearson residuals were calculated for each county $d$. In contrast, under the unit-level model we obtained Pearson residuals for each domain $d$ and covariate class $l$. The conditional distribution of $y_{dl}$ in the right panel is supposed to be binomial with parameters $(n_{dl}, p_{dl})$. The plots do not demonstrate any any serious departures from normality that could indicate a model misspecification. 

The estimates of the fixed effects are given in the main document. As we have already mentioned, since we do not conduct a causality analysis, they should be interpreted with caution.  Assuming area-level Poisson-gamma model, unemployment and young age are associated with higher poverty rates, whereas higher level of studies or living in a small municipality is associated with lower poverty rates. The variance parameter estimate is $\hat{\delta}^P=2.48$. Under the area-level Poisson-lognormal model the sign of education is negative, but this covariate together with living in a small municipality are not significant. In addition, an estimated value of varaince parameter is $\hat{\delta}^B=0.32$. On the other hand, under the unit-level binomial model, the signs of all covariates are positive, and the variance parameter estimate is $\hat{\delta}^B=0.35$. Even though the focus of this analysis is to provide a good prediction for poverty rates, and the changes of sings are less important for us, one would tend to favour the area-level Poisson-gamma model following the intuition about positive and negative associations in this context. The choice of this model specification is also supported by the plot with Pearson residuals which does not exhibit any issues related to model misspecification.

\begin{figure}[htb]
	\centering
	\includegraphics[width=0.9\textwidth]{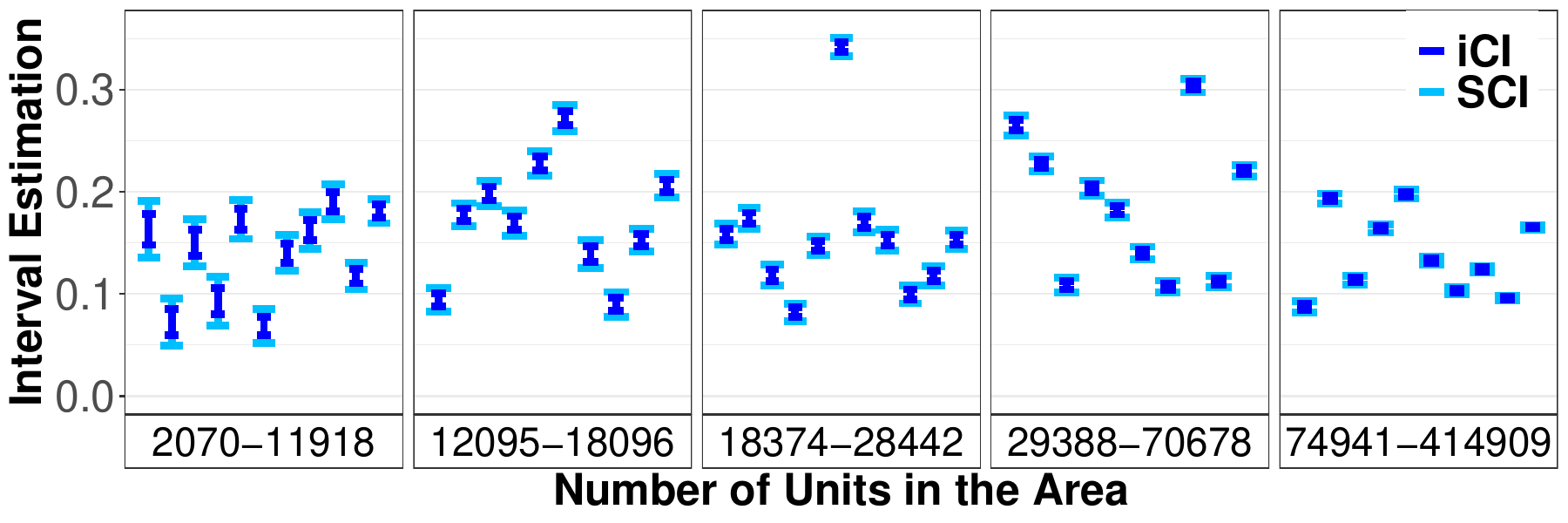}
	\vspace*{-1em}
	\captionof{figure}{$95\%$ iCI and bootstrap SCI for EBP poverty rates in counties of Galicia under the area-level Poisson-lognormal model.}
	\label{fig:plot_order_pl}
\end{figure}

\begin{figure}[htb]
	\centering
	\includegraphics[width=0.9\textwidth]{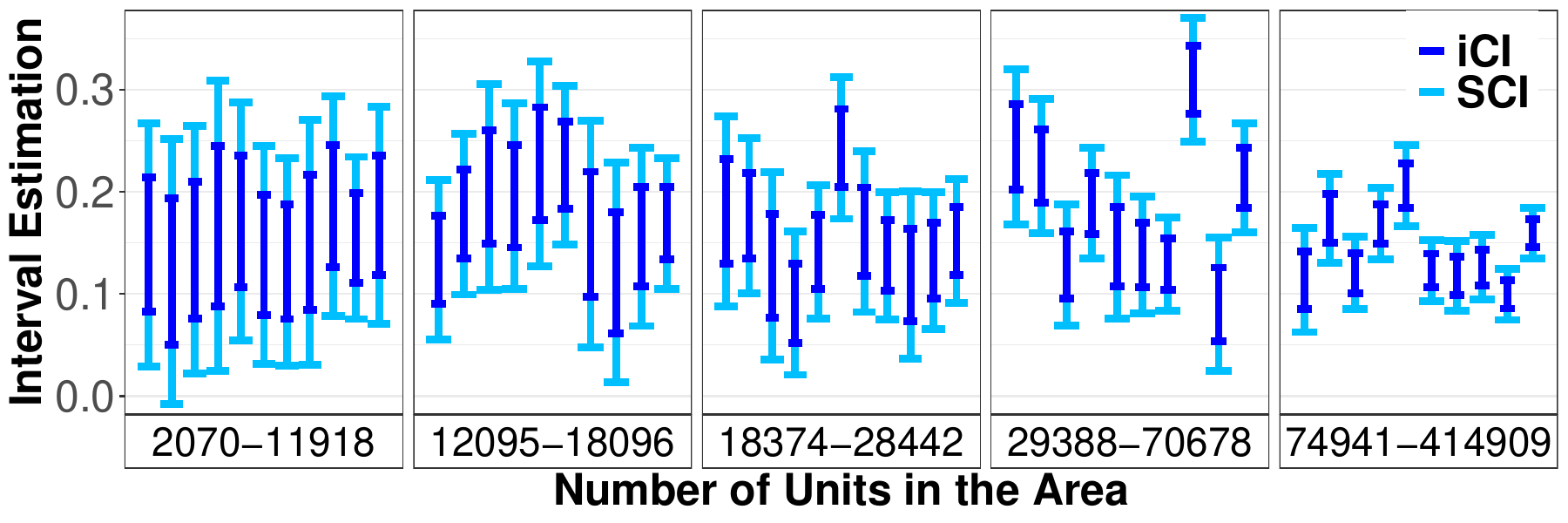}
	\vspace*{-1em}
	\captionof{figure}{$95\%$ iCI and bootstrap SCI for EBP poverty rates in counties of Galicia under the unit-level logit model.}
	\label{fig:plot_order_unit}
\end{figure}

Figures \ref{fig:plot_order_pl} and \ref{fig:plot_order_unit} present bootstrap iCI and SCI constructed using $\hat \sigma(\hat{\zeta}_d) = mse_B$ under the area-level Poisson-lognormal model and unit-level logit respectively. We can draw the same conclusions as in case of the analogous plot under the area-level Poisson-gamma model in Section 5 of the main article.

\begin{figure}[htb]
	\centering
	\includegraphics[width=0.9\textwidth]{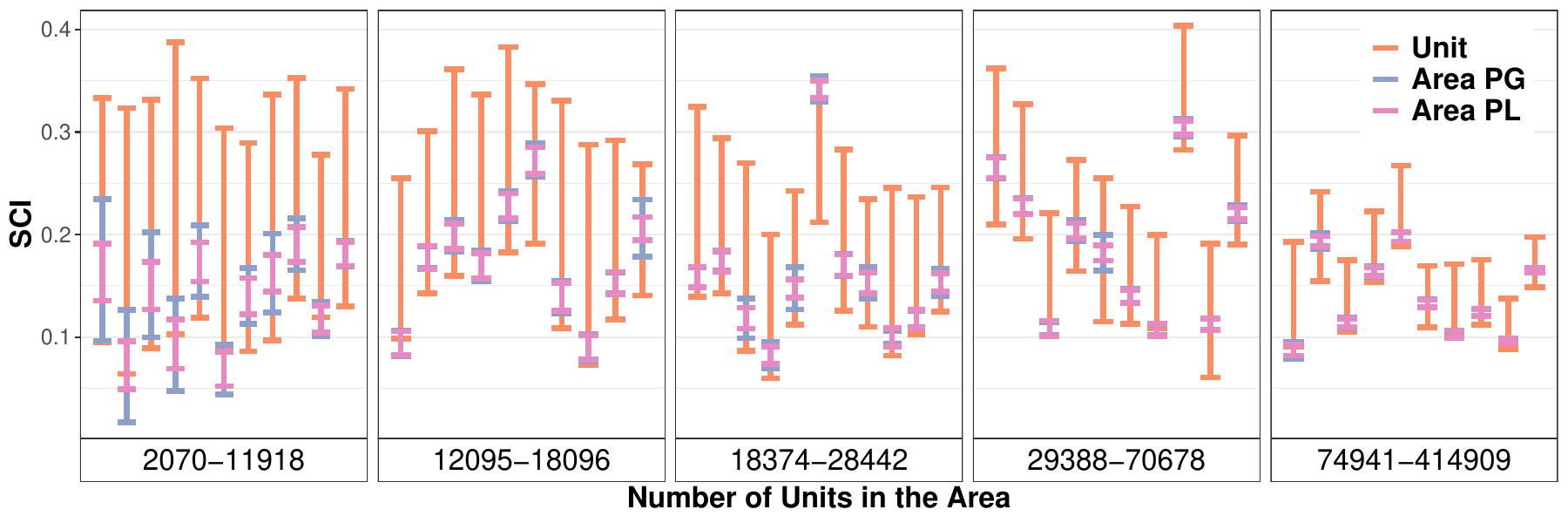}
	\vspace*{-1em}
	\captionof{figure}{$95\%$ bootstrap SCI under area- and unit-level models.}
	\label{fig:plot_order_compar}
\end{figure}

Figure \ref{fig:plot_order_compar} presents SCI bootstrap estimates for poverty rates under area-level Poisson-gamma (PG), Poisson-lognormal (PL) and unit-level binomial (Unit) model. Similarly as for the simulations in the main document, the intervals under the unit-level model are much wider due to much larger variability.

\bibliography{ref2}

\begin{thebibliography}{}

\bibitem[Abramowitz and Stegun, 1966]{abramowitz1966handbook}
Abramowitz, M. and Stegun, I. (1966).
\newblock Handbook of mathematical functions.
\newblock {\em Am. J. Phys.}, 34(2):177--177.

\bibitem[Bianconcini, 2014]{bianconcini2014asymptotic}
Bianconcini, S. (2014).
\newblock Asymptotic properties of adaptive maximum likelihood estimators in
  latent variable models.
\newblock {\em Bernoulli}, 20(3):1507--1531.

\bibitem[Boubeta et~al., 2016]{boubeta2016empirical}
Boubeta, M., Lombard{\'\i}a, M.~J., and Morales, D. (2016).
\newblock Empirical best prediction under area-level {P}oisson mixed models.
\newblock {\em Test}, 25(3):548--569.

\bibitem[Boubeta et~al., 2017]{boubeta2017poisson}
Boubeta, M., Lombard{\'\i}a, M.~J., and Morales, D. (2017).
\newblock Poisson mixed models for studying the poverty in small areas.
\newblock {\em Comput. Statist. Data Anal.}, 107(1):32--47.

\bibitem[Breslow and Clayton, 1993]{breslow1993approximate}
Breslow, N.~E. and Clayton, D.~G. (1993).
\newblock Approximate inference in generalized linear mixed models.
\newblock {\em J. Am. Stat. Assoc.}, 88(421):9--25.

\bibitem[Cameron and Trivedi, 2013]{cameron2013regression}
Cameron, A.~C. and Trivedi, P.~K. (2013).
\newblock {\em Regression analysis of count data}.
\newblock CUP.

\bibitem[Chambers et~al., 2012]{chambers2012m}
Chambers, R., Salvati, N., and Tzavidis, N. (2012).
\newblock M-quantile regression for binary data with application to small area
  estimation.
\newblock {\em Centre for Statistical and Survey Methodology, University of
  Wollongong, Working Paper}, (12):1--24.

\bibitem[Chandra et~al., 2012]{chandra2012}
Chandra, H., Chambers, R., and Salvati, N. (2012).
\newblock Small area estimation of proportions in business surveys.
\newblock {\em J. Stat. Comput. Simul.}, 82(6):783--795.

\bibitem[Chandra et~al., 2019]{chandra2019small}
Chandra, H., Chambers, R., and Salvati, N. (2019).
\newblock Small area estimation of survey weighted counts under aggregated
  level spatial model.
\newblock {\em Surv. Methodol.}, 45(1):31--59.

\bibitem[Chatterjee et~al., 2008]{chatterjee2008parametric}
Chatterjee, S., Lahiri, P., and Li, H. (2008).
\newblock Parametric bootstrap approximation to the distribution of {EBLUP} and
  related prediction intervals in linear mixed models.
\newblock {\em Ann. Statist.}, 36(3):1221--1245.

\bibitem[Chen et~al., 2015]{chen2015}
Chen, S., Jiang, J., and Nguyen, T. (2015).
\newblock {Observed Best Prediction for Small Area Counts}.
\newblock {\em J. Surv. Stat. Methodol.}, 3(2):136--161.

\bibitem[De~Bruijn, 1981]{de1981asymptotic}
De~Bruijn, N.~G. (1981).
\newblock {\em Asymptotic methods in analysis}.
\newblock Courier Corporation.

\bibitem[DiCiccio and Efron, 1996]{diciccio1996}
DiCiccio, T.~J. and Efron, B. (1996).
\newblock Bootstrap confidence intervals.
\newblock {\em Statist. Sci.}, 11(3):189--228.

\bibitem[Erciulescu and Fuller, 2014]{erciulescu2014parametric}
Erciulescu, A.~L. and Fuller, W.~A. (2014).
\newblock Parametric bootstrap procedures for small area prediction variance.
\newblock In {\em Proceedings of the Survey Research Methods Section}.

\bibitem[Franco and Bell, 2015]{franco2015borrowing}
Franco, C. and Bell, W.~R. (2015).
\newblock Borrowing information over time in binomial/logit normal models for
  small area estimation.
\newblock {\em Stat. Trans.}, 16(4):563--584.

\bibitem[Ganesh, 2009]{ganesh2009simultaneous}
Ganesh, N. (2009).
\newblock Simultaneous credible intervals for small area estimation problems.
\newblock {\em J. Multivariate Anal.}, 100(8):1610--1621.

\bibitem[Ghosh et~al., 1998]{ghosh1998}
Ghosh, M., Natarajan, K., Stroud, T. W.~F., and Carlin, B.~P. (1998).
\newblock Generalized linear models for small-area estimation.
\newblock {\em J. Am. Stat. Assoc.}, 93(441):273--282.

\bibitem[Hall and Maiti, 2006]{hall2006parametric}
Hall, P. and Maiti, T. (2006).
\newblock On parametric bootstrap methods for small area prediction.
\newblock {\em J. R. Statist. Soc. B}, 68(2):221--238.

\bibitem[Hidiroglou and You, 2016]{hidiroglou2016comparison}
Hidiroglou, M.~A. and You, Y. (2016).
\newblock Comparison of unit level and area level small area estimators.
\newblock {\em Surv. Methodol.}, 42(42):41--61.

\bibitem[Hobza and Morales, 2016]{hobza2016empirical}
Hobza, T. and Morales, D. (2016).
\newblock Empirical best prediction under unit-level logit mixed models.
\newblock {\em J. Off. Stat.}, 32(3):661--692.

\bibitem[Hobza et~al., 2018]{hobza2018small}
Hobza, T., Morales, D., and Santamar{\'\i}a, L. (2018).
\newblock Small area estimation of poverty proportions under unit-level
  temporal binomial-logit mixed models.
\newblock {\em Test}, 27(2):270--294.

\bibitem[Huber et~al., 2004]{huber2004estimation}
Huber, P., Ronchetti, E., and Victoria-Feser, M.-P. (2004).
\newblock Estimation of generalized linear latent variable models.
\newblock {\em J. R. Statist. Soc. B}, 66(4):893--908.

\bibitem[Huber, 1964]{huber1964robust}
Huber, P.~J. (1964).
\newblock Robust estimation of a location parameter.
\newblock {\em Ann. Math. Statist.}, 35(1):73--101.

\bibitem[Jiang, 1998a]{jiang1998asymptotic}
Jiang, J. (1998a).
\newblock Asymptotic properties of the empirical {BLUP} and {BLUE} in mixed
  linear models.
\newblock {\em Stat. Sin.}, 8(1):861--885.

\bibitem[Jiang, 1998b]{jiang1998consistent}
Jiang, J. (1998b).
\newblock Consistent estimators in generalized linear mixed models.
\newblock {\em J. Am. Stat. Assoc.}, 93(442):720--729.

\bibitem[Jiang, 2003]{jiang2003empirical}
Jiang, J. (2003).
\newblock Empirical best prediction for small-area inference based on
  generalized linear mixed models.
\newblock {\em J. Stat. Plan. Inference}, 111(1-2):117--127.

\bibitem[Jiang, 2007]{jiang2007linear}
Jiang, J. (2007).
\newblock {\em Linear and generalized linear mixed models and their
  applications}.
\newblock Springer Science \& Business Media.

\bibitem[Jiang, 2013]{jiang2013subset}
Jiang, J. (2013).
\newblock The subset argument and consistency of {MLE} in {GLMM}: {A}nswer to
  an open problem and beyond.
\newblock {\em Ann. Statist.}, 41(1):177--195.

\bibitem[Jiang and Lahiri, 2001]{jiang2001empirical}
Jiang, J. and Lahiri, P. (2001).
\newblock Empirical best prediction for small area inference with binary data.
\newblock {\em Ann. Inst. Stat. Math.}, 53(2):217--243.

\bibitem[Joe, 2008]{joe2008accuracy}
Joe, H. (2008).
\newblock Accuracy of {L}aplace approximation for discrete response mixed
  models.
\newblock {\em Comput. Statist. Data Anal.}, 52(12):5066--5074.

\bibitem[Kramlinger et~al., 2018]{kramlinger2018marginal}
Kramlinger, P., Krivobokova, T., and Sperlich, S. (2018).
\newblock Marginal and conditional multiple inference in linear mixed models.
\newblock {\em arXiv:1812.09250}.

\bibitem[Lawless, 1987]{lawless1987negative}
Lawless, J.~F. (1987).
\newblock Negative binomial and mixed {P}oisson regression.
\newblock {\em Can. J. Stat.}, 15(3):209--225.

\bibitem[Leadbetter et~al., 2012]{leadbetter2012extremes}
Leadbetter, M.~R., Lindgren, G., and Rootz{\'e}n, H. (2012).
\newblock {\em Extremes and related properties of random sequences and
  processes}.
\newblock Springer Science \& Business Media.

\bibitem[Lele et~al., 2010]{lele2010}
Lele, S.~R., Nadeem, K., and Schmuland, B. (2010).
\newblock Estimability and likelihood inference for generalized linear mixed
  models using data cloning.
\newblock {\em J. Am. Stat. Assoc.}, 105(492):1617--1625.

\bibitem[L{\'o}pez-Vizca{\'\i}no et~al., 2015]{lopez2015small}
L{\'o}pez-Vizca{\'\i}no, E., Lombard{\'\i}a, M.~J., and Morales, D. (2015).
\newblock Small area estimation of labour force indicators under a multinomial
  model with correlated time and area effects.
\newblock {\em J. R. Statist. Soc. A}, 178(3):535--565.

\bibitem[Molina et~al., 2007]{molina2007small}
Molina, I., Saei, A., and Lombard{\'\i}a, M.~J. (2007).
\newblock Small area estimates of labour force participation under a
  multinomial logit mixed model.
\newblock {\em J. R. Statist. Soc. A}, 170(4):975--1000.

\bibitem[Namazi-Rad and Steel, 2015]{namazi2015level}
Namazi-Rad, M.-R. and Steel, D. (2015).
\newblock What level of statistical model should we use in small area
  estimation?
\newblock {\em Aust NZ J. Stat.}, 57(2):275--298.

\bibitem[Naylor and Smith, 1982]{naylor1982applications}
Naylor, J.~C. and Smith, A.~F. (1982).
\newblock Applications of a method for the efficient computation of posterior
  distributions.
\newblock {\em J. R. Statist. Soc. C}, 31(3):214--225.

\bibitem[Pinheiro and Bates, 1995]{pinheiro1995approximations}
Pinheiro, J.~C. and Bates, D.~M. (1995).
\newblock Approximations to the log-likelihood function in the nonlinear
  mixed-effects model.
\newblock {\em J. Comput. Graph. Stat.}, 4(1):12--35.

\bibitem[Pratesi, 2016]{pratesi2016analysis}
Pratesi, M. (2016).
\newblock {\em Analysis of poverty data by small area estimation}.
\newblock John Wiley \& Sons.

\bibitem[Rabe-Hesketh et~al., 2005]{rabe2005maximum}
Rabe-Hesketh, S., Skrondal, A., and Pickles, A. (2005).
\newblock Maximum likelihood estimation of limited and discrete dependent
  variable models with nested random effects.
\newblock {\em J. Econom.}, 128(2):301--323.

\bibitem[{Reluga} et~al., 2019]{reluga2019}
{Reluga}, K., {Lombard{\'\i}a}, M.~J., and {Sperlich}, S.~A. (2019).
\newblock {Simultaneous inference for mixed and small area parameters}.
\newblock {\em arXiv:1903.02774}.

\bibitem[Romano et~al., 2008]{romano2008control}
Romano, J.~P., Shaikh, A.~M., and Wolf, M. (2008).
\newblock Control of the false discovery rate under dependence using the
  bootstrap and subsampling.
\newblock {\em Test}, 17(3):417--442.

\bibitem[Saei and Taylor, 2012]{saei2012labour}
Saei, A. and Taylor, A. (2012).
\newblock Labour force status estimates under a bivariate random components
  model.
\newblock {\em J. Indian. Soc. Agric. Stat.}, 66(1):187--201.

\bibitem[Scealy, 2010]{scealy2010small}
Scealy, J. (2010).
\newblock Small area estimation using a multinomial logit mixed model with
  category specific random effects.
\newblock {\em Research paper, Australian Bureau of Statistics}.

\bibitem[Torabi, 2012]{torabi2012likelihood}
Torabi, M. (2012).
\newblock Likelihood inference in generalized linear mixed models with two
  components of dispersion using data cloning.
\newblock {\em Comput. Statist. Data Anal.}, 56(1):4259--4265.

\bibitem[Tuerlinckx et~al., 2006]{tuerlinckx2006statistical}
Tuerlinckx, F., Rijmen, F., Verbeke, G., and De~Boeck, P. (2006).
\newblock Statistical inference in generalized linear mixed models: A review.
\newblock {\em Br. J. Math. Stat. Psychol.}, 59(2):225--255.

\bibitem[Tzavidis et~al., 2015]{tzavidis2015robust}
Tzavidis, N., Ranalli, M.~G., Salvati, N., Dreassi, E., and Chambers, R.
  (2015).
\newblock Robust small area prediction for counts.
\newblock {\em Stat. Methods. Med. Res.}, 24(3):373--395.

\end{thebibliography}

\end{document}